%% file: main.tex
\documentclass[a4paper,onecolumn,superscriptaddress,11pt,%
				unpublished,%
				allowfontchageintitle,%
				]{quantumarticle}
\pdfoutput=1

\input{preamble.tex}

\begin{document}
\title{\LARGE ZX-calculus for the working quantum computer scientist}
\date{\today}

\author{John van de Wetering}
\email{john@vdwetering.name}
\homepage{http://vdwetering.name}
\affiliation{Radboud Universiteit Nijmegen}
\affiliation{Oxford University}

\begin{abstract} 
  The ZX-calculus is a graphical language for reasoning about quantum computation that has recently seen an increased usage in a variety of areas such as quantum circuit optimisation, surface codes and lattice surgery, measurement-based quantum computation, and quantum foundations.
  The first half of this review gives a gentle introduction to the ZX-calculus suitable for those familiar with the basics of quantum computing. The aim here is to make the reader comfortable enough with the ZX-calculus that they could use it in their daily work for small computations on quantum circuits and states.
  The latter sections give a condensed overview of the literature on the ZX-calculus. We discuss Clifford computation and graphically prove the Gottesman-Knill theorem, we discuss a recently introduced extension of the ZX-calculus that allows for convenient reasoning about Toffoli gates, and we discuss the recent completeness theorems for the ZX-calculus that show that, in principle, all reasoning about quantum computation can be done using ZX-diagrams.
  Additionally, we discuss the categorical and algebraic origins of the ZX-calculus and we discuss several extensions of the language which can represent mixed states, measurement, classical control and higher-dimensional qudits.
\end{abstract} 

\tableofcontents

\listoffigures

\maketitle

\section{Introduction}

A \emph{ZX-diagram} is a graphical representation of a linear map between qubits reminiscent of a  conventional quantum circuit diagram. For instance, we can write the circuit on the left below as the ZX-diagram on the right:
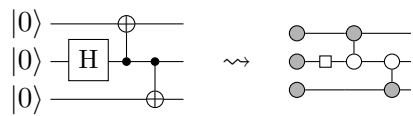
\[\input{ghz-circuit.tikz} \quad \rightsquigarrow \quad \input{ghz-circuit-zx-solo.tikz}\]

ZX-diagrams come equipped with a compact set of graphical rewrite rules that allow us to diagrammatically reason about linear maps. For instance, we can prove diagrammatically that the above circuit implements a GHZ state:
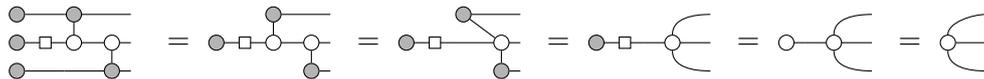
\begin{equation*}
\input{ghz-circuit-zx-nolabel.tikz}
\end{equation*}
The graphical language consisting of ZX-diagrams and their rewrite rules is called the \emph{ZX-calculus}. 

Since the introduction of the ZX-calculus in 2008 by Coecke and Duncan~\cite{CD1,CD2}, the ZX-calculus has been used for a wide variety of tasks related to quantum computing.
One of the first use-cases was in measurement-based quantum computation (MBQC).
Graph states have a simple representation as ZX-diagrams~\cite{DP2} and hence the one-way model and related protocols can be described easily~\cite{kissinger2017MBQC,wetering-gflow}. 
In more recent years the ZX-calculus has been used successfully in quantum circuit optimisation~\cite{cliffsimp,deBeaudrap2020Techniques,kissinger2019tcount,phaseGadgetSynth}; quantum circuit equality validation~\cite{kissinger2019tcount,Lemonnier2020hypergraph}; and as a tool for reasoning about surface code quantum computing~\cite{horsman2011quantum} and lattice surgery~\cite{deBeaudrap2020Paulifusion,Gidney2019efficientmagicstate,horsman2017surgery,hanks2019effective}. 
It has also seen use in the design and verification of quantum error correcting codes~\cite{chancellor2016coherent,duncan2013verifying,EPTCS266.10,SZXCalculus}; elucidating the structure of SAT and \#SAT problems~\cite{debeaudrap2020tensor}; reasoning about certain condensed matter systems~\cite{east2020akltstates}; and in describing natural language processing tasks on a quantum computer~\cite{coecke2020foundations}.
In 2017 it was shown that certain rule sets for the ZX-calculus are \emph{complete}, meaning that any equality between linear maps can be proven entirely diagrammatically~\cite{HarnyCompleteness,SimonCompleteness,HarnyAmarCompleteness}. This means that in principle all reasoning about quantum computation can be done inside the ZX-calculus.

The goal of this paper is threefold.
The first is that after reading this paper the reader should be comfortable enough with the notation and conventions of the ZX-calculus to be able to read any research paper using it. The second goal is to serve as a review of the literature on the ZX-calculus. 
There are well over a hundred papers and theses that use the ZX-calculus%
\footnote{A list of all publications involving the ZX-calculus is maintained at \url{https://zxcalculus.com/publications}.}, 
yet no concise overview of what has been done and what is known exists.
Note that in this review we focus on general results regarding the ZX-calculus. We will not go into detail on domain-specific results except where illustrative (such as in Section~\ref{sec:clifford} where we demonstrate the simplification of Clifford circuits).

The third and perhaps most ambitious goal is to make the reader comfortable enough with ZX-diagrams that it becomes another tool in their arsenal for their day-to-day reasoning about quantum computing. 
Most researchers will regularly use some ad hoc reasoning about quantum circuits and matrices as part of their work. Many of those computations can be replaced by the ZX-calculus in a way that is faster, less error-prone and more amenable to automation.
To aid in accomplishing this goal, this paper will spend quite some time explaining how to construct ZX-diagrams for common operators, explain how the rewriting works, and what the motivation is behind certain rewrites.

\subsection{Overview of the paper}

This paper is divided into two parts.
Sections~\ref{sec:circuits-vs-diagrams}--\ref{sec:example-derivations} give a general introduction to ZX-diagrams and the ZX-calculus. This part should be read sequentially.
Sections~\ref{sec:clifford}--\ref{sec:extensions} each cover a different topic and can be read more or less independently from each other.

In Section~\ref{sec:circuits-vs-diagrams} we informally introduce ZX-diagrams and graphical rewriting by comparing it to standard quantum circuit notation.
Then in Section~\ref{sec:zxdiagrams} we introduce ZX-diagrams in full generality.
Section~\ref{sec:zx-calculus} introduces the rewrite rules of the ZX-calculus. These rules are introduced one by one including motivation and examples. The entire set of rules is presented at the end in Figure~\ref{fig:zx-rules}.
Section~\ref{sec:example-derivations} serves as a repository of examples of how the ZX-calculus can be used in practice, covering for instance the simplification of quantum circuits, and the proving of correctness of protocols like quantum teleportation.

The review portion of the paper starts with Section~\ref{sec:clifford} wherein we discuss stabiliser states and Clifford computation using the ZX-calculus. We give a proof sketch that the ZX-calculus is complete for Clifford computation and we give a diagrammatic proof of the Gottesman-Knill theorem, i.e.~that Clifford computation can be classically efficiently simulated. We furthermore show how measurement-based quantum computation can be represented in the ZX-calculus and how ZX-diagrams can be transformed back into circuits.

In Section~\ref{sec:zx-category} we review the origins of the ZX-calculus in categorical quantum mechanics and we recall formalisations of several concepts that are commonly used in an intuitive way.

Section~\ref{sec:ZH} discusses Toffoli gates and other multi-linear Boolean logic gates and how they can be represented in the ZX-calculus. To do this, we recall the \emph{ZH-calculus}, an alternative graphical calculus, that we can view as an extension of the ZX-calculus.

A question that was the topic of many early works on the ZX-calculus is whether the ZX-calculus is \emph{complete}. That is, whether the rules of the ZX-calculus suffice to prove any equation between linear maps that is true. This question was settled in a series of papers published in 2017-2018. We discuss those results in Section~\ref{sec:completeness}.

Finally, in Section~\ref{sec:extensions} we discuss several ways in which the domain of the ZX-calculus can be extended. We recall how adding a generator to represent the process of discarding allows the representation of mixed and decohered quantum states, and we discuss how the interaction between classical and quantum systems can be represented by `doubling' the quantum process. We end the section with a look into how the ZX-calculus, that acts solely on 2-dimensional qubits, can be modified to function as a language for higher-dimensional qudits.

The appendices give compact overviews of the generators and (derived) rewrite rules of the ZX-calculus.

\subsection{Tools and resources}

This text is intended for readers already comfortable with the basics of quantum computing: circuit notation, qubits, Pauli matrices, phase gates, CNOTs, Clifford unitaries, etc. If the reader is at an earlier stage of learning quantum computing, the author highly recommends the book \emph{Picturing Quantum Processes} by Coecke and Kissinger~\cite{CKbook}. This book assumes only basic familiarity with linear algebra and mathematical reasoning and builds up all of quantum mechanics and quantum computing with graphical tools that concludes with the ZX-calculus.

If after reading this review paper the reader wishes to learn more about the ZX-calculus there are a couple of ways to proceed. The ZX-calculus website \url{https://zxcalculus.com} collects a number of resources surrounding the ZX-calculus. In particular, its publications page\footnote{\url{https://zxcalculus.com/publications}} is a searchable list of all the papers, theses and preprints that have appeared that use the ZX-calculus or associated graphical languages.

There is some software that assists in working with ZX-diagrams. 
\emph{Quantomatic} was the first software that could handle rewriting of ZX-diagrams~\cite{kissinger2015quantomatic}. It is however no longer actively developed, and because it is intended as a general tool for graphical languages, and not just the ZX-calculus, it can be somewhat cumbersome to use. 
A more modern software library is \emph{PyZX}, an open-source Python library that has a number of tools for the conversion between quantum circuits and ZX-diagrams and the automated simplification of ZX-diagrams~\cite{pyzx}\footnote{Disclaimer: the author is the creator of PyZX and a main developer of it.}.
Many people writing papers containing ZX-diagrams use \emph{Tikzit} to typeset them.\footnote{\url{https://tikzit.github.io/}} 
Tikzit generates output that can be compiled in Latex with the \emph{tikz} library and can save many hours in tedious tikz manipulation.
The reader can for instance take a look at the latex source for this paper (available via the ArXiv) in order to see how the diagrams are constructed in tikz.

\section{Quantum circuits vs ZX-diagrams}\label{sec:circuits-vs-diagrams}

In this section we give a brief informal introduction to ZX-diagrams and graphical rewriting. This will serve as motivation for the more in-depth construction of general ZX-diagrams in Section~\ref{sec:zxdiagrams}.

\subsection{From quantum circuits to ZX-diagrams}

Quantum circuit notation has proven very useful for presenting short quantum computations. They usually consist of some horizontal wires, with some \emph{gates} attached to these wires:
\ctikzfig{example-circuit}
The quantum computing community has mostly agreed on the naming conventions of most standard gates: $X$, $Y$, and $Z$ are the corresponding Pauli matrices; The $S$ gate is the $\frac\pi2$ Z-rotation; the $T$ gate is the $\frac\pi4$ Z-rotation; adding a dagger $\dagger$ gives the inverse gate; the $H$ gate is the Hadamard gate; a black dot is a control for some unitary; and a `plus' is a NOT gate, meaning that a black dot connected to a a plus is a CNOT gate. Furthermore, gate names like $U$ are used to denote some custom or black-box unitary that needs to be specified in the paper.

There are some common extensions to this notation, which allow the specification of ancillae, measurement, classical control or post-selection, but some elements continue to be enforced: each horizontal line represents a single qubit, and time flows from left to right.

The translation from the quantum circuit above (minus the `black-box' $U$) to its equivalent ZX-diagram is straightforward. Indeed, we can simply use the following translation table:
\begin{equation}\label{eq:translation-table}
\input{translation-table.tikz}
\end{equation}
And hence we arrive at the following ZX-diagram (again, not including $U$):
\ctikzfig{example-circuit-ZX}
It might not be immediately clear what the benefit of this translation is. Why would this notation be better than the standard quantum circuit notation?
A first hint becomes apparent when we consider the well-known fact that \emph{Z phase gates commute through the control of a CNOT}. For instance, using circuit notation we have the following equality:
\ctikzfig{commute-T-CNOT}
When dealing with quantum circuits one has to simply remember that this is the case. For this one case this is not a problem, but there are of course dozens of useful circuit identities.

The same identity in the ZX-calculus becomes:
\begin{equation}\label{eq:commute-T-CNOT}
\input{commute-T-CNOT-ZX.tikz}
\end{equation}

Although it might not look very different, it is actually part of a simple yet powerful pattern that holds in the ZX-calculus: \emph{dots of the same colour commute through one another}. Here the `colour' of a dot is either white (as is the case for the S gate, the T gate, and the control of a CNOT), or grey (as is the case for the X gate or the target of a CNOT).
Hence, instead of remembering some set of ad hoc circuit identities we can just remember the simple rule that if the dots have the same color, we can move them past one another.

This is in fact not the full story. An even more fundamental set of identities is true in the ZX-calculus: whenever we have two dots of the same color that are connected by a wire we can \emph{fuse} them:
\begin{equation}
\input{commute-T-CNOT-ZX-fuse.tikz}
\end{equation}
This last diagram does not correspond to any standard type of gate, but it is useful: it immediately gives the truth of Eq.~\eqref{eq:commute-T-CNOT} because the right-hand side is manifestly symmetric.

\subsection{States in the ZX-calculus}
We can represent not just unitaries in the ZX-calculus, but any linear map between qubits. Let us demonstrate this with some simple calculations involving single-qubit states.
We have the following evident equality in circuit notation:
\begin{equation}\label{eq:ket0-CNOT}
\input{ket0-CNOT.tikz}
\end{equation}
In words: if we apply a qubit prepared in the $\ket{0}$ computational basis state to the control of a CNOT gate, this is the same as doing nothing.
In the ZX-calculus a qubit prepared in the $\ket{0}$ state is represented by a grey dot:
\begin{equation}\label{ket0-ZX}
\input{ket0-ZX.tikz}
\end{equation}
Hence, the left-hand side of Eq.~\eqref{eq:ket0-CNOT} becomes:
\begin{equation}
 \input{ket0-CNOT-ZX.tikz}
\end{equation}
We can prove Eq.~\eqref{eq:ket0-CNOT} by applying a following simple rule in the ZX-calculus: \emph{a dot with a single wire copies through a dot of the opposite colour}:
\begin{equation}\label{eq:ket0-copy-ZX}
	\input{ket0-copy-ZX.tikz}
\end{equation}
Indeed, we can now calculate:
\begin{equation}\label{eq:ket0-copy-proof}
 \input{ket0-CNOT-ZX-pf.tikz}
\end{equation}
The first equality is just~\eqref{eq:ket0-copy-ZX}. Note that even though~\eqref{eq:ket0-copy-ZX} has two output wires, we are still `allowed' to apply the rewrite rule in this case. This is because wires have no directionality to them in a ZX-diagram. We discuss this property in more detail in Section~\ref{sec:symmetries}. 

The second equality in~\eqref{eq:ket0-copy-proof} applies a version of the rule we discussed in the previous section that \emph{dots of the same colour fuse}. The last rule is new, but simple enough: \emph{dots with 2 wires can be removed}:
\begin{equation}
\input{id-removal.tikz}
\end{equation}
The reader might also be aware of the following circuit identity:
\begin{equation}\label{eq:ketplus-CNOT}
\input{ketplus-CNOT.tikz}
\end{equation}
Here the $\ket{+}$ state denotes the standard $X$ basis state $\ket{+}=\frac{1}{\sqrt{2}}(\ket{0}+\ket{1})$.
The $\ket{+}$ state also has a simple representation in the ZX-calculus:
\begin{equation}\label{ketplus-ZX}
\input{ketplus-ZX.tikz}
\end{equation}
We can now give a proof of Eq.~\eqref{eq:ketplus-CNOT} in the ZX-calculus entirely analogously to that of Eq.\eqref{eq:ket0-CNOT}:
\begin{equation}
 \input{ketplus-CNOT-ZX-pf.tikz}
\end{equation}
Again, a dot with a single wire copies through a dot of the opposite colour, dots of the same colour fuse, and dots with 2 wires can be removed.
That we get essentially the same proof here is an instance of a more general principle that holds in the ZX-calculus: \emph{any identity continues to hold when the colours (white and grey) are interchanged}.
We discuss this property in more detail in the next section.

\subsection{Hadamards and colour changing}\label{sec:hadamard-colour}

As shown in the translation table~\eqref{eq:translation-table}, we have some special notation for the Hadamard gate in ZX-diagrams:
\begin{equation}
\input{had-ZX.tikz}
\end{equation}
We give it a distinct shape (a square instead of a circle) to denote that it behaves differently than the white and grey dots we have encountered in the previous sections. Indeed, while the dots can have any number of wires coming in and out of them, a Hadamard, being a single-qubit unitary, always has exactly two wires. There are two important identities associated to the Hadamard gate in the ZX-calculus.

The first shouldn't come as a surprise: the Hadamard gate is self-inverse, $HH=\id$, and hence two boxes in a row should cancel.
\begin{equation}\label{eq:had-had-cancel}
	\input{had-had-cancel.tikz}
\end{equation}
The second expresses a generalisation of a well-known property of the Hadamard gate: conjugating the Pauli $Z$ with Hadamards we get the $X$ gate: $HZH = X$. In terms of ZX-diagrams:
\begin{equation}
\input{HZH-X.tikz}
\end{equation}
This generalises to the following identity:
\begin{equation}\label{eq:colour-change}
	\input{colour-change.tikz}
\end{equation}
Here $\alpha$ can be any phase $\alpha\in \R$.
The `$\cdots$' mean that there can be any number of inputs and outputs. Whenever we write an equation with such a varying number of inputs and outputs we are assuming that the number of inputs on the left-hand side matches the number of inputs on the right-hand side and similarly for the outputs (as otherwise the identity would not make sense). The number of inputs or outputs can also be zero, so we also get for instance:
\begin{equation}
\input{ketplus-had-ket0.tikz}
\end{equation}
Recalling the definition of $\ket{+}$ and $\ket{0}$ in ZX (Eqs.~\eqref{ket0-ZX} and \eqref{ketplus-ZX}) we see that this simply states that $H\ket{+} = \ket{0}$.

Using the self-inverse rule Eq.~\eqref{eq:had-had-cancel} we can also rephrase the `Hadamard conjugation' rule of Eq.~\eqref{eq:colour-change} as a commutation rule that tells us how to `push' a Hadamard gate through a dot:
\begin{equation}\label{eq:colour-change-push}
	\input{colour-change-push.tikz}
\end{equation}
In a slogan: \emph{a Hadamard gate is commuted through a dot by changing its colour}.

As an application of the rules concerning Hadamards we can find a compact representation of a CZ gate in the ZX-calculus.
Recall that a CZ gate is equal to a CNOT gate conjugated by Hadamards on the target qubit:
\begin{equation}
\input{CZ-as-CNOT.tikz}
\end{equation}
Translating the right-hand side to a ZX-diagram and simplifying using the Hadamard colour-change rule and the self-inverse rule we find:
\begin{equation}\label{eq:CZ-in-ZX}
	\input{CZ-as-CNOT-ZX.tikz}
\end{equation}
What is interesting about this is that the rightmost diagram is symmetric in the two qubits, as the CZ gate indeed is, and that we did not need to introduce any new generators or concepts to the ZX-calculus to represent it.
In fact, we already have all ingredients to represent any linear map between qubits: ZX-diagrams are \emph{universal} (see Section~\ref{sec:universality}).

\section{ZX-diagrams}\label{sec:zxdiagrams}

In this section we will introduce the building blocks of ZX-diagrams and discuss some of their main properties.
We start by defining the basic generators of ZX-diagrams, the \emph{spiders}, in Section~\ref{sec:spiders}. Then in Section~\ref{sec:composition} we will see how to compose these. We discuss the symmetries of ZX-diagrams in Section~\ref{sec:symmetries} which will allow us to treat ZX-diagrams as undirected graphs. Then Sections~\ref{sec:scalars}--\ref{sec:universality} discuss various topics culminating in a proof that ZX-diagrams are in fact universal and hence can represent any linear map between qubits.

\subsection{Spiders}\label{sec:spiders}

The first type of generator is what we referred to as a `white dot' in the previous section. The more common name for this is the \emph{Z-spider}. A Z-spider can have any number of inputs (wires coming in from the left) and outputs (wires coming out of the right) and has the following interpretation as a linear map:
\begin{equation}\label{eq:Z-spider-def}
\input{Zsp-a.tikz} \ \ :=\ \ \ketbra{0 \cdots 0}{0 \cdots 0} + e^{i \alpha} \ketbra{1 \cdots 1}{1 \cdots 1},
\end{equation}
Why it is called a `spider' should be clear from the way the diagram looks. This is the Z-spider because it is defined with respect to the eigenbasis of the $Z$ matrix, $\ket{0}$ and $\ket{1}$.

Similarly, what we referred to as a `grey dot', is more commonly called an \emph{X-spider}. This is defined as above, but with the $Z$ eigenbasis replaced by the $X$ eigenbasis~$\ket{+}$ and~$\ket{-}$:
\begin{equation}
\input{Xsp-a.tikz} \ \ :=\ \ \ketbra{+ \cdots +}{+ \cdots +} + e^{i \alpha} \ketbra{- \cdots -}{- \cdots -}.
\end{equation}

As it is instructive, let us explicitly write out some of the matrices that these spiders are equal to.
Recall that:
\begin{equation*}
	\ket{0} \ = \ \begin{pmatrix}1\\ 0 \end{pmatrix} \quad\quad \ket{1} \ = \ \begin{pmatrix}0\\ 1 \end{pmatrix} \quad\quad \ket{+} \ = \ \frac{1}{\sqrt{2}}\begin{pmatrix}1\\ 1 \end{pmatrix} \quad\quad \ket{-} \ = \ \frac{1}{\sqrt{2}}\begin{pmatrix}1\\ -1 \end{pmatrix}
\end{equation*}
For the 1-input, 1-output $Z$-spider with a phase $\alpha$ we hence have:
\begin{equation}\label{eq:Z-a}
\input{Z-a.tikz} \ \ =\ \ \ketbra{0}{0} + e^{i \alpha} \ketbra{1}{1}
\ \ = \ \ \begin{pmatrix} 1 & 0 \\ 0& 0\end{pmatrix}\ +\ \begin{pmatrix} 0 & 0 \\ 0& e^{i\alpha}\end{pmatrix}\ \  =\ \  \begin{pmatrix} 1 & 0 \\ 0& e^{i\alpha}\end{pmatrix}
\end{equation}
This matrix is the $R_Z(\alpha)$ phase gate that rotates the Bloch sphere by an angle $\alpha$ over the $Z$ axis. 
In particular, for $\alpha=\pi$ we get the $Z$ Pauli matrix, which indeed matches with our translation table~\eqref{eq:translation-table}.
We can do a similar calculation for the 1-input, 1-output $X$-spider with a phase $\alpha$:
\begin{equation}\label{eq:X-a}
\input{X-a.tikz} \ \ =\ \ \ketbra{+}{+} + e^{i \alpha} \ketbra{-}{-}
\ \ = \ \ \frac12\begin{pmatrix} 1 & 1 \\ 1& 1\end{pmatrix}\ +\ \frac12 e^{i\alpha}\begin{pmatrix} 1 & -1 \\ -1& 1\end{pmatrix}\ \  =\ \  \frac12\begin{pmatrix} 1+e^{i\alpha}&   1-e^{i\alpha} \\ 1-e^{i\alpha} & 1+e^{i\alpha}\end{pmatrix}
\end{equation}
This is the familiar $R_X(\alpha)$ phase gate, and for $\alpha=\pi$ we get the Pauli $X$ matrix.

It is customary to drop the label inside the spider when the phase $\alpha$ is $0$:
\[\hfill \input{Zsp.tikz} \ \ =\ \ \ketbra{0 \cdots 0}{0 \cdots 0} + \ketbra{1 \cdots 1}{1 \cdots 1} \hfill\]
\[\hfill \input{Xsp.tikz} \ \ =\ \ \ketbra{+ \cdots +}{+ \cdots +} + \ketbra{- \cdots -}{- \cdots -} \hfill\]

In particular, we have:
\begin{equation}
\input{Z-id.tikz}\ \ = \ \ \input{X-id.tikz} \ \ = \ \ \begin{pmatrix}1&0\\0&1\end{pmatrix}
\end{equation}
In a ZX-diagram we present an identity matrix like the one above by a empty piece of wire, and hence we have the following equation.
\[\input{Z-id.tikz}\ \ = \ \ \input{id.tikz} \ \ = \ \ \input{X-id.tikz}\]
Indeed, it is this equation we used in the proof~\eqref{eq:ket0-copy-proof}.

In the previous section we claimed we could represent the Pauli $Z$ and $X$ basis states using spiders. We can now verify that this is indeed the case:
\begin{align}\label{eq:kets}
\input{ket0.tikz}\ \ &=\ \ \ket{+} + \ket{-} \ = \sqrt{2}\ket{0}
\qquad\qquad
\input{ketplus.tikz}\ \ =\ \ \ket{0} + \ket{1} \ = \sqrt{2}\ket{+} \\
\input{ket1.tikz}\ \ &=\ \ \ket{+} - \ket{-} \ = \sqrt{2}\ket{1}
\qquad\qquad\!
\input{ketminus.tikz}\ \ =\ \ \ket{0} - \ket{1} \ = \sqrt{2}\ket{-}
\end{align}
We see that we get the right states, but with a wrong scalar factor. Just as we can ignore global phases in quantum circuits, we can usually ignore global non-zero scalar factors in ZX-diagrams and that is in fact what we will do throughout this paper. For a more thorough discussion on scalars in ZX-diagrams see Section~\ref{sec:scalars}.

Traditionally, Z-spiders are written as \emph{green} dots and X-spiders as \emph{red} dots.\footnote{The reason for this is that when Bob Coecke and Ross Duncan developed the ZX-calculus in 2007 they had a whiteboard with red and green markers [Duncan, personal communication].} To accommodate colour-blindness and printing we choose not to do this, and instead to follow~\cite{CKbook} and write Z-spiders as white dots and X-spiders as grey dots.
If one wishes to use the traditional green/red colours, the author recommends using the colour-blind friendly versions listed on the ZX-calculus website.\footnote{\url{http://zxcalculus.com/accessibility.html}} 
Note that the green/red or white/grey colours are not universally used: Gidney et al.~\cite{Gidney2019efficientmagicstate} use black dots to represent Z-spiders and white dots to represent X-spiders (in order to match it more closely with the translation of a CNOT into the ZX-calculus).

\subsection{Composition}\label{sec:composition}

A ZX-diagram can be built iteratively by composing other ZX-diagrams either horizontally, by connecting the output wires of the first to the input wires of the second, or vertically, simply by `stacking' the diagrams to create the tensor product. The base case of this inductive construction starts with the Z- and X-spiders.

To make more clear how this works exactly, let us work through an explicit construction of the CNOT gate.
The first ingredient we need is the phaseless Z-spider with 1 input and 2 outputs:
\begin{equation}\label{eq:Zsp-3}
\input{Zsp-3.tikz}\ \ = \ \ \begin{pmatrix}1&0\\0&0\\0&0\\0&1\end{pmatrix}
\end{equation}
Its matrix has 2 columns and 4 rows. The matrix of a general ZX-diagram with $n$ inputs and $m$ outputs will have $2^n$ columns, and $2^m$ rows.
In particular, a ZX-diagram with no inputs is a column vector, and hence a quantum state, while a ZX-diagram with no outputs is a row vector, and hence an effect. If it has no inputs or outputs then it is a size $1\times 1$ matrix, i.e.~just a complex number.

Suppose now that we wish to horizontally compose the spider~\eqref{eq:Zsp-3} with an identity, which has matrix diag$(1,1)$. The way we calculate the result is with the Kronecker product:
\[\input{Zsp-3-id.tikz}\ \ = \ \ \begin{pmatrix}1&0\\0&0\\0&0\\0&1\end{pmatrix}\ \otimes\ \begin{pmatrix}1&0\\0&1\end{pmatrix} \ \ 
= \ \ \begin{pmatrix}1&0&0&0\\0&1&0&0\\0&0&0&0\\0&0&0&0\\0&0&0&0\\0&0&0&0\\0&0&1&0\\0&0&0&1\end{pmatrix}\]
Let us now calculate the matrix of the other ingredient we need:
\[\input{Xsp-3-id.tikz}\ \ = \ \ \begin{pmatrix}1&0\\0&1\end{pmatrix}\ \otimes\ \frac{1}{\sqrt{2}}\begin{pmatrix}1&0&0&1\\0&1&1&0\end{pmatrix} \ 
= \ \ \frac{1}{\sqrt{2}}\begin{pmatrix}1&0&0&1&0&0&0&0\\0&1&1&0&0&0&0&0\\0&0&0&0&1&0&0&1\\0&0&0&0&0&1&1&0\end{pmatrix}\]
Now to make a CNOT we need to horizontally compose these two subdiagrams:
\begin{equation}\label{eq:CNOT-composition}
\input{CNOT-composition.tikz}
\end{equation}

To see that this is indeed a CNOT we calculate its matrix.
On the level of the matrix, the horizontal composition of diagrams corresponds to matrix multiplication:
\[\frac{1}{\sqrt{2}}\begin{pmatrix}1&0&0&1&0&0&0&0\\0&1&1&0&0&0&0&0\\0&0&0&0&1&0&0&1\\0&0&0&0&0&1&1&0\end{pmatrix} \begin{pmatrix}1&0&0&0\\0&1&0&0\\0&0&0&0\\0&0&0&0\\0&0&0&0\\0&0&0&0\\0&0&1&0\\0&0&0&1\end{pmatrix}
\ \ = \ \ \frac{1}{\sqrt{2}}\begin{pmatrix}1&0&0&0\\0&1&0&0\\0&0&0&1\\0&0&1&0 \end{pmatrix}\]
Up to a scalar factor of $\sqrt{2}$ this is indeed the matrix of the CNOT gate.

If this type of matrix calculation seems tedious: that is precisely the point why we want to use the ZX-calculus. 
We will see that many identities which would normally be proven by working with matrices can be replaced with graphical reasoning.

In the previous section we wrote the CNOT gate as 
\begin{equation}
\input{CNOT-ZX.tikz}
\end{equation}
which does not match the diagram~\eqref{eq:CNOT-composition}. In fact, in Section~\ref{sec:spiders} we defined spiders as having inputs, which come from the left, and outputs, which come from the right. So vertical wires aren't even properly defined.

The reason we are allowed to write this wire vertically is because of the following equation (which the reader is invited to verify by showing that the matrix on the right-hand side indeed matches the matrix of the diagram on the left-hand side): 
\begin{equation}
\input{CNOT-symmetry.tikz}
\end{equation}
So whether the middle wire acts as an output of the Z-spider (as in the left-hand side), or as an input to the Z-spider (as in the right-hand side) does not change the matrix it represents. This means that we can write a vertical wire without ambiguity, because we can resolve it either way, and the matrix will be the same.
This is an example of a broader set of symmetries that ZX-diagrams have that we will study in the next section.

For those readers familiar with tensor networks it might be helpful to note that ZX-diagrams are in fact just tensor networks. Indeed, considering them as tensor networks, our notation matches the standard notion of a graphical tensor network as introduced by Penrose~\cite{Penrose}.
Considered as a tensor network, each wire in a ZX-diagram corresponds to a 2-dimensional index and a wire between two spiders denotes a tensor contraction. The Z- and X-spiders are then defined as:
\begin{align*}
\left( \  \input{Zsp-nolegs.tikz} \  \right)_{i_1...i_m}^{j_1...j_n} & =
{ \begin{cases}
1 & \textrm{ if } i_1 = ... = i_m = j_1 = ... = j_n = 0 \\  
e^{i \alpha} & \textrm{ if } i_1 = ... = i_m = j_1 = ... = j_n = 1 \\
0 & \textrm{ otherwise} 
\end{cases}}
\end{align*}
\begin{align*}
\left( \  \input{Xsp-nolegs.tikz} \  \right)_{i_1...i_m}^{j_1...j_n} & =
{ \frac{1}{\sqrt{2}} \cdot 
\begin{cases}
1 + e^{i \alpha} & \textrm{ if } \bigoplus_k i_k \oplus \bigoplus_l j_l = 0 \\  
1 - e^{i \alpha} & \textrm{ if } \bigoplus_k i_k \oplus \bigoplus_l j_l = 1
\end{cases}}
\end{align*}
where $i_k, j_l$ range over $\{0,1\}$ and $\oplus$ is addition modulo~2 (i.e.~XOR). While ZX-diagrams are `just' tensor networks, this glosses over the most important property of ZX-diagrams which sets it apart from the tensor networks that are used in for instance condensed matter physics. Namely, that we have a set of diagrammatic rewrite rules that act on the tensor network itself.

We have been writing our diagrams with the inputs on the left and the outputs on the right in order to mimic conventional quantum circuit notation. This is however not a fixed convention for ZX-diagrams and some authors write ZX-diagrams vertically. While this is not an issue per se, what is more inconvenient is that some authors prefer time flowing upwards, while others have their time flowing downwards. Whether wires coming in from the bottom are inputs or outputs hence varies per paper. The reader is advised to check carefully for each paper which convention is used, for instance by looking at the linear map interpretation of a spider with a single wire going up (if it is a column vector then time is flowing upwards; if it is a row vector, then time is flowing downwards).

\subsection{Symmetries}\label{sec:symmetries}

An important reason ZX-diagrams are a useful representation is because of the number of symmetries that are present in the generators.
To state these symmetries we need to introduce a couple of more generators of ZX-diagrams.

A generator that has a natural circuit counterpart is the \emph{swap}:
\begin{equation}
	\input{swap.tikz} \ \ = \ \ \begin{pmatrix}1&0&0&0\\0&0&1&0\\0&1&0&0\\0&0&0&1\end{pmatrix}
\end{equation}
The swap generators interact with spiders the way you would expect. Namely, that you can `slide' a spider along the wires, such as in the equation commuting a CNOT gate trough a swap below:
\begin{equation}\label{eq:CNOT-SWAP}
\input{CNOT-SWAP.tikz}
\end{equation}
We will not write down the full set of equations that are associated to the swap gate, and instead will let the reader's intuition do the work. An important point to note though is that we are not working in a `braided' setting, and hence the swap is self-inverse:
\begin{equation}
\input{swapswap.tikz}
\end{equation}

The spiders are symmetric tensors. This means that they are invariant under swapping of some of their wires:
\begin{equation}
	\input{spider-swaps.tikz}
\end{equation}
These symmetries hold for all phases $\alpha$, for spiders with any number of input and output wires, and for the swapping of any of their inputs or outputs.

Swapping of wires is not the only symmetry ZX-diagrams possess: it also allows you to swap inputs with outputs. In order to state this symmetry we need to introduce two other generators, that we write as bended pieces of wire:
\begin{equation}\label{eq:cup-cap-def}
	\input{cup.tikz} \ \ =\  \ \begin{pmatrix}1\\0\\0\\1\end{pmatrix} \qquad \qquad \qquad\input{cap.tikz} \ \ =\ \ \begin{pmatrix}1&0&0&1\end{pmatrix}
\end{equation}
The first of these, the state, we call the \emph{cup}, while the second, the effect, we call the \emph{cap}. The astute reader might recognise these matrices as the maximally entangled 2-qubit state, also known as the Bell state, and respectively the Bell effect.
The reason we write these generators as pieces of wire is because, just as with the swap, they act like pieces of wires. In particular, they satisfy the \emph{yanking equations}:
\begin{equation}\label{eq:yank-equations}
\input{yank-equations.tikz}
\end{equation}
Just as we can slide spiders along the wires of a swap gate (cf.~\eqref{eq:CNOT-SWAP}), we can slide a spider along the wires of cups and caps. For instance:
\begin{equation}
\input{spider-cap.tikz}
\end{equation}
The general symmetries for the spiders with respect to the cups and caps can be expressed as follows:
\begin{equation}\label{eq:spider-cupcap}
\input{spider-cupcap.tikz}
\end{equation}

The might just look like some meaningless bending of wires \emph{and that is exactly the point}. 
While the cup and cap generator represent specific matrices and must be treated as such when calculating the matrix of a ZX-diagram, when working with just the ZX-diagram we can treat cups and caps simply as pieces of wires and bend them in any way we see fit. We let the notation do the work for us.

In fact, now that we have introduced the cups and caps we can state a particularly useful property that ZX-diagrams enjoy:
\begin{quote}
	\emph{Only connectivity matters}
\end{quote}
What this means is that a ZX-diagram can be deformed arbitrarily by moving the generators around in the plane, bending and unbending the wires, and as long as the order of the inputs and outputs of the diagram are preserved, it represents the same matrix.
Another way to state this is that if two ZX-diagrams contain the same number of Z- and X-spiders, have the same phases, are connected by the same number of wires in each diagram, and the inputs and outputs of the diagrams are connected in the same order to the same spiders, then the ZX-diagrams represent the same matrix.
Again another way to state this is that we can treat ZX-diagrams as undirected multigraphs (i.e.~graphs that allow more than one edge between vertices), with some additional data on the vertices. In this representation the spiders are the vertices, and each vertex is labelled by a type (Z or X), a phase and a list of inputs and outputs it is connected to.

To make more clear the type of freedom this entails, \emph{only connectivity matters} means that the following ZX-diagrams all represent the same matrix.
\begin{equation}
\input{deformation-example.tikz}
\end{equation}
With any of these ZX-diagrams, if we wish to calculate the matrix they represent, we would first have to express it as a composition of tensor products of compositions of tensor products (and so on) of Z-spiders, X-spiders, identities, swaps, cups and caps. Alternatively, viewing the ZX-diagram as a tensor network, we can contract the tensors of the spiders according to the connectivity of the diagram using whatever tensor contraction algorithm we see fit.

\subsection{Scalars}\label{sec:scalars}
In the diagram deformation example above, the zero-arity X-spider with a $-\frac\pi2$ phase can be moved completely freely throughout the diagram. This is because a zero-arity spider, or rather, any ZX-diagram with zero inputs and outputs, represents a \emph{scalar}, a single complex number. For instance, we have:
\begin{equation}\label{eq:scalars-ZX}
\input{scalars.tikz}
\end{equation}
If we have a pair of scalar subdiagrams in a ZX-diagram then in the calculation of the matrix these combine by multiplying together the numbers they represent.
Note that using just combinations of the scalar diagrams of \eqref{eq:scalars-ZX} we can then represent any complex number as a scalar ZX-diagram.%
\footnote{This boils down to showing we can represent any complex number as a multiplication of $\frac{1}{\sqrt{2}}$, $\sqrt{2}e^{i\alpha}$ and $1+e^{i\alpha}$. It suffices to show we can represent any complex number $z$ with $\abs{z}\leq 2$, because then for any number with larger absolute value we can first rescale it by multiplying with $\frac{1}{\sqrt{2}}$ repeatedly. We can choose an $\alpha$ such that $\abs{z} = \abs{1+e^{i\alpha}}$. Hence $\frac{z}{1+e^{i\alpha}} = e^{i\beta}$ for some $\beta$. Hence $z = (1+e^{i\alpha})(\sqrt{2}e^{i\beta})\frac{1}{\sqrt{2}}$.}

In this paper and in many other work dealing with the ZX-calculus scalar factors are often dropped. This is for the same reason that physicists will sometimes work with unnormalised quantum states: it is simply inconvenient and sometimes unnecessary to know the exact scalar values. Note that dropping the scalar should only be done if the scalar is non-zero, since otherwise the entire diagram should cancel in an appropriate way.

A particular case where non-zero scalar factors can be ignored is when dealing with ZX-diagrams representing unitary quantum circuits (we in fact already did this in Section~\ref{sec:spiders} in the representation of the CNOT gate, where we suppressed a factor of $\frac{1}{\sqrt{2}}$). The reason we can ignore those factors is because we can always easily retrieve the correct factor: the matrix $M$ is proportional to a unitary, and hence $MM^\dagger = \lambda \id$ for some $\lambda\in \R_{>0}$. The correct scalar (up to an innocuous global phase) is then $1/\sqrt{\lambda}$. We can efficiently calculate this value $\lambda$ by composing the ZX-diagram with its adjoint and simplifying it until it reduces to the identity.  This is efficient, because in the adjoint the order of the gates is flipped, and hence we can cancel each gate with its inverse to reveal the constant of proportionality of the diagram representing each gate and multiplying these factors together.

A situation where knowing the exact scalar value is important is when calculating an amplitude, for instance when contracting a circuit with a particular input state and output effect.

We can often write the scalar correction just as a number next to the diagram, instead of representing it as a scalar ZX-diagram. For instance, instead of writing
\begin{equation}
\input{CNOT-scalar.tikz}
\end{equation}
to represent a scalar-accurate CNOT gate, we can write
\begin{equation}
\input{CNOT-scalar2.tikz}
\end{equation}
In this paper, we will ignore global scalar factors everywhere. Hence `=' should be read as `equal up to global non-zero scalar'.

\subsection{Adjoints, transpose and conjugate}

There are some useful global operations on ZX-diagrams that correspond to well-known matrix operations.

First of all, if we take a ZX-diagram and negate all the phases in the spiders, so that for instance $\frac{\pi}{2}$ is mapped to $-\frac{\pi}{2} \equiv \frac{3\pi}{2}$, and $\pi$ is mapped to $-\pi\equiv \pi$, the matrix of the resulting diagram is the \emph{conjugate} of the matrix of the original diagram (recall that the conjugate of a complex matrix is simply the pointwise complex conjugate of each the elements of the matrix).

The transpose of a matrix is also easily represented by ZX-diagrams: we use cups and caps to change every input into an output and vice versa.
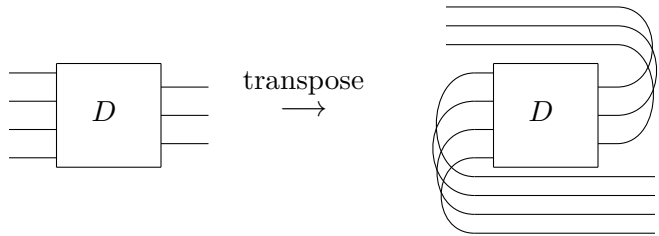
\ctikzfig{make-transpose}
Note that these `crossovers' of the wires are necessary to ensure that the first input is mapped to the first output, instead of to the last output. If we apply this procedure to the CNOT gate we can verify that the CNOT gate is indeed self-transpose:
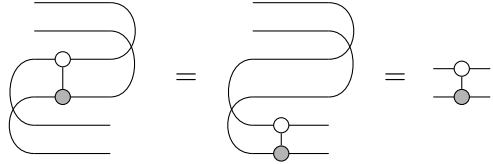
\ctikzfig{CNOT-transpose}
Here in the first step we slid the spiders along the wires to the bottom, and in the second step we applied the first yanking equation~\eqref{eq:yank-equations}.

The adjoint of a matrix is constructed by taking the transpose and conjugating the matrix. Hence, the adjoint of a ZX-diagram is constructed by interchanging the inputs and outputs using cups and caps as described above, and then negating all the phases of the spiders in the diagram.

\subsection{Hadamards}
In Section~\ref{sec:circuits-vs-diagrams} we used some special notation for the Hadamard gate:
\begin{equation*}
	\input{had.tikz}\ \ =\ \ \frac{1}{\sqrt{2}}\begin{pmatrix}1&1\\1&-1\end{pmatrix}
\end{equation*}
We can either decide to add this `Hadamard box' as a new generator to the ZX-calculus or by seeing it as a \emph{derived} generator defined in terms of the other generators. We will take the latter route.%
\footnote{For practical purposes it makes no difference whether the Hadamard is a generator or is simply syntactic sugar. However, when studying completeness of the ZX-calculus (cf.~Section~\ref{sec:completeness}) or when defining non-standard interpretations of ZX-diagrams, it is relevant how the Hadamard is defined.}

Recall that any single-qubit unitary gate is equal (up to global phase) to a rotation of the Bloch sphere over some axis. Hence, by decomposing the rotation using Euler angles we can represent any single-qubit unitary with a Z-rotation followed by an X-rotation, followed once again by a Z-rotation:
\ctikzfig{euler-decomposition}
In the particular case of $U=H$, we can take $\alpha=\beta=\gamma=\frac\pi2$:
\begin{equation}\label{eq:had-euler1}
\input{had-euler1.tikz}
\end{equation}

The global phase of $e^{-i\frac\pi4}$ is included in order to make the equation exactly equal. If one is willing to accept these global scalar factors as an inherent part of the ZX-calculus this definition is fine. On a formal level it is however desirable to have a representation that does not need scalar corrections:
\begin{equation}\label{eq:had-gadget}
\input{had-gadget.tikz}
\end{equation}

Note that these are by no means the only ways to represent the Hadamard gate using spiders. There is in fact an entire family of representations that will also be useful to note:
\begin{equation*}
	\input{had-euler2.tikz}
\end{equation*}
Using any of these definitions it is easy to show that the Hadamard box in the ZX-calculus is self-transpose:
\begin{equation}
\input{had-transpose.tikz}
\end{equation}
This means that, just like with the spiders, the orientation of the Hadamard is not important, which is why we are allowed to draw it as a simple box. Also just like with the spiders, we can slide Hadamards along wires as intuition would suggest.

\subsection{Universality}\label{sec:universality}

We have now seen that quite a number of quantum gates and constructions can be represented in the ZX-calculus. This raises the following question: exactly which linear maps can be represented in the ZX-calculus. The answer to this is simple: \emph{everything}. Here, `everything' means any complex matrix of size $2^n\times 2^m$ for some $n$ and $m$.

We have already seen how to represent an arbitrary complex number using a scalar ZX-diagram (Section~\ref{sec:scalars}). We have also seen that we can represent arbitrary Z and X rotations of the Bloch sphere (Eqs.~\eqref{eq:Z-a} and \eqref{eq:X-a}). These rotations generate all single-qubit unitaries (up to global phase, which we can reconstruct by adding some scalar diagram). We have also seen that we can construct CNOT gates (see~\eqref{eq:CNOT-composition}). 
The single-qubit unitaries together with the CNOT gate form a universal gate set, and hence any unitary on an arbitrary number of qubits can be written as a composition of these gates.
So we see that we can indeed represent any unitary between qubits as a ZX-diagram.

For any $n$-qubit normalised quantum state $\ket{\psi}$ there exists a $n$-qubit unitary $U$ such that $U\ket{0\cdots0} = \ket{\psi}$. As we can represent both $\ket{0\cdots 0}$ and $U$ as a ZX-diagram, this means that we can represent an arbitrary normalised quantum state as a ZX-diagram. By composing it with the appropriate scalar diagram we can rescale this to a vector of arbitrary norm.

Now given some linear map $L$ from $n$ qubits to $m$ qubits we can transform it into a $(n+m)$-qubit state by the Choi-Jamio\l{}kowski isomorphism:
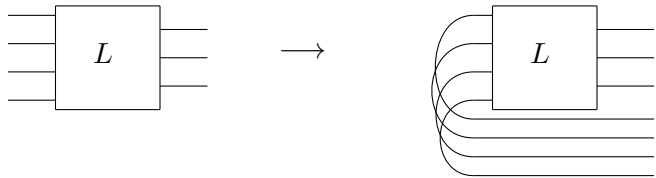
\ctikzfig{choi-jamiolkowski}
This state can be represented as a ZX-diagram, and hence by bending back the wires using the yanking equation~\eqref{eq:yank-equations} we get a representation of $L$ as a ZX-diagram.

The property of ZX-diagrams that they can represent any linear map is known as the \emph{universality} of the ZX-calculus. Together with two other properties we will encounter later, soundness and completeness, they form the backbone of the usefulness of ZX-diagrams.

Note that just because we \emph{can} represent any linear map using a ZX-diagram, this does not mean that such representations will necessary be `nice'. It is an ongoing challenge to find good representations of useful linear maps in the ZX-calculus. For instance, in Section~\ref{sec:ZH} we will introduce a new (derived) generator that allows us to more easily represent Toffoli gates in the ZX-calculus, since the native representation is rather cumbersome.

\section{The ZX-calculus}\label{sec:zx-calculus}
In the previous section we introduced ZX-diagrams as essentially a graphical notation for complex matrices of size $2^n\times 2^m$. In this section we will see how we can actually do calculations with ZX-diagrams, transforming it from notation into a language. Each subsection here introduces new rewrite rules and demonstrates how these rewrite rules can be used to prove well-known properties of quantum circuits and states. We summarise all the rules in Figure~\ref{fig:zx-rules} in Section~\ref{sec:complete-calculus}.

\subsection{Spider fusion and identity removal}

The most fundamental of all the graphical rewrite rules allowed in the ZX-calculus is \emph{spider fusion}. This says that when two spiders of the same colour are connected by one or more wires, that we can \emph{fuse} these two spiders together. When spiders are fused, their phases are added together. In terms of diagrams:
\begin{equation}\label{eq:spider-fusion}
\input{spider-fusion-Z.tikz} \qquad\qquad\qquad
\input{spider-fusion-X.tikz}
\end{equation}
Note that we interpret the numbers $\alpha$ and $\beta$ as phases $e^{i\alpha}$ and $e^{i\beta}$, and hence this addition is assumed to be modulo $2\pi$.

We already saw a few instances of spider fusion in Section~\ref{sec:circuits-vs-diagrams}, such as when proving that a T gate commutes through the control of a CNOT:
\ctikzfig{commute-T-CNOT-ZX-fuse2}
Flipping the colours, we can use spider fusion to show that a X gate commutes through the  target of a CNOT gate:
\ctikzfig{commute-X-CNOT-fuse}
The adding of the phases essentially generalises the fact that two rotations of the Bloch sphere in the same direction add together:
\begin{equation}\label{eq:phases-add}
\input{phases-add.tikz}
\end{equation}

Recalling that in the ZX-calculus $\ket{0} = \input{ket0.tikz}$ and $\ket{1} = \input{ket1.tikz}$ (cf.~\eqref{eq:kets}), we also see that spider fusion gives us that applying the Pauli $X$ gate to $\ket{0}$ gives $\ket{1}$:
\begin{equation}
\input{ket0-pi.tikz}
\end{equation}
An equation like $Z\ket{-} = \ket{+}$ is given similarly:
\begin{equation}
\input{ketminus-pi.tikz}
\end{equation}
Note here that we have $\pi+\pi = 2\pi \equiv 0$.

In fact, the spider fusion equations~\eqref{eq:spider-fusion} states something much stronger then is evident from these particular cases: any family of linear maps satisfying an analogous set of equations is completely characterised by an orthonormal basis~\cite{coecke2013new}, cf.~Section~\ref{sec:frobenius}.

Another rewrite rule we already saw is \emph{identity removal}:
\begin{equation}\label{eq:id-removal}
\input{id-removal.tikz}
\end{equation}

This can be interpreted in several ways. Considering a 1-input 1-output spider as a phase gate over either the Z- or X-axis, this rule says that a rotation by 0 degrees does nothing. Combining it with spider fusion it says that the inverse of a rotation by $\alpha$ is a rotation by $-\alpha$:
\ctikzfig{alpha-min-alpha}
Composing the diagrams of \eqref{eq:id-removal} with a cup it says we can make a Bell state by either taking a maximally entangled state over the Z basis or over the X basis:
\ctikzfig{id-removal-cup}
Identity removal also allows us to get rid of self-loops on spiders:
\begin{equation}\label{eq:self-loop-removal}
\input{self-loop-removal.tikz}
\end{equation}
Here in the first step we applied the rule in reverse to \emph{add} an identity. Then we fused the two spiders that are now connected by two wires.

\subsection{The copy rule and \texorpdfstring{$\pi$}{pi}-commutation}

The second set of rules of the ZX-calculus we will look at concerns the interaction of the Pauli X and Z gates and their eigenstates $\ket{0}, \ket{1}, \ket{+}, \ket{-}$ with the spiders. Recall from Eq.~\eqref{eq:kets} that (up to global non-zero scalar factors):
\begin{align*}
\input{ket0.tikz}\ \ &=\ \ \ket{0}
\qquad\qquad
\input{ketplus.tikz}\ \ =\ \ \ket{+} \\
\input{ket1.tikz}\ \ &=\ \ \ket{1}
\qquad\qquad\!
\input{ketminus.tikz}\ \ =\ \ \ket{-} \\
\input{X-pi.tikz} \ \ &=\ \ X 
\qquad \quad
\input{Z-pi.tikz} \ \ =\ \ Z
\end{align*}

The spider fusion rule already covers the interaction of a couple of these maps and states. Indeed we can use the spider fusion rule to show that a Z gate commutes through a Z-spider:
\ctikzfig{spider-Z-commute}
Spider fusion also shows what happens when a $\ket{+}$ or $\ket{-}$ state is applied to a Z-spider:
\ctikzfig{spider-ket0-fuse}
By flipping all the colours we then also know that an X gate commutes through an X spider and that the $\ket{0}$ and $\ket{1}$ fuse into an X spider.

A more interesting question is what happens to an X gate when it encounters a Z-spider (or vice versa, a Z gate that encounters an X-spider).
Consider a Z-spider with just a single input and no phase. Its linear operator is then given by $\ketbra{0 \cdots 0}{0} + \ketbra{1 \cdots 1}{1}$. Hence, if we apply an X gate to the input the resulting linear operator is 
$$\ketbra{0 \cdots 0}{0}X + \ketbra{1 \cdots 1}{1}X = \ketbra{0 \cdots 0}{1} + \ketbra{1 \cdots 1}{0},$$ 
as an X gate interchanges $\ket{0}$ and $\ket{1}$. This is easily seen to be equivalent to a Z-spider \emph{followed} by an X gate on each of the outputs:
$$\ketbra{0 \cdots 0}{1} + \ketbra{1 \cdots 1}{0} = (X\otimes\cdots\otimes X)\ketbra{1 \cdots 1}{1} + (X\otimes\cdots\otimes X)\ketbra{0 \cdots 0}{0}.$$
Hence, in terms of diagrams, we have:
\begin{equation}\label{eq:Z-spider-X-copy}
\input{Z-spider-X-copy.tikz}
\end{equation}
We will refer to this as a \emph{$\pi$-copy rule}.
By applying the appropriate cups and caps it should be clear that such a $\pi$-copy rule continues to hold regardless of how many inputs and outputs the spider has.
But what about when the spider has a non-zero phase?
Using spider fusion in reverse to \emph{unfuse} the phase we can reduce this to a simpler question:
\begin{equation}
\input{Z-spider-X-copy-phase.tikz}
\end{equation}
Hence, we need to resolve what happens when we apply an X gate to a $\ket{0}+e^{i\alpha}\ket{1}$ state:
\[X(\ket{0}+e^{i\alpha}\ket{1}) = X\ket{0}+e^{i\alpha}X\ket{1} = \ket{1} + e^{i\alpha} \ket{0} = e^{i\alpha}(\ket{0} + e^{-i\alpha}\ket{1}).\]
Ignoring the global phase $e^{i\alpha}$ we can write this in terms of a diagram:
\begin{equation}
\input{alpha-X-phase.tikz}
\end{equation}
The most generic case, together with its colour-flipped counterpart is then:
\begin{equation}\label{eq:pi-copy-generic}
\input{pi-copy-generic.tikz}
\end{equation}

Similar copy rules hold for the eigenstates of the X and Z operators. Indeed, applying a $\ket{0}$ to the 1-input Z-spider $\ketbra{0 \cdots 0}{0} + e^{i\alpha}\ketbra{1 \cdots 1}{1}$ it is clear that we get the output $\ket{0\cdots 0}$:
\begin{equation}\label{eq:ket0-spider-copy}
\input{ket0-spider-copy.tikz}
\end{equation}
The same works for $\ket{1}$ (although we then do get a global phase of $e^{i\alpha}$ that we will ignore):
\begin{equation}
\input{ket1-spider-copy.tikz}
\end{equation}

By introducing a Boolean variable $a$ that is either $0$ or $1$ we can represent the last two equations as a single parametrised equation:
\begin{equation}\label{eq:keta-spider-copy}
\input{keta-spider-copy.tikz}
\end{equation}
Here the $a\pi$ phase is either zero (when $a=0$), or $\pi$ (when $a=1$), and hence the input represents either a $\ket{0}$ or a $\ket{1}$.

As before, the same equation but with the colours flipped also continues to hold, which gives rules for copying $\ket{+}$ and $\ket{-}$ states through X-spiders. We will refer to all these rules as \emph{state-copy rules}. These state-copy rules only hold when the spider being copied has a phase of $0$ or $\pi$. For any other phase the analogue of~\eqref{eq:keta-spider-copy} does not hold.

Note that the orientation of the wires in the rewrite rules we have derived is irrelevant, as we can pre-compose and post-compose each rule with `cups and caps' to change inputs to outputs and vice versa. For instance, for the `reverse' of the state-copy rule:
\begin{equation}
\input{gen-copy-reverse.tikz}
\end{equation}

At this point the reader might wonder why we focus so much on X and Z, and not on Y. The reason for this is that Y can be presented as a composition of X and Z, and hence the rules relating to the Y-eigenbasis can be derived from the equations we have covered here. In principle, we could have worked with the `XY'-calculus instead of the ZX-calculus, but there is an important reason to prefer Z and X over Y. 
Namely, the Z and X eigenbases are \emph{self-conjugate} meaning that each of the vectors in the basis is equal to its own componentwise complex conjugate, for instance: $\overline{\ket{0}} = \ket{0}$. 
This is not the case for the Y eigenbasis $\ket{\pm i} := \ket{0} \pm i\ket{1}$. As a result, the `Y-spider' does not enjoy the symmetry between inputs and outputs of~\eqref{eq:spider-cupcap} that the Z- and X-spider do.

\subsection{Formal rewriting and soundness}
At this point it might be helpful to say a bit more about the formal nature of the rewriting we are doing.
The rewrite rules we have given above are of the form $D_1=D_2$, where $D_1$ and $D_2$ are some ZX-diagrams with the same number of inputs and outputs.
We then used these rewrite rules on \emph{larger} diagrams that contain either $D_1$ or $D_2$ as subdiagrams. This is because when we are asserting $D_1=D_2$, we are actually asserting an entire family of equalities. We are for instance also asserting that $D_1\otimes E = D_2\otimes E$ for any other ZX-diagram $E$, and $D\circ D_1 = D\circ D_2$ for any composable ZX-diagram $D$. 

In a sense we have the following implication:
\begin{equation}\label{eq:boxed-rewriting}
	\scalebox{0.85}{\input{boxed-rewriting.tikz}}
\end{equation}
I.e.~if $D_1$ appears as a subdiagram of some larger `ambient' diagram $D'$, then $D_1$ can be replaced by $D_2$ inside of $D'$. 

An important question to settle here is whether this is \emph{sound}. A rewrite rule in a language is called sound when it preserves the underlying semantics. In the setting of ZX-diagrams, the language is the set of diagrammatic rewrite rules of the ZX-calculus, and the semantics of a ZX-diagram is the linear map it represents. So a rule in the ZX-calculus is sound when the diagrams on each side of the equation represent the same matrix. For a single equation like $D_1=D_2$ soundness is easy to check: just calculate the matrix on both sides. But how can we be sure that employing this rewrite rule in a larger context as in~\eqref{eq:boxed-rewriting} is still sound, as this concerns an infinite family of rewrite rules?

To make the following discussion more clear, we will adopt the convention of the literature on the ZX-calculus to denote the matrix of a ZX-diagram $D$ by $\intf{D}$. The soundness of a rule $D_1=D_2$ is then witnessed by the equation $\intf{D_1}=\intf{D_2}$.

By definition of how we calculate the matrix of a ZX-diagram, we have $\intf{D_1\otimes D_2} = \intf{D_1}\otimes \intf{D_2}$ and $\intf{D_1\circ D_2} = \intf{D_1}\intf{D_2}$ (where this latter juxtaposition represents matrix multiplication).
Hence, if we have $\intf{D_1}=\intf{D_2}$, then we also have $\intf{D_1\otimes E} = \intf{D_1}\otimes \intf{E} = \intf{D_2}\otimes \intf{E} = \intf{D_2\otimes E}$, and hence $D_1\otimes E = D_2\otimes E$ is also sound. The analogous result for composition with $E$ also holds. 
As ZX-diagrams are built iteratively by repeated tensor products and composition, this shows that \eqref{eq:boxed-rewriting} is indeed sound.

In terms of category theory, we can frame $\intf{\cdot}$ as a strong monoidal functor from the category of ZX-diagrams to the category of complex matrices of size $2^n$, see Section~\ref{sec:zx-category} for more details. The type of local rewriting we do can be understood as an instance of \emph{double-pushout rewriting}~\cite{bonchi2020string}.

\subsection{Colour changing}
As has been noted several times in the preceding sections, when we have derived some rule in the ZX-calculus, an analogous rule with the colours interchanged also holds. This follows from the rules concerning the Hadamard gate that we introduced in Section~\ref{sec:hadamard-colour}. Let us repeat these rules here for ease of reference:
\begin{equation}\label{eq:had-rules}
	\input{had-had-cancel.tikz}\qquad\qquad\qquad \input{colour-change-push.tikz}
\end{equation}
The first of these rules states that the Hadamard gate is self-inverse, while the second says that commuting a Hadamard through a spider changes its colour (reflecting the fact that the Hadamard interchanges the eigenbasis of the Z and X gate).
Note that the Hadamard commutation rule also holds for spiders that have no inputs, so for instance:
\begin{equation}\label{eq:Z-state-to-X}
\input{Z-state-to-X.tikz}
\end{equation}

Let us give an example to demonstrate how the two Hadamard rules~\eqref{eq:had-rules} imply the colour-inverse of any other rule we have.
For instance, suppose we want to prove the colour-inverse of Eq.~\eqref{eq:Z-spider-X-copy}:
\begin{equation}\label{eq:X-copy-repeat}
\input{Z-spider-X-copy.tikz}
\end{equation}
We start with the left-hand side of the colour-swapped version, and then insert two Hadamards on the input by applying the self-inverse rule in reverse:
\ctikzfig{colour-swap-example1}
We then commute one of the Hadamards all the way to the outputs:
\ctikzfig{colour-swap-example2}
We now see the left-hand side of the original (not colour-swapped) equation~\eqref{eq:X-copy-repeat}, and hence we can apply it:
\ctikzfig{colour-swap-example3}
Now it remains to commute the Hadamards back to the left, and to cancel the resulting double Hadamard:
\ctikzfig{colour-swap-example4}
We have hence succeeded in proving the colour-inverse of Eq.~\eqref{eq:X-copy-repeat}:
\ctikzfig{X-spider-Z-copy}

\subsection{The bialgebra and Hopf rule}\label{sec:bialgebra-hopf}

The previous sets of rules all have a very distinct topologically intuitive character: spider fusion allows you to fuse adjacent spiders of the same colour; identity removal and the Hadamard self-inverse rule allow you to remove certain vertices; the $\pi$-copy, state-copy and colour-change rule allow you to commute certain generators through spiders by copying them to the other side.
It is therefore relatively easy (after you gain some practice) to spot where they can be applied and what the effect of their application will be on the structure of the rest of the diagram.

The last major rule of the ZX-calculus takes more time to work with intuitively, although it does correspond to a natural and crucial property of the interaction of the Z- and X-spider.

Before we introduce the rule, let us give some motivation.
Treating the $\ket{0}$ and $\ket{1}$ states as Boolean bits, we can view the phaseless 1-input, $n$-output Z-spider as the COPY gate that copies the bit $0$ to $0\cdots0$ and the bit $1$ to $1\cdots 1$. Indeed, this is exactly what Eq.~\eqref{eq:keta-spider-copy} states. Analogously, we can view the phaseless 2-input, 1-output X-spider as the XOR gate:
\ctikzfig{XOR-states}
The XOR gate and the COPY gate have a natural commutation relation. Indeed, first XORing bits, and then copying the outcome is the same as first copying each of the bits and then XORing the resulting pairs:
\begin{equation}\label{eq:XOR-COPY-bialgebra}
\input{XOR-COPY-bialgebra.tikz}
\end{equation}

This equation says that the XOR \emph{algebra} and the COPY \emph{coalgebra} together form a \emph{bialgebra} (for more about these algebraic structures see Section~\ref{sec:frobenius}). This is why we refer to the analogous equation in the ZX-calculus as the \emph{bialgebra rule}:
\begin{equation}\label{eq:bialgebra-rule}
\input{bialgebra-rule.tikz}
\end{equation}
This rule is essential to many proofs in the ZX-calculus. As a demonstration of its utility, let us prove the well-known `three CNOTs make a swap' circuit identity. This proof will also reveal the need for one additional rule that can actually be derived with the bialgebra rule.
We start with the three CNOTs and deform the diagram to demonstrate where the bialgebra rule can be applied:
\begin{equation}
\input{three-CNOTs-proof1.tikz}
\end{equation}
Now, apply Eq.~\eqref{eq:bialgebra-rule} from right-to-left, and then deform the diagram again to bring it to a cleaner form:
\begin{equation}
\input{three-CNOTs-proof2.tikz}
\end{equation}
To finish the proof it now remains to show that two CNOTs applied in succession cancel each other:
\begin{equation}
\input{CNOT-CNOT.tikz}
\end{equation}
The first step towards doing this might seem clear, namely, we fuse the adjacent spiders of the same colour:
\begin{equation}\label{eq:CNOT-CNOT-fuse}
\input{CNOT-CNOT-fuse.tikz}
\end{equation}

But now we are seemingly stuck.

The solution comes from another equation between the Boolean gates XOR and COPY. When we first apply an XOR and then a COPY, we can commute the XOR by essentially `copying' it through the COPY gate. But what happens when we first apply a COPY \emph{and then} an XOR gate to the two outputs of the COPY: the input $0$ first gets copied to $00$ and then XORed to $0$, while the input $1$ first gets copied to $11$ and then XORed to $0$. So regardless of the input, we output a $0$. Generalising the notion of a XOR gate to allow for a 0-input XOR gate that just outputs its unit (namely $0$), and a 0-output COPY gate that discards its output we can write this relation diagrammatically as:
\begin{equation}
\input{XOR-COPY-hopf.tikz}
\end{equation}
An algebra (XOR) and a coalgebra (COPY) satisfying this equation are known together as a \emph{Hopf algebra}. This is why we refer to the following analogous equation between the Z- and X-spider as the \emph{Hopf rule}:
\begin{equation}\label{eq:hopf-rule}
\input{hopf-rule.tikz}
\end{equation}
Operationally, this equation can be interpreted as saying that the Z-basis and X-basis correspond to \emph{complementary} observables, where having maximal information about the Z observable gives minimal information about the X observable~\cite[Section~9.2]{CKbook}.
In this same vein the bialgebra rule~\eqref{eq:bialgebra-rule} says that the observables are \emph{strongly complementary}~\cite{coecke2012strong}.

It turns out that complementarity can be derived from strong complementarity using some clever deformation of the diagram and the rules from the previous sections~\cite{duncan2016interacting}:
\ctikzfig{hopf-proof}
In this derivation, we first deformed the diagram, then we introduced identities, we unfused some spiders, applied the bialgebra rule, copied a state twice, and in the last equation we ignored the dangling scalar diagram as discussed in Section~\ref{sec:scalars}.

The Hopf rule is exceedingly useful. Indeed, combining it with some spider fusion we can use it to show that we can always cancel pairs of connections between spiders of opposite colours:
\begin{equation}\label{eq:wire-pair-cancel}
\input{wire-pair-cancel.tikz}
\end{equation}
So if two spiders of opposite colour are connected by $n$ wires, then this can be reduced to $n~\text{mod}~2$ wires.
In particular, we can finish the proof that 2 CNOTs cancel each other out, and thus that 3 CNOTs make a SWAP:
\begin{equation}
\input{CNOT-CNOT-fuse2.tikz}
\end{equation}

We can extend the bialgebra rule~\eqref{eq:bialgebra-rule} to spiders of arbitrary arity:
\begin{equation}\label{eq:bialgebra-rule-many}
\input{bialgebra-rule-many.tikz}
\end{equation}
That is, for any two connected phaseless spiders of different colours we can apply the bialgebra rule, resulting in a fully connected bipartite graph as on the right-hand side.
For $n=m=2$ this is exactly Eq.~\eqref{eq:bialgebra-rule}. For $n=1$ or $m=1$ this follows in a trivial manner by adding and removing identities (cf.~Eq.~\eqref{eq:id-removal}).
When $n=0$ or $m=0$ this is just the state-copy rule, i.e.~(the colour-swapped) Eq.~\eqref{eq:ket0-spider-copy}.
Now suppose $n=2$ and $m=3$, we will show how to derive its bialgebra rule using just spider fusion and the $n=2,m=2$ bialgebra rule:
\begin{equation}
\input{bialgebra-rule-2-3.tikz}
\end{equation}
We leave it as an exercise for the reader how to derive the cases for larger $n$ and $m$ (hint: use induction or see \cite[Theorem~9.71]{CKbook}).

There are two common mistakes people make when using the bialgebra rule.
The first is that the bialgebra rule~\eqref{eq:bialgebra-rule} only works when the phases on the spiders are zero. When the phases are $\pi$ a modification is possible by combining it with the $\pi$-copy rules~\eqref{eq:pi-copy-generic}:
\begin{equation}
	\input{bialgebra-rule-pi.tikz}
\end{equation}
But when one of the spiders contains some arbitrary phase $\alpha$, the result will be more complicated:
\begin{equation}
	\input{bialgebra-phase.tikz}
\end{equation}
The additional bit of diagram we get is called a \emph{phase gadget} and is studied in more detail in Section~\ref{sec:phase-polynomial}.

The second common mistake is that people apply the rule~\eqref{eq:bialgebra-rule} from right-to-left without paying attention to how many outputs the spiders have, wrongfully equating the following diagrams:
\begin{equation}
	\input{bialgebra-wrong.tikz}
\end{equation}
The correct way to apply the bialgebra rule here is to first unfuse the spider:
\begin{equation}
	\input{bialgebra-wrong-corrected.tikz}
\end{equation}

\subsection{The complete calculus}\label{sec:complete-calculus}

We now have all the main ingredients for reasoning with ZX-diagrams using the ZX-calculus, so let us summarise what we have already seen. The rules of the ZX-calculus are presented in Figure~\ref{fig:zx-rules}.

\begin{figure}
\centering
\input{ZX-rules.tikz}
\caption[Rules of the ZX-calculus]{
A convenient presentation of the ZX-calculus. These rules hold for all $\alpha, \beta \in \R$, and due to \HadamardRule and \HHRule all rules also hold with the colours interchanged. Note `...' should be read as `0 or more', hence the spiders on the left-hand side of \SpiderRule are connected by one or more wires.
The letters stand respectively for spider-$(\bm{f})$usion, $(\bm{h})$adamard, $(\bm{id})$entity, $(\bm{hh})$-cancellation, $(\bm{\pi})$-commute, $(\bm{c})$opy, and $(\bm{b})$ialgebra. Note that these rules are only correct up to non-zero scalar.
}
\label{fig:zx-rules}
\end{figure}

The Hadamard is defined as any one of the following diagrams (that can be shown to be equivalent using the rules of Figure~\ref{fig:zx-rules}):
\begin{equation*}
	\input{had-euler2.tikz}
\end{equation*}

Besides these concrete rules we have several `meta-rules':
\begin{itemize}
	\item Only connectivity matters: diagrams can be deformed in any way, as long as the order of the inputs and outputs of the whole diagram is preserved.
	\item Each of the rules in Figure~\ref{fig:zx-rules} also holds with the inputs and outputs interchanged.
	\item Every rule also holds with the colours (white and grey) interchanged.
	\item Every rule holds with all the phases negated.
\end{itemize}

We have also seen several combined and derived rules:
\begin{itemize}	
 	\item The rules \CopyRule and \BialgRule can be combined and generalised to the following rule:
	\ctikzfig{bialgebra-rule-many}
	\item The rules of Figure~\ref{fig:zx-rules} imply the Hopf rule:
	\ctikzfig{hopf-rule}
	\item The rules \PiRule and \CopyRule can be combined to give a more generic state-copy rule:
	\ctikzfig{keta-spider-copy}
\end{itemize}

Note that in the remainder of the paper we will often label steps in a derivation with the rule names of Figure~\ref{fig:zx-rules} to denote which are applied. See for example~\eqref{eq:ghz-circuit-zx}.

The rules we present in Figure~\ref{fig:zx-rules} are not a minimal set. A smaller set of equivalent rules can be found in for instance~\cite[Definition~9.108]{CKbook} and a ruleset that contains the correct scalars can be found in~\cite{BackensSimplified}.

\section{Some example derivations}\label{sec:example-derivations}

Let us now use the ZX-calculus to reason about several small examples that demonstrate its use in reasoning about quantum computing. This will also demonstrate several best practices when trying to simplify a diagram using the ZX-calculus.
A reader wishing to try out some diagrammatic proving themselves could take a look at Appendix~\ref{app:circuit-identities} and try to prove these circuit identities by hand (although it might be useful to read Section~\ref{sec:phase-polynomial} first).

\subsection{GHZ-preparation circuit}\label{sec:ghz}
The following quantum circuit implements the GHZ state $\ket{000}+\ket{111}$:
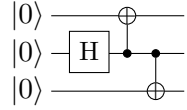
\ctikzfig{ghz-circuit}
We can verify this by translating it into a ZX-diagram and simplifying it using the rules of Figure~\ref{fig:zx-rules}:
\begin{equation}\label{eq:ghz-circuit-zx}
\input{ghz-circuit-zx.tikz}
\end{equation}

By the definition of the Z-spider, this last diagram is equal to $\ket{000}+\ket{111}$ which is indeed the 3-qubit GHZ state.
As is commonly the case for simplifying ZX-diagrams, we start with fusing adjacent spiders of the same colour using \SpiderRule and removing identities using \IdRule. Once this is done we have a diagram that contains a state: a spider with just one leg. Many rules and simplifications deal with states and hence this is an obvious next target for simplification. It is connected to a Hadamard, so the only thing we can do with it is change its colour using \HadamardRule. The final step is another spider fusion.

\subsection{Pauli pushing}

Knowing how the Pauli operators commute through a circuit (or a more general computation) is important for several areas in quantum computation, such as when verifying the functioning of an error correcting code, implementing a measurement-based quantum computation, or calculating the stabiliser tableau of a Clifford circuit.
The ZX-calculus makes it easy to remember the rules for how the Pauli's propogate. Indeed, the only rules required are spider fusion~\SpiderRule and the $\pi$-copy rule~\CopyRule (and occasionally an identity will need to be removed with~\IdRule).

For instance, suppose we wish to know how the Pauli operator $X\otimes Z$ commutes through the following circuit:
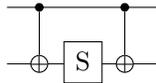
\ctikzfig{circuit-pauli-commuting}
We write it as a ZX-diagram with the Pauli's X and Z on the left, and start pushing them to the right:
\ctikzfig{circuit-pauli-zx}
Note that we have used~\SpiderRule both to commute spiders of the same colour through each other (like the Pauli Z through the control of the CNOTs, or the Pauli Z through the S gate), as well as to combine the phases of the Pauli's in order to cancel out the phases.

We see that we end up with the Pauli~$Y\otimes \,\id$. Of course, this process was quite lengthy, because every rewrite step is shown. Once you get comfortable with this process, many of the steps can be combined and it becomes easy to see where each of the Pauli's ends up.

\subsection{Magic state injection}
The following circuit implements the well-known \emph{magic state injection} protocol, where a $\ket{T}:=\ket{0}+e^{i\pi/4}\ket{1}$ magic state is pushed onto a qubit using a measurement and potential correction:
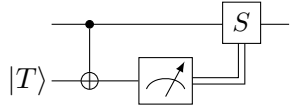
\ctikzfig{magic-state-injection}
Here the double wire represents a classical measurement outcome being fed forward into the $S$ gate, so that the $S$ gate is only applied if the measurement outcome corresponded to $\bra{1}$.

We can represent this in the ZX-calculus by introducing a Boolean variable $a$ to represent the measurement outcome. We can then easily prove its correctness:
\begin{equation}
\input{magic-state-injection-zx.tikz}
\end{equation}
Most of this proof is spider fusion, but there are a couple of interesting bits. The step labelled \PiRule actually consists of two branches, because if $a=0$, then the grey dot can be removed using \IdRule so that the phase on the white dot remains $\frac\pi4$, while if~$a=1$, then~\PiRule is indeed applied, and the phase flips to $-\frac\pi4$. In the step after that we make the observation that $(-1)^a\frac\pi4 = \frac\pi4 -a\frac\pi2$.

The usage of a variable to denote a measurement outcome is a simple `hack' to deal with classically controlled circuits, and is used quite often when doing measurement-based quantum computation in the ZX-calculus~\cite{kissinger2017MBQC,wetering-gflow}.
For more complicated scenario's we can also adopt a variant of the ZX-calculus where we allow both classical and quantum wires to exist~\cite{coecke_paquette_pavlovic_2009,coecke2016categorical}. We discuss this in more detail in Section~\ref{sec:doubling}.

\subsection{Teleportation}\label{sec:teleportation}
The standard state-teleportation protocol consists of two parties, Alice and Bob, that share a maximally entangled state $\ket{00}+\ket{11}$. Alice does some quantum operations, measures her states, sends the classical measurement outcomes to Bob, and Bob does some quantum corrections based on those outcomes. At the end Bob has the state that Alice started out with. We can represent this as a quantum circuit as follows:
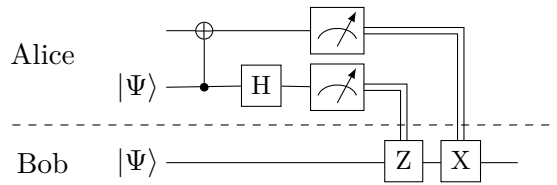
\ctikzfig{teleport-state}
Here the $\ket{\Psi}$ label is to denote that the bottom two qubits start in the maximally entangled Bell state. By representing the measurement outcomes by Boolean variables $a$ and $b$ we can again represent this in the ZX-calculus and prove the correctness of the protocol:
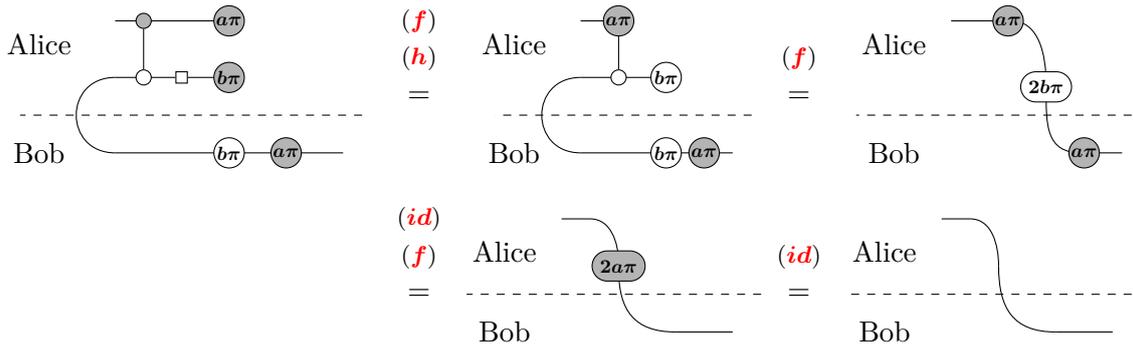
\ctikzfig{teleport-state-zx}
Here we used the fact that $2b\pi$ is either $2\pi$ or $0$ which are both $0$ modulo $2\pi$, and hence the spider can be removed using \IdRule.

\subsection{Detecting entanglement}
Let us now do a more involved calculation with the ZX-calculus, demonstrating how one goes about systematically simplifying a diagram.

Suppose you are given the following (somewhat randomly chosen) quantum circuit:
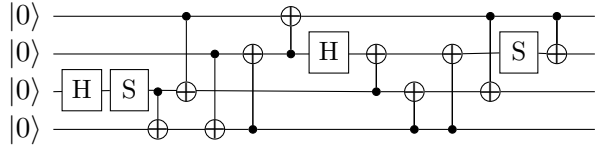
\ctikzfig{circuit-entanglement}
Suppose further that one wishes to know which qubits are entangled after applying this circuit. 
Calculating the state directly gives $\frac12(1,0,0,0,i,0,0,0,0,0,0,1,0,0,0,i)^T$. While a trained eye might be able to spot that the 2nd qubit is unentangled, and that the remainder form a GHZ-state, at first glance this is altogether not obvious. Let us do a simplification of the same circuit in the ZX-calculus.
In this simplification we have a couple of steps marked (*), (**), (***). These will be explained after.
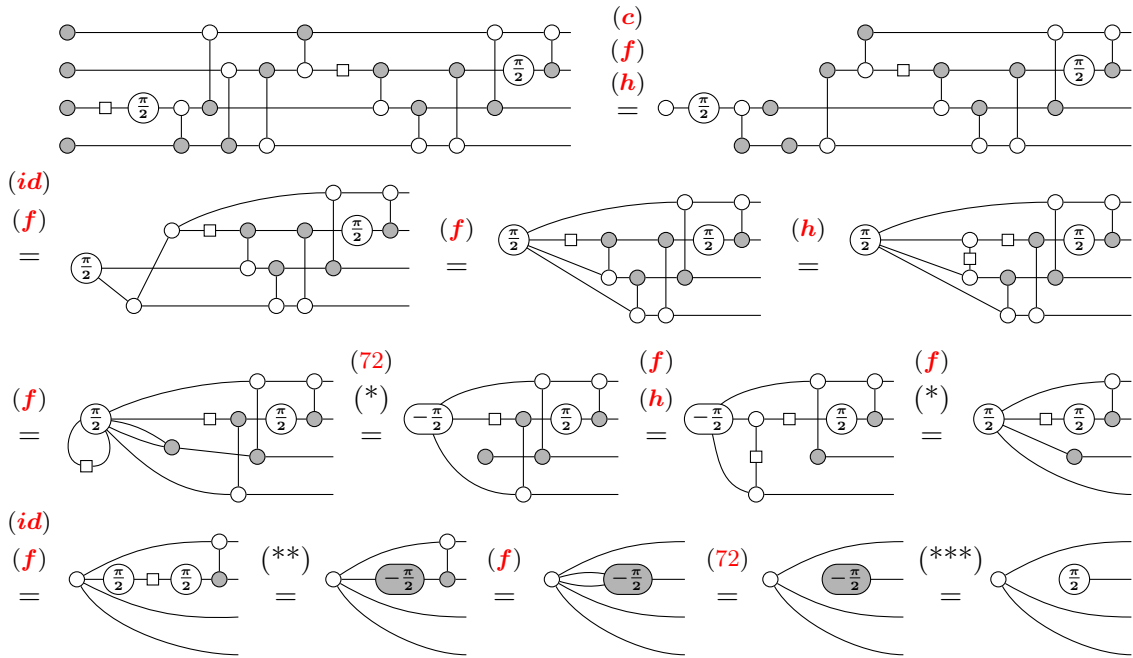
\ctikzfig{circuit-entanglement-zx}
If this is one of the first times encountering ZX-diagrams it might not necessarily be clear what is happening here, but we are in fact following a simple algorithm. 
Most of the steps follow the general pattern that we always try to fuse spiders of the same colours with \SpiderRule (though we don't do that here for all spiders at the same time to prevent cluttering), always remove identities with \IdRule, always copy states through spiders with \CopyRule, and always remove parallel wires between spiders of opposite colours using the Hopf rule \HopfRule. Hadamards are moved out of the way with \HadamardRule when no other obvious move is available.
We have marked three other types of rewrite rules (*), (**) and (***). These all rely on the particular decomposition \eqref{eq:had-euler1} of the Hadamard gate into its Euler angles \input{had-euler1.tikz}, where we ignore the global phase:
\begingroup
\allowdisplaybreaks
\begin{align}
	(*):&\quad \input{had-self-loop.tikz} \label{eq:had-self-loop-removal}\\
	(**):&\quad \input{had-absorb-angles.tikz}\\[0.2cm]
	(***):&\quad \input{S-state-equality.tikz}\label{eq:s-state-equality}
\end{align}
\endgroup
Note that in the last step of the proof of (*) we dropped the scalar $\frac\pi2$ X-spider. The equation (*) is particularly useful because it allows us to get rid of Hadamard self-loops, which is almost always the prudent thing to do in a simplification. 
The equation (**) is an example of how a series of Clifford phase gates can often be combined together into a simpler series of phase gates.
The equation (***) relates two ways to write the eigenstate $\ket{i}:=\frac{1}{\sqrt{2}}(\ket{0}+i\ket{1})$ of the Pauli operator $Y$ (up to global phase). Namely, the left-hand side says $\ket{i} = R_X(-\frac\pi2)\ket{0}$ while the right-hand side says $\ket{i} = R_Z(\frac\pi2)\ket{+}$. An analogous equation holds for $\ket{-i}:= \frac{1}{\sqrt{2}}(\ket{0}-i\ket{1})$, which boils down to flipping the signs of the phases in the spiders.

The size of this circuit is nearing the limit of what is still comfortable to rewrite manually. In fact, the software PyZX~\cite{pyzx} was used to verify and help with the rewriting here. This entire simplification can be done automatically and nearly instantaneously with PyZX using one of its built-in simplification strategies. We presented the derivation here in detail to show how one would go about systematically simplifying a ZX-diagram.
A more formal approach to simplifying a (Clifford) diagram is presented in Section~\ref{sec:simp-clifford}.

\subsection{Phase polynomials and phase gadgets}\label{sec:phase-polynomial}

\emph{Phase polynomials} are a class of unitaries that have proven useful in quantum circuit optimisation and verification~\cite{amy2016t,AmyVerification,amy2018cnot,heyfron2018efficient}.
Phase polynomials are the unitaries generated by circuits containing CNOT gates and Z-phase rotations. The action of such a unitary on a computational basis state can be represented as $\ket{\vec{x}} := \ket{x_1x_2\cdots x_n} \mapsto e^{if(x_1,x_2,\ldots,x_n)}\ket{A\vec{x}}$ where $f:\mathbb{B}^n\rightarrow \R$ is a semi-Boolean function and $A:\mathbb{B}^n\rightarrow \mathbb{B}^n$ is a \emph{linear} Boolean function (i.e.~one that can be represented by a CNOT circuit~\cite{markov2008optimal}).
The interesting part of a phase polynomial is given by its diagonal action, the part that is given by the semi-Boolean function $f$. We will write $U_f$ for the unitary that implements the diagonal action $U_f\ket{\vec{x}} = e^{if(\vec{x})}\ket{\vec{x}}$.

Each semi-Boolean function arising in this manner can be written as a linear combination of XOR terms. An example of such a decomposition would be $f(x_1,x_2,x_3) = \frac{\pi}{4}(x_1\oplus x_2) + \frac{\pi}{2}x_3 + \pi(x_1\oplus x_2\oplus x_3)$. Writing $f=\sum_i f_i$ where each $f_i$ is a single XOR term we then see that $U_f = \prod_i U_{f_i}$, where the order in the product is irrelevant as each $U_{f_i}$ commutes with one another.

It turns out that the unitary $U_f$ where $f$ consists of a single XOR term has a natural representation as a ZX-diagram. Namely:
\begin{equation}\label{eq:phase-gadget-unitary}
  \input{phase-gadget-unitary.tikz} \ \  = \ \ U_f \text{ where }\  f(x_1,\ldots,x_n) = \alpha (x_1\oplus\cdots\oplus x_n).
\end{equation}
We will refer to a ZX-diagram of this shape and a unitary $U_f$ as \emph{phase gadgets}, as they look like small diagrammatic gadgets that add a phase to the state if the parity of the state is odd.
We can relate this representation to a more common circuit construction of phase gadgets where we build a `ladder of CNOTs' to construct the appropriate parity, apply a phase, and then uncompute with the opposite ladder of CNOTs~\cite{phaseGadgetSynth}:
\begin{equation}\label{eq:phasegadget}
\input{phase-gadget-circ.tikz}
\end{equation}
For 2 qubits we can prove this correspondence easily:
\begin{equation}
\input{MS-simplify.tikz}
\end{equation}
For more qubits we prove by induction (which we leave as an exercise for the reader).
The action of a phase gadget is symmetric on its qubits. While this is clear in the representation of a phase gadget as a ZX-diagram, this is not the case in its circuit representation~\eqref{eq:phasegadget}. This asymmetry in the representation requires circuit optimisation procedures based on circuits to build dedicated methods for handling phase gadgets and phase polynomials~\cite{nam2018automated}, while in the ZX representation the symmetry is evident.

A straightforward calculation shows that the 2 qubit unitary presented in this calculation is equal (up to a global phase) to $\exp(-i\frac{\alpha}{2} Z\otimes Z)$. Such a unitary is sometimes called an \emph{Ising-type interaction} and is (up to a global basis change) the unitary that is implemented by the natural 2-qubit interaction in ion trap quantum computers, the \emph{M\o{}lmer-S\o{}rensen interaction}~\cite{molmersorensen1999}. It turns out that any circuit can be written as a product of gates of the form $\exp(i\alpha \vec P)$ where $\vec P$ is some tensor product of Pauli's~\cite{Litinski2019gameofsurfacecodes}. Each such `exponentiated Pauli' can be written efficiently as a ZX-diagram~\cite{phaseGadgetSynth}.

For the unitaries in Eq.~\eqref{eq:phase-gadget-unitary} we have $U_{f_1}U_{f_2} = U_{f_1+f_2}$. In particular, if $f_1$ and $f_2$ contain identical XOR terms and have phases $\alpha$ and $\beta$, then the resulting $f=f_1+f_2$ has the same XOR term but with phase $\alpha+\beta$. This relation implies that two phase gadgets, presented as ZX-diagrams, with exactly the same connectivity should be able to fuse together. This is readily shown in the ZX-calculus:
\begin{equation}\label{eq:phase-gadget-fusion}
	\input{gf-proof.tikz}
\end{equation}

That the phases of phase gadgets with identical XOR terms fuse lies at the heart of several quantum circuit optimisation routines~\cite{nam2018automated,amy2014polynomial} including one using the ZX-calculus~\cite{kissinger2019tcount}.

Instead of writing a phase polynomial as a sum of XOR terms, we can also write a phase polynomial as a sum of AND terms. For instance, the CCZ gate implements the diagonal action $\ket{xyz} \mapsto e^{i\pi (x\cdot y\cdot z)}$. This translation between XOR and AND representations of phase polynomials is used (either implicitly or explicitly) in several other circuit optimisation methods and has close links to Reed-Muller decoding and 3-tensor factorisation~\cite{amy2016t,heyfron2018efficient,deBeaudrap2020Techniques}. We discuss this transition between XOR and AND in more detail in Section~\ref{sec:fourier}.

This marks the end of the introductory part of this paper. The remaining sections each cover separate topics related to the ZX-calculus that can mostly be read independently.

\section{Clifford computation}\label{sec:clifford}

The Clifford unitaries are those unitaries generated by compositions of CNOT, Hadamard and S gates. Equivalently, Clifford unitaries are precisely those unitaries $U$ that map a `Pauli string' $\vec{P} = P_1\otimes \cdots \otimes P_n$ to another Pauli string under the conjugation $U\vec{P}U^\dagger$.
The states that can be reached starting from $\ket{0\cdots 0}$ and applying a Clifford unitary to it are the \emph{Clifford states} (also known as the \emph{stabiliser states}, because they can be characterised as those states that are stabilised by families of Pauli strings).

Even though Clifford states and unitaries do not allow universal quantum computation, they are still incredibly useful in a wide variety of tasks, such as in error correcting codes, measurement-based quantum computation and numerous protocols such as quantum key distribution, teleportation or dense coding.
As is well-known, computation with Clifford states and unitaries can be efficiently classically simulated. This is the content of the Gottesman-Knill theorem~\cite{aaronsongottesman2004}.

For our purposes, it will also be useful to introduce the notion of a \emph{Clifford linear map}. These are the linear maps that can be produced by composing Clifford states, Clifford unitaries, and $\bra{0}$, i.e.~the post-selection for the $\ket{0}$ state.

It turns out that Clifford linear maps are precisely those linear maps that can be presented by ZX-diagrams containing only phases that are multiples of $\frac{\pi}{2}$. Indeed, $\ket{0}$, $\bra{0}$, CNOT, Hadamard and S are are all representable by ZX-diagrams that contain only multiples of $\frac{\pi}{2}$ phases, and hence this is also true for their compositions. The converse is also not too hard to see.

This \emph{Clifford fragment} of the ZX-calculus, where spiders are only allowed to carry a phase that is a multiple of $\frac\pi2$, turns out to be particularly well-behaved. In this section we will study this fragment and reprove some well-known properties of Clifford states and Clifford computation, including a derivation that each Clifford state is equal to a graph state with local Cliffords and a graphical proof of the Gottesman-Knill theorem.
Along the way we will briefly sketch how measurement-based quantum computation can be represented in the ZX-calculus (Section~\ref{sec:MBQC}) and we discuss the question of how a ZX-diagram can be transformed back into a circuit (Section~\ref{sec:circuit-extraction}).

\subsection{Graph states}\label{sec:graph-state}

A particularly useful subset of the Clifford states are the \emph{graph states}. For each simple undirected graph $G=(V,E)$, we can define its corresponding graph state $\ket{G}$ as 
$$\ket{G}:= \prod_{(v,w)\in E} \text{CZ}_{v,w} \ket{+}^{\otimes \lvert V\rvert}.$$
I.e.~we prepare the $\ket{+\cdots +}$ state, where the number of qubits is equal to the number of vertices in the graph, and then for every edge in $G$ we apply a CZ gate between the corresponding qubits.%
\footnote{Graph states are also sometimes defined implicitly as the unique stabiliser state stabilised by a particular set of Pauli strings defined in terms of the graph.}

Recall from~\eqref{eq:CZ-in-ZX} that a CZ gate is represented in the ZX-calculus as follows:
\begin{equation}\label{eq:CZ}
	\text{CZ} \ = \ \input{CZ.tikz} 
\end{equation}
As furthermore a $\ket{+}$ state is a Z-spider with a single output, the translation of a graph into a graph state is particularly straightforward~\cite{DP1} in the ZX-calculus. An example:
\begin{equation}\label{eq:graph-state-ex}
\input{graph-state-ex.tikz}
\end{equation}

In words: every vertex becomes a Z-spider with an output attached to it. Each edge in the graph becomes an edge between the corresponding spiders, and this edge has a Hadamard box on it.

For graphs with many edges, writing Hadamard boxes on each edge might become cumbersome. To remedy this issue, \cite{cliffsimp} introduced a new notation for a \emph{Hadamard-edge}: a blue dotted edge instead of a black edge. To stick with our black and white colour scheme, we will instead write this as a grey dotted edge:
\begin{equation}\label{eq:def-had-edge}
\input{blue-edge-def.tikz}
\end{equation}

Any Clifford state is equal to a \emph{Graph State with Local Cliffords} (GSLC)~\cite{vandennest2004graphical}. Hence, given a $n$-qubit Clifford state $\ket{\psi}$, we can find a $n$-vertex graph $G$, and local Cliffords $U_1,\ldots, U_n$ such that $\ket{\psi} = (\bigotimes_{i=1}^n U_i) \ket{G}$.
A local Clifford in this context is a single-qubit Clifford unitary, i.e.~a unitary that can be written as a composition of Hadamard and S gates. In terms of ZX-diagrams, such a unitary can always be written as a composition of three phase gates where the phases are multiples of $\frac\pi2$:
\begin{equation}
\input{local-Clifford.tikz}
\end{equation}
Here, $k_i,k_i'\in \mathbb{Z}$, and the `colouring' ZXZ or XZX can be chosen based on what is more useful in context.
Indeed, our definition of the Hadamard gate in terms of spiders~\eqref{eq:had-euler1} was of this form.

For example, composing the graph state of~\eqref{eq:graph-state-ex} with some local Cliffords could give:
\ctikzfig{graph-state-local-clifford}

\subsection{Graph-like diagrams}\label{sec:graph-like}
In the previous sections we worked with general ZX-diagrams that can contain Z-spiders, X-spiders and Hadamard gates, connected in any valid manner.
For some purposes, especially when it comes to automated rewriting, it is however useful to reduce to a slightly more restricted type of ZX-diagram that we call a \emph{graph-like diagram}~\cite{cliffsimp}. Before giving an explicit description of those diagrams, let us give a simple algorithm that transforms an arbitrary ZX-diagram into a graph-like diagram:
\begin{enumerate}
	\item Convert all X-spiders into Z-spiders by pushing out Hadamards (by using \HadamardRule of Figure~\ref{fig:zx-rules}).
	\item Cancel all adjacent pairs of Hadamard gates using \HHRule.
	\item Fuse all connected spiders using \SpiderRule.
	\item View the remaining Hadamard gates between spiders as the Hadamard-edges of~\eqref{eq:def-had-edge}.
	\item Remove all self-loops using \eqref{eq:self-loop-removal} and \eqref{eq:had-self-loop-removal}.
	\item If two spiders are connected by multiple Hadamard-edges, we remove these using a variation of the Hopf-rule:
	\begin{equation}\label{eq:remove-double-edge}
		\input{double-had-edge.tikz}
	\end{equation}
\end{enumerate}

What remains after these simplifications is a ZX-diagram that contains only Z-spiders (no X-spiders), and the only connections between these Z-spiders are by single Hadamard-edges. Diagrams that have this property we call graph-like.
Hence, a graph-like diagram is fully specified by a simple graph (describing the spiders and the Hadamard-edge connectivity), a list of phases for the vertices, and an ordered list specifying to which vertices the inputs and outputs are connected (and whether this connection is a regular or Hadamard-edge).%
\footnote{There is actually a subtlety here that we are choosing to ignore. Namely, that an input wire could also be an output wire, without any spider appearing on it. We can either deal with this as a special case, or modify our algorithm so that it introduces identity spiders with \IdRule to prevent this from happening.}
As an example, we can rewrite the following circuit to a graph-like diagram as follows:
\begin{equation}\label{eq:graph-like-ex}
	\input{graph-like-ex.tikz}
\end{equation}

Graph states can be seen as a special case of graph-like diagrams, where all the phases are zero, and every spider is connected to a unique output (and the diagram has no inputs).

\subsection{Measurement-based quantum computation}\label{sec:MBQC}

An interesting way to think about ZX-diagrams is as a description of a \emph{measurement-based quantum computation} (MBQC).
The circuit model is only one of the ways in which we can present a quantum computation. In this model we start with a simple input state (usually $\ket{0\cdots 0}$), then apply a sequence of unitary gates, each usually only affecting a small number of qubits, and finally measure some of the qubits.
In this model we could say that the `actual computation' happens through the application of gates, as the input state and the measurements are often the same for entirely different computations.
In contrast, in MBQC the computation happens through the clever application of particular measurements. In the most well-studied model of MBQC, the \emph{one-way model}~\cite{MBQC1}, some highly entangled graph state is prepared, often in some standard way that doesn't depend on the computation to be performed, and then single qubit measurements are applied, where later measurements can depend on previous outcomes. 
Crucially, each qubit can be measured in a different measurement basis, which allows us to inject phases into the state and to shape the entanglement in a controlled way. As measurements are fundamentally non-deterministic we need ways to correct for `wrong' measurement outcomes in order to implement a deterministic computation.

We can write a computation in the one-way model of MBQC straightforwardly in the ZX-calculus. Indeed, the development of the ZX-calculus was originally motivated by the desire to describe MBQC in a better way~\cite{DuncanMBQC}. We give here a brief description. For a thorough treaty, see~\cite{wetering-gflow}. 

We have already seen how to represent a graph state in the ZX-calculus in Section~\ref{sec:graph-state}. It hence remains to show how to represent measurements in the ZX-calculus. In the one-way model, measurements are usually restricted to measurement bases that live on the three principal axes of the Bloch sphere. 
For instance, on the XY-plane we have measurement bases of the form $\ket{\pm_{XY,\alpha}} := \ket{0} \pm e^{i\alpha}\ket{1}$, 
while measurements on the YZ-plane are $\ket{\pm_{YZ,\alpha}} = \ket{+} \pm e^{i\alpha}\ket{-}$ (XZ-plane measurements are defined similarly, but we won't need these here). 
Here we consider the `$+$' state to correspond to the `desired' outcome of the measurement, 
while the `$-$' state corresponds to the `wrong' outcome that needs to be corrected in order to implement the desired outcome.
These measurement outcomes, represented as effects, correspond to simple ZX-diagrams:
\begin{align*}
	&\bra{+_{XY,\alpha}} \ = \ \input{bra-Z-alpha.tikz} \qquad \qquad\quad \bra{-_{XY,\alpha}} \ = \ \input{bra-Z-alpha-pi.tikz} \\
	&\bra{+_{YZ,\alpha}} \ = \ \input{bra-X-alpha.tikz} \qquad \qquad\quad \bra{-_{YZ,\alpha}} \ = \ \input{bra-X-alpha-pi.tikz}
\end{align*}
A computation in the one-way model is then represented by for instance:
\begin{equation}
	\input{MBQC-example.tikz}
\end{equation}
Here we took the graph state from~\eqref{eq:graph-state-ex} and measured each of its 4 qubits. The Boolean variables $a$, $b$, $c$ and $d$ represent each of the measurement outcomes and are the output of this computation. In this example we had pre-determined measurement phases (respectively $\frac\pi2$, $0$, $\frac\pi4$ and $\pi$), but in general each of these phases, and even whether we use the XY-plane or YZ-plane, can depend on earlier measurement outcomes.

For a complete calculation in the one-way model we have a fixed input and measure all the qubits and hence this is represented by a scalar ZX-diagram, i.e.~a diagram with no inputs or outputs. It is however often helpful to think of `fragments' of a computation that can be composed together to create more complicated computations. These fragments have inputs and outputs just like a quantum circuit would.
For instance, a \emph{measurement fragment} that implements a CNOT gate is given by the following diagram:
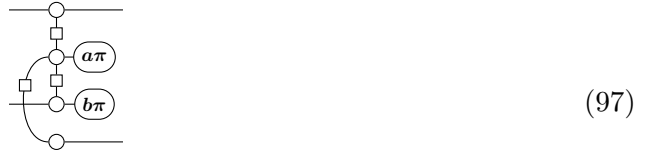
\begin{equation}\label{eq:MBQC-CNOT}
	\input{MBQC-CNOT.tikz}
\end{equation}
Here the top qubit is both an input and an output and hence isn't measured. The third qubit is an input, but is not an output and hence is measured. The fourth qubit is an output and hence also isn't measured. This fragment implements a CNOT gate up to a known Pauli error that depends on the measurement outcomes $a$ and $b$. We can calculate this Pauli error by simplifying the diagram:
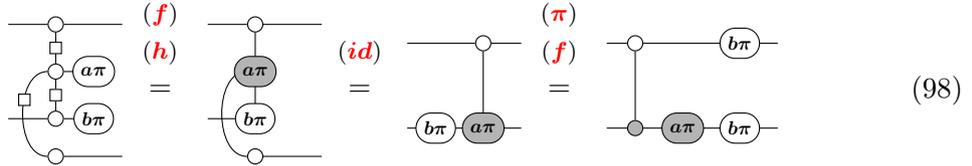
\begin{equation}
	\input{MBQC-CNOT-simp.tikz}
\end{equation}
Hence we end up with the circuit $(Z^b\otimes (Z^bX^a))\circ \text{CNOT}$. Since the error is a known Pauli, later measurements can be adapted to absorb these errors so that the overall effect of this measurement fragment is the application of a CNOT gate. This process of pushing Pauli errors outside of the pattern is known as \emph{feed-forward} in the MBQC literature.

The diagram~\eqref{eq:MBQC-CNOT} is very close to being a graph-like ZX-diagram. Indeed, the only thing needed to make it graph-like is to fuse the measurement effects into the spider it is attached to. If the fragment also contained YZ-plane measurements, then we could make the diagram graph-like by changing the colour of the resulting X-spiders using \HadamardRule.

It turns out that the converse is also true: any graph-like ZX-diagram can be described as a measurement fragment. For instance, we can convert the graph-like diagram of~\eqref{eq:graph-like-ex} to a measurement fragment as follows:
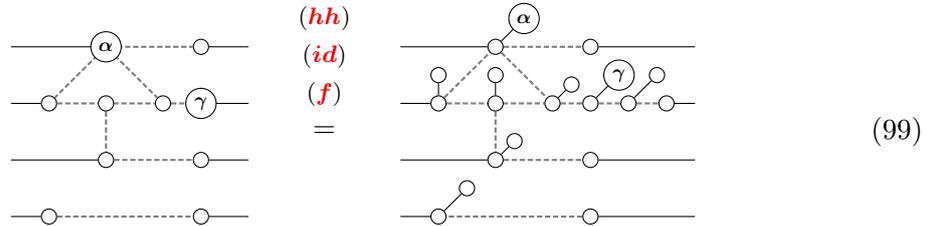
\begin{equation}
	\input{graph-like-to-MBQC-form.tikz}
\end{equation}
Here we introduce additional identity spiders and measurements, because the $\gamma$ phase happens on an output qubit, which can't be measured itself. Note that this pattern represents a `post-selected' measurement fragment where we assume that the measurement outcome is always the correct one, so that we don't have to deal with measurement errors that have to be fed forward. 
It is in general not possible to write a (graph-like) ZX-diagram as a measurement pattern in such a way that measurement errors can be taken into account. Indeed, as a ZX-diagram can represent an arbitrary linear map and deterministic measurement patterns are always isometries, we would not expect this to be the case. Consider for instance the extreme case where the ZX-diagram has inputs but no outputs: there would be nowhere for the errors to be fed forward into.

\subsection{Local complementation and pivoting}

As mentioned in Section~\ref{sec:graph-state}, each Clifford state can be represented as a graph state composed with some local Cliffords. This graph state is however not unique: we can `absorb' some of the local Cliffords by changing the graph. The main way to do this is by applying \emph{local complementations}  of the graph.

Let $G$ be a graph, and $u$ a vertex in $G$. The \emph{local complementation of $G$ about $u$}, commonly written as $G\star u$, is the graph which has the same vertices and edges as $G$, except that the neighbourhood of $u$ is complemented, i.e.~two neighbours $v$ and $w$ of $u$ are connected in $G\star u$ if and only if they are \emph{not} connected in $G$~\cite{kotzig}.
For example:
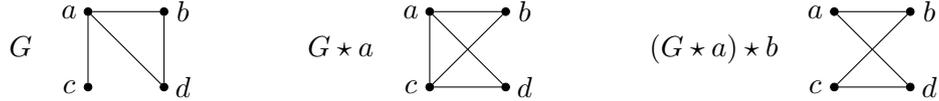
\begin{equation*}
G\quad\input{graph1-lab.tikz}\qquad\qquad G\star a\quad\input{graph1-lab-1.tikz}\qquad\qquad (G\star a) \star b\quad\input{graph1-lab-2.tikz}
\end{equation*}
Interestingly, if we have a graph state $\ket{G}$, we can get a graph state $\ket{G\star u}$ just by applying some local Cliffords on the original graph state~\cite{vandennest2004graphical}.
In fact, it turns out that a graph state $\ket{G_1}$ can be transformed to a graph state $\ket{G_2}$ using local Clifford unitaries if and only if there is a series of local complementations that transforms $G_1$ into $G_2$~\cite{elliott2008graphical}.

The transformation of a graph state via local complementation can be represented in the ZX-calculus as follows:
\begin{equation}\label{eq:local-comp-zx}
\input{local-comp-ex.tikz}
\end{equation}
Here $N(u)$ denotes the neighbourhood of the vertex $u$ that we do the local complementation about.
This equation was originally proved in the ZX-calculus in Ref.~\cite{DP1}. A more accessible proof is provided in Ref.~\cite[Prop.~9.125]{CKbook}.

Given a connected pair of vertices $(u,v)$ in $G$, we define the \emph{pivot of $G$ along $uv$}, written as $G\wedge uv$, as the graph $G\star u\star v\star u$. Note that it turns out that this graph is equal to $G\star v \star u \star v$, and hence the ordering of $u$ and $v$ is not important.

On a graph, pivoting consists in exchanging $u$ and $v$, and complementing the edges
between three particular subsets of the vertices: the common neighbourhood of $u$ and $v$ (i.e.~$N_G(u)\cap N_G(v)$), the exclusive neighbourhood of $u$ (i.e.~$N_G(u)\setminus (N_G(v)\cup \{v\})$), and exclusive neighbourhood of $v$ (i.e.~$N_G(v)\setminus (N_G(u)\cup \{u\})$). Schematically:
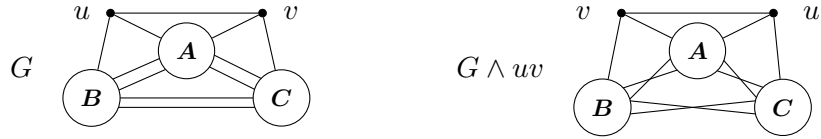
\[G \quad\input{pivot-L.tikz}\qquad\qquad \quad G\wedge uv \quad\input{pivot-R.tikz}
\]

As a pivot is just a series of local complementations, it can also be performed on a graph state by the application of a particular set of local Cliffords. Indeed, in terms of ZX-diagrams, we have:
\begin{equation}\label{eq:pivot-zx}
\input{pivot-desc.tikz}
\end{equation}
I.e.~we can perform a pivot on the graph state by applying a Hadamard to the vertices we pivot along, and applying a Z gate to the vertices in their common neighbourhood.

\subsection{Simplification of Clifford ZX-diagrams}\label{sec:simp-clifford}

In Section~\ref{sec:graph-like} we introduced the notion of graph-like diagrams. Any ZX-diagram can be transformed into a graph-like diagram, but if the original diagram was Clifford, then the resulting diagram can actually be simplified considerably more using variations of the local complementation and pivoting rule described above.

In particular, using the local complementation rule~\eqref{eq:local-comp-zx}, we can prove the following simplification~\cite{cliffsimp}:
\begin{equation}\label{eq:lc-simp}
\input{lc-simp.tikz}
\end{equation}
In words: if we have a spider (here marked on the left-hand side by a $*$) with a $\pm\frac{\pi}{2}$ phase in a graph-like diagram that is \emph{internal}, i.e.~that is not connected to inputs or outputs but only to other spiders, then we can remove it from the diagram by complementing the connectivity on its neighbourhood and changing some phases.%
\footnote{In the equation~\eqref{eq:lc-simp} we display a fully connected graph on the right-hand side, but if there were already edges present between some of the spiders, then the resulting double edges would be removed by~\eqref{eq:remove-double-edge}, and hence this indeed results in a local complementation}

Note that in a Clifford ZX-diagram each spider has a phase $k\frac\pi2$ for some $k\in\mathbb{Z}$. Using the rule above repeatedly on a graph-like Clifford diagram we can remove \emph{all} internal spiders with a $\pm \frac\pi2$ phase. Hence, the only internal spiders, those not connected to an input or output, that remain are those that have a $0$ or $\pi$ phase.
Most of these internal spiders can also be removed, by using a variation on the pivot rule~\eqref{eq:pivot-zx}:
\begin{equation}\label{eq:pivot-simp}
\input{pivot-simp.tikz}
\end{equation}
Here, the marked spiders on the left-hand side are internal connected spiders with a $0$ or $\pi$ phase. On the right-hand side, these spiders are removed, at the cost of complementing their neighbourhood in the manner described by the pivot rule, and introducing some phases.
Using~\eqref{eq:pivot-simp} repeatedly we can remove the remaining internal spiders that are connected to at least one other internal spider. The remaining internal spiders are then those that carry a $0$ or $\pi$ phase and are connected only to boundary spiders. It turns out that these internal spiders can also be removed by using a different variation of the pivot rule; see~\cite{cliffsimp}.

Hence, given a Clifford ZX-diagram, we can first convert it into a graph-like diagram, and then by repeatedly applying~\eqref{eq:lc-simp} and~\eqref{eq:pivot-simp} remove all internal spiders so that only the boundary spiders are left.

In particular, if our starting diagram was a state (that is, it has no inputs), then the resulting diagram is very close to being a graph state with local Cliffords (the only complication coming from the fact that multiple outputs could be connected to the same spider, a problem that is easily solved by introducing some dummy spiders using \IdRule).
This means that this procedure gives a straightforward diagrammatic method to convert a Clifford state into GSLC form.

Furthermore, if the starting diagram has no inputs or outputs, so that it represents a scalar, for instance if we are calculating an amplitude $\bra{0\cdots 0}C\ket{0\cdots 0}$ of some Clifford circuit $C$, then our procedure removes \emph{all} spiders that have any connection to another spider. 
The resulting diagram hence consists of a bunch of arity-0 spiders, that each carry a phase of $0$ (representing a scalar $2$; cf.~\eqref{eq:scalars-ZX}) or a phase $\pi$ (representing the scalar $0$). 
Since this simplification procedure runs in polynomial time on the number of spiders in the original diagram, this shows that we can calculate amplitudes of Clifford circuits efficiently. We have hence given a diagrammatic proof of the Gottesman-Knill theorem.%
\footnote{There are a couple of complications we are skipping over here. Namely, we have not presented scalar-accurate rewrite rules. So to actually use our method for amplitude calculation, one would also have to take into account the scalar accumulated after each applied rewrite rule. Additionally, the Gottesman-Knill theorem also states that you can efficiently calculate \emph{marginal} probabilities as well, which our procedure in its current form does not do. This can however be fixed by using the \emph{doubling construction} of~\cite{CKbook} that we describe in Section~\ref{sec:doubling} that allows the representation of non-pure quantum processes in the ZX-calculus.}

There is one more consequence of this simplification procedure. There is an efficient algorithm for recognizing whether two GSLC states are equal, which consists of a series of local complementations that transforms one into the other~\cite{elliott2008graphical}. 
So suppose we are given two Clifford ZX-diagrams that implement the same matrix. As our procedure reduces Clifford diagrams to something GSLC-like, we could use this algorithm to find a series of local complementations that transforms the first reduced diagram into the second. 
As all our rewrite rules to get to this point were derived starting from the rules in Figure~\ref{fig:zx-rules}, we hence see that the rules of Figure~\ref{fig:zx-rules} suffice to transform any Clifford diagram into any other that implements the same matrix. Hence, the ZX-calculus with the rules of Figure~\ref{fig:zx-rules} is \emph{complete} with respect to the Clifford fragment. This result, using a related method, was first proven by Backens~\cite{BackensCompleteness}.

While we have used the local complementation rules here to simplify Clifford diagrams, they are also useful to simplify non-Clifford diagrams. Indeed, these rules were first introduced in~\cite{cliffsimp} to simplify non-Clifford circuits and with some additional variations on the pivoting rule, they were used in~\cite{wetering-gflow} to simplify MBQC patterns.

\subsection{From ZX-diagrams to circuits}\label{sec:circuit-extraction}

We started this paper with the observation that a quantum circuit can be transformed into a ZX-diagram in a straightforward way (simply by translating each of the gates in turn). The converse question, of how a ZX-diagram can be transformed into an equivalent circuit, is much harder and is known as the \emph{circuit extraction} problem~\cite{cliffsimp}.

Let us first note that the general case is not solvable: as a quantum circuit is unitary and ZX-diagrams can represent arbitrary linear maps, most ZX-diagrams are not
writable as circuits. There are then two ways in which we can approach the issue of translating from ZX-diagrams to circuits: we can either expand the scope of what we consider a quantum circuit, or we can limit the type of ZX-diagrams we try to translate.

Following the first approach, a way to let quantum circuits represent arbitrary linear maps between qubits is to allow ancillae and post-selected measurements. With these allowances, a translation from ZX-diagrams to circuits is straightforward. Indeed, each Z-spider can be presented as a series of CNOT gates with some qubits prepared in the $\ket{0}$ or $\ket{+}$ state or post-selected to the $\bra{0}$ or $\bra{+}$ effect:
\begin{equation}
	\input{spiders-as-circuits.tikz}
\end{equation}
In this approach, phases on spiders become phase gates, spiders with no output wires become post-selections and spiders with no input wires become ancilla preparations. Higher-arity spiders can be decomposed into lower-arity ones by unfusing the spiders and hence can also be transformed into post-selected circuits.

For the second approach we need to restrict to ZX-diagrams that represent unitary computations. Given a ZX-diagram with the promise that it implements a unitary matrix there is a straightforward way to turn it into an equivalent circuit: calculate its matrix and synthesise a circuit using any synthesis algorithm. Unfortunately this method is likely to take time exponential in the number of inputs and outputs and hence isn't very practical. The real question then is whether there is an \emph{efficient} algorithm that transforms a given ZX-diagram that is promised to be unitary into a circuit. 
It is unlikely that such an algorithm exists, because the promise that it is unitary is a global property of the diagram: it can still mean that large chunks of the diagram represent a non-unitary linear map that is somehow canceled out by a later portion of the diagram. Hence, such an algorithm would somehow have to be using knowledge about the global structure of the diagram to nonlocally construct a circuit out of it. It is unclear how this would be possible without the ability to calculate which unitary the diagram implements so that it would implicitly calculate the matrix of the diagram.
Similarly, it is unlikely that there is an efficient algorithm for determining whether a given ZX-diagram implements a linear map that is (proportional to) a unitary, as in the 0-input, 0-output case this would mean we could efficiently determine whether an amplitude of a circuit is zero or not, which comes awfully close to being a \textbf{QMA}-complete~\cite{bookatz2012qmacomplete} problem.

The only hope then for an efficient algorithm is for us to know more about the \emph{local} structure of the diagram. 
The current best-known algorithm~\cite{wetering-gflow} can extract a circuit out of any ZX-diagram that has a \emph{gflow}, a combinatorial property for graph-like ZX-diagrams and MBQC patterns that corresponds to the possibility of consistently feed-forwarding Pauli errors from measurement outcomes at the locations of each of the spiders~\cite{GFlow,DP2}. The description and correctness of the algorithm is a bit involved so let us demonstrate its functioning in the simplest case: extracting a circuit from Clifford normal forms.

The algorithm described in Section~\ref{sec:simp-clifford} removes all internal spiders from a Clifford ZX-diagram. All the remaining spiders are hence connected to at least one input or output. The resulting diagram then looks something like this:
\begin{equation}
	\input{gslc-normal-form.tikz}
\end{equation}
I.e.~we have a left row of spiders connected to inputs and a right row of spiders connected to outputs, with a possible Hadamard on the input/output wire and a phase on the spiders that is a multiple of $\frac\pi2$. The spiders on the left and on the right are connected in some bipartite way by Hadamard-edges described by a biadjacency matrix $\mathcal{P}$. Additionally, there can be arbitrary Hadamard-edges between the spiders on the same side.

The first step in transforming this into a circuit is to unfuse the Hadamard-edges connected on the same side into CZ gates (cf.~\eqref{eq:CZ}), and the phases into Z-phase gates. By changing the colour of the spiders on the right-hand side the middle part of the diagram becomes something that corresponds to a classical linear map:
\begin{equation}
	\input{gslc-normal-form2.tikz}
\end{equation}
Indeed, the Z-spiders on the left-hand side of $\mathcal P$ copy computational basis states, which are then fed into the X-spiders on the right which act as XORs. Hence, the output of this part of the diagram can be described in terms of parities of the inputs. For instance: $\ket{x_1, x_2, x_3, x_4} \mapsto \ket{x_1 \oplus x_2, x_1 \oplus x_3, x_4, x_3}$. Such an operation is known as a \emph{linear Boolean} function. These linear maps can always be described by a circuit of CNOT gates~\cite{markov2008optimal}. The way this works is that the biadjacency matrix $\mathcal P$ is seen as a matrix over the finite field $\mathbb F_2$. This matrix is then reduced to the identity using a Gaussian elimination algorithm. Each row operation needed to do this elimination corresponds to a CNOT gate in the final circuit. After this middle part of the diagram is synthesised into a CNOT circuit, we have indeed succeeded in rewriting this Clifford diagram into a circuit.

The circuit extraction algorithm of~\cite{cliffsimp} generalises this approach to graph-like ZX-diagrams that have internal spiders. The Gaussian elimination then does not fully reduce the matrix, but just a part of it, which allows one or more internal spiders to become a boundary spider. This is then repeated until all spiders are `consumed' and the diagram is transformed into a circuit. 
The caveat here is that the Gaussian elimination must be able to reduce at least one row of the matrix to make progress. This is where the promise that the ZX-diagram has a gflow is needed. 
The algorithm of~\cite{wetering-gflow} further generalises this approach by being able to deal with phase gadgets (cf.~Section~\ref{sec:phase-polynomial}) that are present in the diagram.

\section{ZX-diagrams, categorically}\label{sec:zx-category}
The ZX-calculus originated from the computer science department of Oxford University and investigations into the relations between category theory and quantum theory. As a result, many papers using the ZX-calculus are written for computer scientists and often use language from category theory. 
This might be hard to parse if one has a different background, say, in physics.
In this section we therefore give a brief summary of some of the conventions and definitions related to category theory one can encounter reading ZX-calculus papers.
A reader not interested in the formal background of the ZX-calculus can safely skip to Section~\ref{sec:ZH}. 
A reader looking for more formal background is invited to consider Vicary and Heunen's book on categorical quantum mechanics~\cite{heunen2019categories}.

\subsection{Category theory background}

The ZX-calculus is often cast as a particular type of \emph{category}. A category is a mathematical structure that consists of \emph{objects} and \emph{morphisms} (the latter are sometimes also called \emph{arrows}) satisfying certain properties~\cite{mac2013categories}. Probably the most familiar category is \textbf{Sets}. In this category the objects are standard mathematical sets, and the morphisms are the functions between these sets.
Another familiar category is \textbf{Vec}$_\C$, where the objects are complex vector spaces and the morphisms are linear maps between these vector spaces.

In a category, given a morphism $f:A\rightarrow B$ from object $A$ to object $B$, and a morphism $g:B\rightarrow C$ from object $B$ to object $C$, we can \emph{compose} these morphisms to get a morphism $g\circ f:A\rightarrow C$. Composition is required to be associative: $h\circ (g\circ f) = (h\circ g)\circ f$. In addition, for each object $B$ we have a special \emph{identity} morphism $\id_B:B\rightarrow B$ that acts as the identity for the composition operation: $\id_B\circ f = f$ and $g\circ \id_B = g$.
These rules for composition and identity are the only things that are required of a standard category.

For ZX-diagrams we cannot only compose sequentially, but also in parallel using a tensor product. A category that has tensor products in some suitable manner is called a \emph{monoidal} category~\cite{selinger2010survey}. The category of ZX-diagrams is in fact \emph{symmetric monoidal}, which means that $A\otimes B$ is always isomorphic to $B\otimes A$ for all objects $A$ and $B$ (in a particularly canonical manner). Monoidal categories play a foundational role in the field of applied category theory. A good introduction for the interested reader is given by Fong and Spivak's book~\cite{fong2019invitation}.

We will call this category of ZX-diagrams \textbf{ZX}. The objects in this category are the natural numbers: the object $n$ corresponds to $n$ parallel wires. A morphism from $n$ to $m$ in this category is a ZX-diagram with $n$ inputs and $m$ outputs. Composition of morphisms is the regular sequential composition of ZX-diagrams. The tensor product of two objects $n$ and $m$ is defined to be $n\otimes m:= n+m$, and the tensor product of two morphisms (i.e.~two ZX-diagrams) is given by stacking the two diagrams on top of one another.

A symmetric monoidal category where the objects are natural numbers and the tensor product is given by addition of the objects is also called a \emph{PROP} (which stands for PROduct and Permutation)~\cite{maclane1965categorical}, and hence sometimes authors will speak of the `PROP of ZX-diagrams'.

ZX-diagrams have cups and caps (see~\eqref{eq:cup-cap-def}) which allows inputs to be transformed into outputs and vice versa. Without going into details, a symmetric monoidal category which has cup and cap-like structure is called \emph{compact closed}~\cite{kelly1980coherence}.
Compact closed category are the fundamental structure in categorical quantum mechanics~\cite{abramsky2004categorical}.

\subsection{Interpretation of a ZX-diagram}

In this paper we intentionally blur the line between a ZX-diagram and the linear map it represents. However, in many papers dealing with more foundational questions about the ZX-calculus it is often desirable to be more strict about which object it is we are dealing with.
In this case, given a ZX-diagram $D$, we denote the linear map it represents by $\intf{D}$. This construction $\intf{\cdot}$ that associates a linear map to a diagram is called the \emph{interpretation} of the diagram.

The interpretation of a ZX-diagram is a linear map between qubits. Such linear maps live in a category that we call \textbf{Qubit}. This is a \emph{subcategory} of \textbf{Vec}$_\C$, meaning that the objects correspond to complex vector spaces and the morphisms are linear maps. In fact, \textbf{Qubit} is a PROP, where we interpret the object $n$ as the vector space $(\C^2)^{\otimes n} \cong \C^{2^n}$. The tensor product $n\otimes m = n+m$ corresponds to the regular tensor product of the vector spaces $\C^{2^n}\otimes \C^{2^m} \cong \C^{2^{n+m}}$.
The morphisms of \textbf{Qubit} are just linear maps $f:\C^{2^n}\rightarrow \C^{2^m}$ (i.e~\textbf{Qubit} is a \emph{full} subcategory of \textbf{Vec}$_\C$).
It turns out that the interpretation gives a type of mapping between \textbf{ZX} and \textbf{Qubit} that is called a \emph{functor}.

A functor $F:\textbf{C}\rightarrow \textbf{D}$ between two categories \textbf{C} and \textbf{D} is something that maps each object $A$ in \textbf{C} to an object $F(A)$ in \textbf{D}. It also maps each morphism $f:A\rightarrow B$ in \textbf{C} to a morphism $F(f):F(A)\rightarrow F(B)$ in \textbf{D}. A functor is required to map identities to identities, $F(\id_A)=\id_{F(A)}$, and to preserve composition, $F(g\circ f) = F(g)\circ F(f)$.
For example, there is the functor $U:\textbf{Vec}_\C\rightarrow \textbf{Sets}$ that sends a vector space $V$ to the set $U(V)=V$ and a linear map $f:V\rightarrow W$ to the same function $U(f) = f$. This functor essentially does nothing except change the context in which we view the objects and morphisms. Such a functor is known as a \emph{forgetful} functor.

The interpretation of a ZX-diagram is a functor $\intf{\cdot}:\textbf{ZX}\rightarrow \textbf{Qubit}$. The objects of both \textbf{ZX} and \textbf{Qubit} are the natural numbers and we simply set $\intf{n} = n$ for all $n\in \N$. As the morphisms of \textbf{ZX} are precisely ZX-diagrams, we set $\intf{D}$ to be the linear map the diagram $D$ implements. 
We need to check that $\intf{\id_n} = \id_n$ and that $\intf{D\circ D'} = \intf{D}\circ \intf{D'}$, but this is immediate from how we defined the interpretation of a ZX-diagram iteratively starting from the atomic generators. 
Note that we also have $\intf{D\otimes D'} \cong \intf{D}\otimes \intf{D'}$, i.e.~that the tensor product of diagrams corresponds to the tensor product of the underlying linear maps. This makes $\intf{\cdot}$ into a \emph{strong monoidal} functor.

In Section~\ref{sec:universality} we showed that for any linear map between qubits we can find a ZX-diagram that represents it, i.e.~that the ZX-calculus is \emph{universal}. In terms of the interpretation functor $\intf{\cdot}$ we can phrase this as stating that for any morphism $f:\C^{2^n}\rightarrow \C^{2^m}$ in \textbf{Qubit} we can find a morphism $D:n\rightarrow m$ in \textbf{ZX} such that $\intf{D} = f$. A functor which has this property, which is reminiscent of the surjectivity of a function, is called \emph{full}.

\subsection{Rewriting and completeness}

In the previous subsection we saw that universality of the ZX-calculus corresponds to fullness of the interpretation functor. For a functor, being full is what being surjective is to a function. An immediate question is then: what about injectivity?
An `injective' functor is called \emph{faithful}: for a faithful functor $F:\textbf{C}\rightarrow \textbf{D}$ an equality $F(f:A\rightarrow B)=F(g:A\rightarrow B)$ implies $f=g$.

The functor $\intf{\cdot}:\textbf{ZX}\rightarrow \textbf{Qubit}$ is \emph{not} faithful. This is because there are multiple different ZX-diagrams that represent the same linear map. This of course does not come as a surprise as we have been constantly rewriting ZX-diagrams with the promise that the linear maps they represent are the same.

The solution to this `problem' is to consider a modified category \textbf{ZX}$_{/\sim}$ where ZX-diagrams that can be transformed into one another by the rewrite rules of Figure~\ref{fig:zx-rules} are identified with one another. The way to do this is to consider the smallest equivalence relation $\sim$ with the following properties:
\begin{itemize}
	\item $D_1\sim D_2$ for each of the equalities $D_1=D_2$ in Figure~\ref{fig:zx-rules}.
	\item If $D_1\sim D_2$, then also $(E\circ D_1\circ C)\sim (E\circ D_2\circ C)$ for all diagrams $C$ and $E$. In other words: if two diagrams are equivalent, then they remain equivalent when some other diagrams are composed with them.
	\item If $D_1\sim D_2$, then also $(E\otimes D_1\otimes C)\sim (E\otimes D_2\otimes C)$ for all diagrams $C$ and $E$.
\end{itemize}
These rules for the equivalence relation $\sim$ correspond precisely to how we have been using ZX-calculus rewrite rules intuitively in the previous sections.

We can now define the morphisms of \textbf{ZX}$_{/\sim}$ to be equivalence classes $[D]_{/\sim}$ of diagrams under the relation $\sim$. Using the properties of $\sim$ above it is straightforward to verify that $[D_1]_{/\sim}\circ [D_2]_{/\sim} := [D_1\circ D_2]_{/\sim}$ gives a well-defined composition of these classes. We can similarly define a tensor product, and hence \textbf{ZX}$_{/\sim}$ is again a PROP.

For the interpretation functor $\intf{\cdot}:\textbf{ZX}_{/\sim}\rightarrow \textbf{Qubit}$ we wish to set $\intf{[D]_{/\sim}} := \intf{D}$. For this to be well-defined we need the value of $\intf{D}$ to not depend on the representative of the equivalence class, and hence we need to check that if $D_1\sim D_2$, then $\intf{D_1}=\intf{D_2}$. As $\sim$ simply asks whether two diagrams are rewritable into one another this property asks that rewriting should preserve the linear map of the diagram, i.e.~that our rewriting system is \emph{sound} with respect to the interpretation $\intf{\cdot}$. Due to the nature of $\sim$ it suffices to check this only for the equalities of Figure~\ref{fig:zx-rules}, for which it is indeed true.

Hence, we have an interpretation functor $\intf{\cdot}:\textbf{ZX}_{/\sim}\rightarrow \textbf{Qubit}$ that maps a class of equivalent ZX-diagrams to the linear map they represent. 
Is this functor faithful? Equivalently, given two diagrams $D_1$ and $D_2$ with $\intf{D_1}=\intf{D_2}$, do we have $D_1\sim D_2$?
This property, that two ZX-diagrams which represent the same linear map should be rewritable into one another is known as \emph{completeness}.
The ruleset we presented in Figure~\ref{fig:zx-rules} is \emph{not} complete (and hence our functor is not faithful). This means that there are in fact ZX-diagrams that represent the same linear map, while there is no set of rewrites that will transform these ZX-diagrams into one another.
Fortunately, there exist extended rulesets that \emph{are} complete. We will encounter such an extended ruleset in Section~\ref{sec:ZH} and we discuss the question of completeness in more detail in Section~\ref{sec:completeness}.
Note that, formally speaking, completeness is a property of the functor $\intf{\cdot}$. We could consider a different interpretation of ZX-diagrams into the category $\textbf{Qubit}$ or even into an entirely different category, and the resulting interpretation \emph{could} be complete (i.e.~the interpretation functor could be faithful). 
In addition, we can also consider interpretation functors for subcategories of \textbf{ZX}, such as \textbf{ZX}$^{\frac{\pi}{2}}$ which consists of ZX-diagrams where all spiders have phases that are multiples of $\frac\pi2$ (i.e.~they are Clifford ZX-diagrams as studied in Section~\ref{sec:clifford}). These interpretation functors can then also have different properties. In the specific case of \textbf{ZX}$^{\frac{\pi}{2}}$, the interpretation functor is faithful (i.e.~the calculus is complete, see Section~\ref{sec:clifford}), but it is of course no longer full, as only Clifford maps can be represented by such diagrams.

\subsection{!-boxes}

In several places throughout this paper we have written ZX-diagrams with `$\cdots$' in them to denote that there is an arbitrary number of wires there. For instance in~\SpiderRule and~\CopyRule in Figure~\ref{fig:zx-rules}. More formally, we can view equations with 'dots` in them as a family of equations parametrised by the number of wires. 
Some issues or confusion may however arise when we have multiple `$\cdots$' in the same equation, such as in~\eqref{eq:bialgebra-rule-many} where the `dots' on the inputs can denote a different number of wires from the dots on the outputs. 

In order to formalise the use of these dots to represent an arbitrary number of wires, the notion of \emph{!-boxes} was introduced~\cite{dixon2009graphical,EPTCS143.5}.
A !-box (pronounced `bang-box') in a ZX-diagram is a box surrounding a part of the diagram that is allowed to fan out arbitrarily. In other words, the contents of a !-box is allowed to be copied an arbitrary number of times (including zero times). For instance:
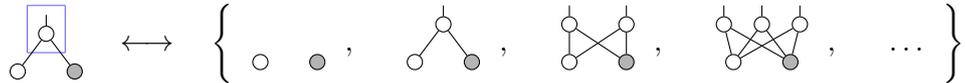
\[ \input{bang-box-example.tikz} \quad \longleftrightarrow \quad
 \left\{
 \ \ \input{bang-box-example0.tikz}\ \ ,\quad
 \ \ \input{bang-box-example1.tikz}\ \ ,\quad
 \ \ \input{bang-box-example2.tikz}\ \ ,\quad
 \ \ \input{bang-box-example3.tikz}\ \ ,\quad
 \ \ \ldots\ \  \right\}
\]
Here the blue box denotes the !-box. This family of diagrams is well-defined, because each of our generators, Z-spiders and X-spiders, are allowed to have an arbitrary number of wires attached to them. Note that we can't have a Hadamard box be attached to something inside a !-box, because a Hadamard box always has arity 2 (unless we generalise it to the multi-arity H-boxes of Section~\ref{sec:ZH}).

We can also write equations of diagrams with !-boxes in them. For instance, we can express \CopyRule as follows:
\[
\input{copy-bangboxed.tikz} \ \ \longleftrightarrow \ \ 
\left\{
\ \input{copy-bb0.tikz}\ , \ \ 
\ \ \input{copy-bb1.tikz}\ ,\ \ 
\ \ \input{copy-bb2.tikz}\ ,\ \ 
\ldots\ 
\right\}
\]
The dashed box in the first equation denotes an empty diagram, corresponding to the !-box being expanded zero times.

Some rules require more than one !-box to be expressed.
For instance, the generalised bialgebra equation~\eqref{eq:bialgebra-rule-many} can be expressed as follows:
\begin{equation}\label{eq:ZX-bialgebra-bb}
	\input{ZX-bialgebra-bb.tikz}
\end{equation}
When writing an equation involving !-boxes it is important that the !-boxes on the left-hand side are matched 1-to-1 to those on the right-hand side. In~\eqref{eq:ZX-bialgebra-bb} it is clear that the left !-boxes of each side of the equation match, because otherwise the equations would not be well-typed (they would have a different number of inputs). Formally, we would have to give a `name' to each !-box so that it is clear which corresponds to which on each side of the equation. In practice, it is often clear how they correspond by their positioning in the diagram.
Consider for instance the spider fusion rule \SpiderRule expressed using multiple !-boxes:
\begin{equation}
	\input{spider-rule-BB.tikz}
\end{equation}

!-boxes can also overlap and be nested (but note that when nested in a rule, they must be nested in the same way on both sides of the equation). 
For instance, we can express (a special case of) the pivot simplification rule~\eqref{eq:pivot-simp} using !-boxes as follows:
\begin{equation}
	\input{pivot-BB.tikz}
\end{equation}
Note that not all repeated structure can be expressed using !-boxes. For instance, the iterated application of a linear map, or the notion of a fully connected graph as in~\eqref{eq:lc-simp}.%
\footnote{This first issue can be fixed through the definition of a different type of box as suggested in~\cite{MillerBakewell2020finite}. For~\eqref{eq:lc-simp} specifically, the issue of a fully connected graph can be sidestepped by viewing Hadamard boxes as H-edges so that they can cross into !-boxes. A fully connected graph can then be represented by a vertex inside a !-box with an H-box looping to the same vertex that goes outside the !-box.}

!-boxes are used in the software Quantomatic~\cite{kissinger2015quantomatic} in order to express families of diagrammatic rewrite rules. While !-boxes are not necessary for many manual derivations, they do sometimes become essential for expressing particularly complicated families of rewrite rules. For instance, normal form diagrams in the ZH-calculus~\cite{backens2018zhcalculus} are naturally expressed using a variation on !-boxes called \emph{annotated !-boxes}, where a !-box is labelled by an element from some finite set and elements inside the !-box are allowed to use this element as a variable.
Complicated rewrite rules such as the graphical Fourier transform, cf.~Section~\ref{sec:fourier} or~\cite{GraphicalFourier2019}, or the hyper-pivot rule~\cite{Lemonnier2020hypergraph} also require (annotated) !-boxes to be expressed concisely.

\subsection{Monoids, comonoids, Frobenius algebras and bialgebras}\label{sec:frobenius}

The central generators of ZX-diagrams, the spiders, have an algebraic structure that is used far beyond just the ZX-calculus. In this section we will give some definitions regarding this structure, each building on the previous.

One of the most basic algebraic structures is the \emph{monoid}. A monoid is a set $M$ with a binary operation $\cdot$ that is associative, $a\cdot(b\cdot c) = (a\cdot b)\cdot c$, and has a unit, $1\cdot a = a = a\cdot 1$. In a monoidal category we can view a monoid as a pair of morphisms $\mu:M\otimes M\rightarrow M$ and $\nu:I\rightarrow M$ satisfying $\mu\circ(\mu\otimes \id_M) = \mu\circ(\id_M\otimes \mu)$ and $\mu\circ(\nu\otimes \id_M) = \id_M = \mu\circ(\id_M\otimes \nu)$. Here $\mu$ represents the multiplication operation, and $\nu$ points towards the identity in $M$ ($I$ here is the monoidal unit of the category, $A\otimes I \cong A$ for all $A$).
We can form a monoid in \textbf{ZX} where $M=1$ by using the Z-spider:
\begin{equation}
	\mu \ =\  \input{Z-mult.tikz} \qquad \qquad \nu\  =\  \input{Z-state.tikz}
\end{equation}

The associativity and unit conditions are then given by spider fusion:
\begin{equation}
\input{spider-monoid.tikz}
\end{equation}
In fact, this is a \emph{commutative} monoid, since:
\begin{equation}\label{eq:spider-monoid-commutative}
\input{spider-monoid-commutative.tikz}
\end{equation}
Of course nothing here is particular to the Z-spider, and this also works for the X-spider.

A notion that is perhaps less familiar, but still equally fundamental is the \emph{comonoid}. Whereas a monoid takes two elements of $M$ and outputs a single element of $M$, a comonoid does the opposite, taking a single element and outputting two. Note that, in general, the word `co' in category theory denotes that you should take the dual definition which is given by switching the direction of all the morphisms involved. In terms of our diagrams this corresponds to horizontally flipping the diagram. Concretely, a comonoid consists of maps $\delta: M\rightarrow M\otimes M$ and $\epsilon: M\rightarrow I$ that satisfy $(\delta\otimes \id_M)\circ \delta = (\id_M\otimes \delta)\circ \delta$ and $(\epsilon\otimes \id_M)\circ \delta = \id_M = (\id_M\otimes \epsilon)\circ\delta$. The map $\delta$ is called the co-multiplication, while $\epsilon$ is called the co-unit. Again, the Z-spider forms an example of this:
\begin{equation}
	\delta \ =\  \input{Z-comult.tikz} \qquad \qquad \epsilon\  =\  \input{Z-effect.tikz}
\end{equation}
And the co-associativity and co-unit equations are again proven using spider fusion:
\begin{equation}
\input{spider-comonoid.tikz}
\end{equation}
Analogous to~\eqref{eq:spider-monoid-commutative}, this comonoid is \emph{co-commutative}.

We now see that the basic Z-spiders of arity 1 and arity 3 form a monoid and a comonoid. This monoid and comonoid structure of the Z-spider interact in a particularly nice way that makes the Z-spider a \emph{Frobenius algebra}.%
\footnote{Frobenius algebras have a long history. The name itself was coined in 1941~\cite{nakayama1941frobeniusean}, with the first categorical definition given in 1969~\cite{lawvere1969ordinal}. For a good modern reference see~\cite{street2004frobenius} or the monograph~\cite{kock2004frobenius}. The first use of Frobenius algebras in describing structure in quantum theory was in 2008~\cite{coecke2008measurements}
}.
A Frobenius algebra is a monoid together with a comonoid that satisfy the \emph{Frobenius conditions}:
\begin{equation}
	\input{frobenius-conditions.tikz}
\end{equation}
As the monoid and comonoid of the Z-spider are (co-)commutative, the structure we get is actually a commutative Frobenius algebra. 
Additionally, the Z-spider also satisfies the following equation:
\begin{equation}
	\input{frobenius-special.tikz}
\end{equation}
Frobenius algebras satisfying these conditions are called \emph{special}.
The Z-spiders hence form a commutative special Frobenius algebra. This Frobenius algebra has one more important property that will require some further definitions.

In the category of ZX-diagrams \textbf{ZX} or the category of finite-dimensional complex vector spaces \textbf{fVec}$_\C$ we have a way of `flipping' a morphism $f:A\rightarrow B$ to a morphism $f^\dagger:B\rightarrow A$. Namely, by taking the adjoint of the linear map in the latter case, and flipping a ZX-diagram horizontally and changing the signs of all the phases in the former case. This $\dagger$ operation satisfies 3 properties of interest: it is involutive, $(f^\dagger)^\dagger = f$, it is contravariant, $(f\circ g)^\dagger = g^\dagger\circ f^\dagger$, and it is the identity on objects, $A^\dagger = A$. A category that has an operation with these properties is called a \emph{dagger-category}~\cite{selinger2007dagger}.

For the Z-spider, the monoid $\mu$ and the comonoid $\delta$ are related by the dagger: $\mu^\dagger = \delta$. Similarly, the unit $\nu$ and counit $\epsilon$ are also related via $\nu^\dagger = \epsilon$. A Frobenius algebra on a dagger-category where the monoid and comonoid are related in this manner is called a dagger-Frobenius algebra.

So to summarise: the Z-spiders form a commutative special dagger-Frobenius algebra. Our entire discussion also holds for the X-spiders, and in fact holds for any family of maps satisfying a version of the spider fusion rule~\SpiderRule and the identity rule~\IdRule. It turns out that a converse statement is also true~\cite{coecke2013new}: if we have a commutative special dagger-Frobenius algebra on a finite-dimensional complex vector space $V$, then there exists an orthonormal basis $\ket{v_1},\ldots\ket{v_n}$ of~$V$ such that the Frobenius algebra is given by the spiders
\begin{equation}
\input{black-spider.tikz}\ \ =\ \ \sum_{i=1}^n\ketbra{v_i\cdots v_i}{v_i\cdots v_i},
\end{equation}
analogously to the definition of the Z-spider in \eqref{eq:Z-spider-def}.
Another way we can view this result is that the spider equation \SpiderRule essentially gives a diagrammatic characterisation of the notion of an orthonormal basis.

The Frobenius algebra structure characterises a single spider, either Z or X. The interaction between the two species of spiders is however also a special case of a well-studied structure.
A \emph{bialgebra} consists of a monoid (that we will suggestively write as the Z-spider) and a comonoid (that we write as the X-spider) satisfying the following equations:
\begin{equation}
	\input{bialgebra-def.tikz}
\end{equation}
Here the dashed box represents an empty diagram. The Z-spider and X-spider evidently form a bialgebra\footnote{In a bialgebra these equations hold on the nose, while in the ZX-calculus they only hold up to scalar. Such a structure might more accurately be called a \emph{scaled} bialgebra.}; see \CopyRule and \BialgRule.
As was the case for the Frobenius structure, they do not just form a bialgebra. As the monoid and comonoid structure are (co-)commutative we can call this a commutative bialgebra, but there is some additional structure.
Given a bialgebra on an object $M$, an \emph{antipode} is a map $a:M\rightarrow M$ that satisfies the \emph{Hopf law}:
\begin{equation}
	\input{hopf-law-antipode.tikz}
\end{equation}
A bialgebra that has an antipode is called a \emph{Hopf algebra}. The Z-spider and X-spider form a Hopf algebra where the antipode is equal to the identity, as is evident from~\eqref{eq:hopf-rule}.

We chose the Z-spider to supply the monoid and the X-spider to supply the comonoid, but we could have chosen to do this the other way around. A pair of Frobenius algebras that is a Hopf algebra in these two ways is variously called a pair of \emph{interacting Frobenius algebras}~\cite{duncan2016interacting}, \emph{interacting Hopf algebras}~\cite{bonchi2017interacting}, \emph{interacting bialgebras}~\cite{bonchi2014interacting} or a \emph{Hopf-Frobenius algebra}~\cite{collins2019hopffrobenius}.

Much of what we said in this section can generalise to linear maps on higher-dimensional spaces (in contrast to the two-dimensional spaces we have been working with). We discuss qudits in more detail in Section~\ref{sec:qudits}.

\section{Toffoli gates and the ZH-calculus}\label{sec:ZH}

In our discussion on the ZX-calculus in Sections~\ref{sec:circuits-vs-diagrams}--\ref{sec:clifford} we have focused on gate sets consisting of single-qubit gates combined with the CNOT gate. This however skips over other widely used multi-qubit gates, in particular Toffoli gates and other multiply-controlled gates.
While the ZX-calculus is very useful when it comes to reasoning about circuits based on CNOT gates and related gates like the CZ or the Ising-type unitaries such as the phase gadget of Section~\ref{sec:phase-polynomial}, it is considerably harder to work with multiply-controlled gates.
This can be pinpointed to the fact that there is no concise way to represent the AND operation on computational basis states $\ket{x}\otimes\ket{y}\mapsto \ket{x\cdot y}$. Note that the important word here is `concise'. As ZX-diagrams are universal (cf.~Section~\ref{sec:universality}) any linear map \emph{can} be represented, but not always in a useful manner.

In this section we will discuss a variation on the ZX-calculus that is called the \emph{ZH-calculus}. This was introduced in 2018 by Backens and Kissinger~\cite{backens2018zhcalculus} and introduces a new generator, the \emph{H-box}, that generalises the Hadamard box we have been using to arbitrary arity, and which allows an easy representation of Toffoli gates and other multiply-controlled gates.
Though the ZH-calculus was introduced as an alternative calculus, we will treat it here simply as an extension of the ZX-calculus that allows us to efficiently reason about a larger class of diagrams.%
\footnote{The H-box is not the only way to simplify the representation of Toffoli gates in the ZX-calculus. Another approach is given by introducing the `triangle' generator that we will encounter in Section~\ref{sec:completeness-oxford}.}

First, in Section~\ref{sec:H-boxes} we introduce H-boxes.
Then in Section~\ref{sec:ZH-calculus} we introduce the additional rewrite rules that allow us to reason about this generator.
As a demonstration of its utility, we use H-boxes in Section~\ref{sec:controlled-unitaries} to construct the graphical representation of several controlled unitaries.
Then in Section~\ref{sec:fourier} we recall the Fourier transform of semi-Boolean functions in order to relate H-boxes to Z- and X-spiders and the phase polynomials of Section~\ref{sec:phase-polynomial}.
Finally, in Section~\ref{sec:toffoli-optimise} we prove diagrammatically some known optimisations of Toffoli gates using H-boxes and the Fourier transform.

\subsection{H-boxes}\label{sec:H-boxes}

As discussed above, using Z- and X-spiders, there is no concise way to represent the coherent AND gate that maps the computational basis state $\ket{x}\otimes \ket{y}$ to $\ket{x\cdot y}$.
If we had such an AND gate, we could implement a Toffoli gate as follows:
\begin{equation}
\input{Toffoli-AND.tikz}
\end{equation}
Indeed, inputting computational basis states into the top two qubits we calculate:
\begin{equation}
\input{Toffoli-AND-calc.tikz}
\end{equation}
Hence, this construction `activates' a NOT gate on the bottom qubit when the top two qubits are in the $\ket{1}$ state. This is exactly the action $\ket{x,y,z} \mapsto \ket{x,y, (x\cdot y)\oplus z}$ of the Toffoli gate.

As discussed in Section~\ref{sec:bialgebra-hopf}, the Z-spider implements a coherent COPY gate when acting on computational basis states, and the X-spider implements an XOR. We might then hope that we could add another spider that implements the AND gate. However, one of the main symmetries present in the Z- and X-spiders does not hold for the AND gate (which can be verified by calculating the associated matrices):
\begin{equation}
\input{AND-not-symmetric.tikz}
\end{equation}
It is however possible to `split up' the AND gate into two components, each of which do have all the symmetries that the spiders have. 
We call these components \emph{H-boxes}, for reasons that will become clear, and they are defined as follows:
\begin{equation}\label{eq:H-spider-def}
 \input{H-spider.tikz} \ \ \ := \ \ \sum a^{i_1\ldots i_m j_1\ldots j_n} \ket{j_1\ldots j_n}\bra{i_1\ldots i_m}
\end{equation}
The sum in this equation is over all $i_1,\ldots, i_m, j_1,\ldots, j_n\in\{0,1\}$ and $a$ is an arbitrary complex number. Hence, an H-box represents a matrix with all entries equal to 1, except the bottom right element, which is equal to $a$.
We have for instance
\begin{equation}
	\input{Hbox-2-1.tikz} \ \ = \ \ 
	\begin{pmatrix}
	1&1&1&1\\
	1&1&1&a
	\end{pmatrix}
	\ \ \ \text{and} \ \ \ 
	\input{Hbox-1-1.tikz} \ \ = \ \ 
	\begin{pmatrix}
	1&1\\
	1&a
	\end{pmatrix}.
\end{equation}
Hence, in particular, when $a=-1$, the 1-input 1-output H-box is just a rescaled Hadamard:
\begin{equation}\label{eq:Hbox-is-had}
	\input{had.tikz} \ \ = \ \ \frac{1}{\sqrt{2}}\ \input{Hbox-1-1-1.tikz}
\end{equation}
The H-boxes are thus a generalisation of Hadamard gates to an arbitrary number of inputs and outputs  (hence, the letter `H'). 
Just as spiders with a zero phase are depicted without a phase label, we will depict H-boxes with a label of $-1$ without any label:
\begin{equation}
	\input{H-spider-nolabel.tikz} \ \ \ :=\  \ \ \input{H-spider-minus.tikz}
\end{equation}
This convention means that the 1-input 1-output H-box with a phase of $-1$ is denoted exactly the same as the Hadamard gate in the ZX-calculus. As these two diagrams represent the same matrix up to a scalar factor of $\frac{1}{\sqrt{2}}$ this is fine as long as one does not care about the exact scalar value. Where this matters we will denote which definition we are using. In this section we will use the definition of H-boxes of \eqref{eq:H-spider-def}.

The linear maps that H-boxes represent have all the symmetries we would expect from a spider:
\ctikzfig{Hbox-symmetries}
Now, as promised, the H-boxes are indeed the components with which we can construct the coherent AND:
\begin{equation}
	\input{AND-from-hbox.tikz} \ \ = \ \ 
	\begin{pmatrix}
		2&2&2&0\\
		0&0&0&2
	\end{pmatrix}
	\ \ = \ \ 2 \ \ \input{AND.tikz}
\end{equation}
We can then easily write down the Toffoli gate:
\begin{equation}
	\input{toffoli-hbox.tikz}
\end{equation}
This simple representation of the AND (requiring just two diagrammatic generators) allows us to efficiently diagrammatically reason about constructions involving the AND, as we will see in the next section.

\subsection{The ZH-calculus}\label{sec:ZH-calculus}

The reason we care about H-boxes is because they come equipped with a new set of rewrite rules called the \emph{ZH-calculus}. 
These rewrite rules can essentially be grouped into two sets. The rules in the first set are motivated by the relationship of the H-boxes to the AND gate, while the second section contains rules that relate H-boxes with different labels.

While ostensibly part of the second set, let us first note the following fundamental relation between an arity-1 H-box and a Z-spider:
\begin{equation}\label{eq:H-state-as-Z}
	\input{H-a-state.tikz} \ \ = \ \ \input{Z-a-state.tikz}
\end{equation}
In particular, taking respectively $\alpha = 0$ and $\alpha = \pi$, we get:
\begin{equation}\label{eq:H-state-unit}
	\input{H-state-1.tikz} \ \ = \ \ \input{Z-state.tikz} \qquad \quad \input{H-state.tikz} \ \ = \ \ \input{Z-pi-state.tikz}
\end{equation}

To understand the first set of rules it will be helpful to use multi-input AND gates:
\begin{equation}\label{eq:AND-mult-hbox}
	\input{AND-multi.tikz} \ \ = \ \ \input{AND-from-hbox-multi.tikz}
\end{equation}
A rule on H-boxes we have already seen is that two Hadamard gates cancel, cf.~\eqref{eq:had-had-cancel}. Using our interpretation of multi-input AND gates~\eqref{eq:AND-mult-hbox} using H-boxes we can gain a different view on this equation. Indeed, using \eqref{eq:AND-mult-hbox} we see that two Hadamard gates in a row correspond to an AND gate with a single input, and this gate is of course the identity.

The first new rule expresses how a sequence of ANDs can be combined into a single multi-input AND:
\begin{equation}
\input{AND-spider-fusion.tikz}
\end{equation}
This rule can be presented a bit more generally as an \emph{H-box fusion} rule:
\begin{equation}\label{eq:Hbox-spider}
	\input{H-spider-rule-phased.tikz}
\end{equation}

An important consequence of this rule is that H-boxes absorb $\ket{1}$ states:
\begin{equation}\label{eq:Hbox-absorb-1}
	\input{Hbox-absorb-1.tikz}
\end{equation}
We will see later in~\eqref{eq:X-H-state-copy} that, in contrast, a $\ket{0}$ `explodes' an H-box.

In Section~\ref{sec:bialgebra-hopf} we saw how the interpretation of the Z- and X-spider as respectively COPY and XOR lead us to the bialgebra rule that allowed us to push (phaseless) Z- and X-spiders through each other. This equation~\eqref{eq:XOR-COPY-bialgebra} involving COPY and XOR holds in exactly the same way when XOR is replaced by AND:
\begin{equation}
\input{AND-COPY-bialgebra.tikz}
\end{equation}
We can directly translate this into a rule involving Z-spiders and H-boxes:
\begin{equation}\label{eq:ZH-bialgebra-AND}
	\input{ZH-bialgebra-AND.tikz}
\end{equation}
By pushing the Hadamard through the Z-spider and cancelling some Hadamards we can also present this in a format that is often more convenient:
\begin{equation}\label{eq:ZH-bialgebra}
	\input{bialgebra-rule-H-many.tikz}
\end{equation}
As in~\eqref{eq:bialgebra-rule-many}, the right-hand side of both of these equations is a fully connected bipartite graph.
Note that as a special case of the second equation (taking $n=0$) we get the following useful state copy rule, which is a counterpart of~\eqref{eq:Hbox-absorb-1}:
\begin{equation}\label{eq:X-H-state-copy}
	\input{X-H-state-copy.tikz}
\end{equation}
Here in the last step we dropped the scalar subdiagram, as it only contributes a (usually irrelevant) non-zero scalar.

Another consequence of this bialgebra rule is that the identification of a $1$-labelled H-box with a Z-spider of~\eqref{eq:H-state-unit} can be generalised to higher arity as follows:
\begin{equation}\label{eq:unit-rule-many}
\input{unit-rule-many-pf.tikz}
\end{equation}

Let us now introduce the last pair of AND-inspired rewrite rules for H-boxes. These are based on the following identities:
\begin{equation}
	\input{AND-COPY-cancel.tikz} \ \ = \ \ \input{id.tikz} \qquad \ \  \input{AND-postselect.tikz}
\end{equation}
The first is quite self-evident: if we copy a value and then AND those values together, it is the same thing as doing nothing to the value. 
The second requires a bit more explanation. It expresses a fact about the possible ways that AND can return $\ket{1}$. Indeed, as a linear map, we can write AND as $\ketbra{0}{00}+\ketbra{0}{01}+\ketbra{0}{01}+\ketbra{1}{11}$, and hence post-selecting the output of AND with $\bra{1}$ we calculate $\bra{1}\!\text{AND} = \bra{11}$.

Writing the ANDs as H-boxes and simplifying the expressions a bit we get the following rewrite rules\footnote{These rules were not present as axioms in the original ZH-calculus paper~\cite{backens2018zhcalculus}. Instead, they follow as consequences of their (O) rule, which is arguably harder to motivate and to use. It is possible to prove (O) using the rules of~\eqref{eq:Hbox-ANDlike-rules}, and so in that sense they are equivalent~\cite{zhcompleteness2020}.}:
\begin{equation}\label{eq:Hbox-ANDlike-rules}
	\input{copy-Hbox.tikz} \ \ = \ \ \input{had.tikz} \qquad \quad \input{Hbox-pi-rule.tikz}
\end{equation}

Note that using~\eqref{eq:H-state-unit} we could also have written the second equation of~\eqref{eq:Hbox-ANDlike-rules} as:
\begin{equation}
	\input{Hbox-pi-rule2.tikz}
\end{equation}

\begin{figure}
\centering
\input{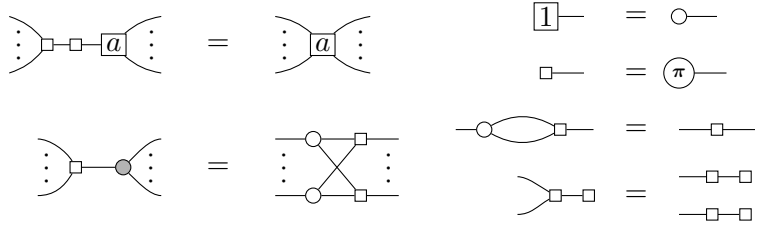}
\caption{
	The phase-free rules of the ZH-calculus.
}
\label{fig:zh-rules}
\end{figure}

The rules introduced so far are summarised in Figure~\ref{fig:zh-rules}.

We have now covered all the `AND inspired' ZH-calculus rules. In fact, these rules, together with some of the ZX-calculus rules of Figure~\ref{fig:zx-rules}, are complete for the Hadamard-Toffoli fragment of the ZH-calculus~\cite{zhphasefree}, where phases on spiders are limited to $0$ and $\pi$ and H-boxes can only be labelled by the default $-1$.
This means that when we have two circuits (or more generally, arbitrary diagrams) that represent the same linear map and that consist solely of Hadamard and Toffoli gates, then by representing these diagrams using spiders and H-boxes and using just the rules of Figure~\ref{fig:zx-rules} and~\ref{fig:zh-rules} we can prove diagrammatically that they are equal.%
\footnote{The complete ruleset presented in~\cite{zhphasefree} is somewhat different from the one we present. That the ruleset we present is complete is shown in a forthcoming paper~\cite{zhcompleteness2020}.}
This is significant because Hadamard+Toffoli circuits are approximately universal for quantum computation~\cite{ShiToffoliHadamard,aharonov2003simple}.
Hence, unlike the rules of Figure~\ref{fig:zx-rules} that are only complete for the Clifford fragment, we could in principle use the rules we have covered up to this point to do arbitrary calculations for the approximately universal model of quantum computation given by the Toffoli+Hadamard gate set.

The second set of rules of the ZH-calculus has to do with how labelled H-boxes interact.
The first two of these rules allow us to perform arithmetic with H-boxes:
\begin{equation}\label{eq:multiply-average}
	\input{multiply-rule-phased.tikz} \qquad\qquad \quad \input{average-rule.tikz}
\end{equation}
These rules are known as the `multiply rule' and the `average rule'~\cite{backens2018zhcalculus}.
When $a$ and $b$ are complex phases, the multiply rule is just an instance of the adding of phases when spiders fuse, cf.~\eqref{eq:H-state-as-Z} and~\eqref{eq:phases-add}.
The average rule has no counterpart in the standard ZX-calculus. In practice, the average rule doesn't turn out to be useful very often. We include it here merely for completeness' sake.

The multiply rule can be generalised to H-boxes of arbitrary arity:
\begin{equation}\label{eq:mult-rule-many}
	\input{multiply-rule-many.tikz}
\end{equation}
I.e.~when two H-boxes are connected to exactly the same set of Z-spiders, then we can fuse the H-boxes together.
With the rules we have seen before, the proof of this generalisation is straightforward:
\begin{equation}
\input{multiply-rule-many-pf.tikz}
\end{equation}
This rule appears quite often in the form where $a=b=-1$, so that $ab=1$ and the H-box decomposes into white spiders according to \eqref{eq:unit-rule-many}.
As an example, we can use this rule to prove that two Toffoli gates reduce to the identity:
\begin{equation}
\input{toffoli-self-inverse.tikz}
\end{equation}

Then there is only one more rule: the \emph{introduction rule}:
\begin{equation}\label{eq:intro-rule}
	\input{intro-rule.tikz}
\end{equation}
It is called the introduction rule, because it allows you to introduce additional edges to an H-box (at the cost of copying the H-box). As do many of the previously introduced rules, it has a generalisation to H-boxes of arbitrary arity:
\begin{equation}\label{eq:intro-rule-many}
	\input{intro-rule-many.tikz}
\end{equation}
Most of the use-cases of this rule are when it is applied from right-to-left. Indeed, it is a close cousin of the multiply rule~\eqref{eq:mult-rule-many}. Both rules target pairs of H-boxes connected to the same set of Z-spiders, although in the case of the introduction rule, they must also differ by a NOT gate on one of the connections, and have the same label.
As an example, we can use the introduction rule to prove that if we apply both a controlled-phase gate, and a NOT-conjugated controlled-phase gate that this reduces to just a simple phase gate:
\begin{equation}
	\input{control-phase-intro-reduce.tikz}
\end{equation}

As noted above, the `AND inspired' rules together with the ZX-calculus rules are complete for diagrams generated by Toffoli and Hadamard gates. When we add these three additional rules, multiply, average and introduction, we get a rule set that is complete for \emph{all} diagrams~\cite{backens2018zhcalculus}. The ZH-calculus, being a superset of the ZX-calculus, is also universal. Hence, we can, in principle, replace all reasoning about qubit linear maps with diagrammatic reasoning.
Whether it is beneficial to do so of course depends on the situation.

\subsection{Controlled unitaries}\label{sec:controlled-unitaries}

A useful feature of the ZH-calculus is that it allows us to quite easily see how to make a controlled-unitary out of a unitary given as a ZH-diagram.
This is perhaps most easily demonstrated by the difference between a CZ and a CCZ gate in the ZH-calculus:
\begin{equation}
	CZ \ \  =\ \  \input{CZ.tikz} \qquad \qquad CCZ \ \ = \ \  \input{CCZ.tikz}
\end{equation}
This suggests a general procedure for adding a control qubit: identify which H-box `activates' the application of your gate, and add another wire to it which connects to a Z-spider on your control qubit.
Sometimes, one has to work a bit to uncover the correct H-box. For instance, to see how a Z gate relates to a CZ, we unfuse its phase:
\begin{equation}
	\input{Z-to-CZ.tikz}
\end{equation}
This procedure also works for making controlled-phase gates if the phase is something other than $\pi$:
\begin{equation}\label{eq:cz-alpha}
	\text{CZ}(\alpha) \ \ = \ \ \input{CZ-alpha.tikz}
\end{equation}
For diagrams containing X-spiders we will usually have to convert these to Z-spiders using \HadamardRule in order to see where we should add the control wire. For instance, to go from a CNOT to a CCNOT (Toffoli):
\begin{equation}
	\input{CNOT-to-tof.tikz}
\end{equation}
Note that we here added a control wire to the `middle' H-box, but left the Hadamards on the qubit wire alone. This is a general rule for constructing a controlled diagram. 
For instance, it might be tempting to define a controlled-Hadamard as follows:
\begin{equation}
	\input{CHAD-fake.tikz}
\end{equation}
While this does indeed implement a Hadamard gate when the control qubit is in the $\ket{1}$ state (although with the wrong normalisation), it does not reduce to the identity when the control qubit is $\ket{0}$:
\begin{equation}
	\input{CHAD-fake-pf.tikz}
\end{equation}
To construct the actual controlled-Hadamard we need to find the `hidden' H-boxes in the Hadamard gate. The way we do this is by using its decomposition into Euler angles:
\begin{equation}\label{eq:had-euler3}
	\input{had-euler3.tikz}
\end{equation}
We can now make each of these phase gates into controlled phase gate using~\eqref{eq:cz-alpha}. 
When transforming this Euler decomposition into its controlled version, the ignorable global phase $e^{-i\frac{\pi}{4}}$ becomes a local phase that must be taken into account. This is in fact another instance of finding the hidden H-boxes of the diagram, as a scalar is just an H-box with zero wires.
We hence get the following transformation:
\begin{equation}\label{eq:had-make-controlled}
	\input{had-make-controlled.tikz}
\end{equation}
Where in the last step we used the identity $e^{i\frac\pi2} = i$.

While this procedure works and gives the correct diagram for a controlled-Hadamard, it is not the most efficient implementation of a controlled-Hadamard. A better version is realised by making the observation that if we only control the middle phase-gate and the global phase of~\eqref{eq:had-euler3} that we get a diagram that implements a Hadamard when the control is $\ket{1}$, and implements an X gate otherwise:
\begin{equation}\label{eq:had-controlled-attempt2}
	\input{had-controlled-attempt2.tikz}
\end{equation}
Hence, to make this a controlled-Hadamard, we need to add an X gate on the target wire to cancel the already existing X gate, but doing this will result in the wrong unitary being implemented when the control is $\ket{1}$. To remedy this error we add another gate to the circuit: a CNOT (i.e.~a controlled-X gate). 
We hence arrive at the final controlled-Hadamard circuit:
\begin{equation}\label{eq:CHAD}
	\input{CHAD.tikz}
\end{equation}
Note that we get the $-\frac\pi2$ X-phase by combining the first $\frac\pi2$ phase of~\eqref{eq:had-controlled-attempt2} with the added $\pi$ phase coming from the $X$ gate.
The gate~\eqref{eq:CHAD} is indeed what one would find for a controlled-Hadamard in a standard textbook (although if one starts with a different Euler decomposition of the Hadamard gate, one might get a CZ gate instead of a CX gate, along with some other permutations of the gates). Note that the `controlled-$i$' gate is usually further decomposed to get a presentation of the controlled-Hadamard gate in a more basic gate set. We will see how to do this decomposition in Section~\ref{sec:fourier}.

It is currently not clear how one would relate~\eqref{eq:CHAD} and the more complicated~\eqref{eq:had-make-controlled} via an intuitive diagrammatic transformation (as the calculus is complete, there is a set of graphical rewrites that transforms one into the other, but this is likely to be a complicated affair).
So how would one find~\eqref{eq:CHAD}? The crucial observation is that only controlling a single phase in the diagram, instead of all three, already resulted in a gate close to the one we desired. The remainder of the construction was then to keep adding simple gates until we get the exact gate we wanted. Experience learns that this method of experimentation and trial-and-error is often successful.

Let us construct one more often-encountered controlled unitary: the controlled-swap.
Our starting point is the implementation of a swap using three CNOTs:
\begin{equation}
	\input{swap-three-cnots.tikz}
\end{equation}
We could make this controlled by transforming each of the CNOTs into a Toffoli. However, just as with the controlled-Hadamard, we realise that if we `deactivate' the middle CNOT, that the outer CNOTs cancel each other, and hence it suffices to add a control to the middle CNOT. We can diagrammatically simplify the resulting gate:
\begin{equation}
	\input{CSWAP-simplify.tikz}
\end{equation}
In the final diagram, if the control qubit is in the $\ket{1}$ state, we can ignore the control wire (because the $\ket{1}$ is absorbed as in~\eqref{eq:Hbox-absorb-1}). The resulting diagram then has the shape of the \emph{butterfly network} commonly used in the literature on linear network coding~\cite{debeaudrap_et_al:LIPIcs:2014:4818,Beaudrap2020quantumlinear}. In this case, the two resulting Hadamards in the middle cancel out, and by inputting states this can easily be verified to implement a swap. On the other hand, if the control is $\ket{0}$, the 3-ary H-box disconnects (cf.~\eqref{eq:X-H-state-copy}), and the diagram collapses to the identity.

\subsection{The graphical Fourier transform}\label{sec:fourier}

As was shown in Section~\ref{sec:universality}, the ZX-calculus is universal, meaning that any linear map between qubits can be represented as a ZX-diagram. This includes H-boxes. So how can an H-box be constructed using Z- and X-spiders?
We have already seen a number of special cases. In~\eqref{eq:H-state-as-Z} we represented 1-ary H-boxes carrying a complex phase as a Z-spider. The Euler decomposition of the Hadamard gate~\eqref{eq:had-euler1} gives a translation for a 2-ary H-box with phase $-1$.
In fact, a more useful translation turns out to be~\eqref{eq:had-gadget} as this generalises in a straightforward way to 2-ary H-boxes carrying an arbitrary complex phase:
\begin{equation}\label{eq:had-gadget-alpha}
	\input{had-gadget-alpha.tikz}
\end{equation}
This identity is already sufficient to decompose the controlled-phase gate that appears in~\eqref{eq:CHAD}.
To see where this identity comes from and how it generalises we need to introduce the notion of a Fourier transform of a semi-Boolean function.

A semi-Boolean function is a function $f:\mathbb B^n\rightarrow \R$, where $\mathbb B = \{0,1\}$ is the Booleans. We can write a semi-Boolean function as a sum of primitive terms in two useful ways.
The first way is to take the primitive function to be XOR:
\[f(\vec x) = \sum_{\vec y\in \mathbb B^n} \lambda_{\vec y}\left(\bigoplus \vec y\cdot \vec x\right).\]
Here the $\lambda_{\vec y}$ are real coefficients that determine $f$, the `multiplication' operation $\cdot$ is the componentwise multiplication of bitstrings, and $\bigoplus$ denotes the XOR of all the bits in a bitstring.
Let us give an example. If $\vec y = 101$, then $\bigoplus \vec y\cdot \vec x = x_1\oplus x_3$. Note that in this sum we are treating the Booleans $1$ and $0$ both as Booleans and as real numbers.
This decomposition contains $2^n$ independent parameters $\lambda_{\vec y}$. As $f$ has $2^n$ possible inputs, we see that each semi-Boolean function can indeed uniquely be written in this way. The phase polynomials we saw in Section~\ref{sec:phase-polynomial} are examples of semi-Boolean functions written as XOR terms.

The second decomposition is to take the primitive function to be AND:
\[f(\vec x) = \sum_{\vec y\in \mathbb B^n} \hat{\lambda}_{\vec y}\prod \vec x^{\vec y}.\]
Here $\vec x^{\vec y}$ is the bitstring $x_1^{y_1}x_2^{y_2}\cdots x_n^{y_n}$ where we set $0^0 = 1$ and $1^0 = 1$. Hence, if $\vec y = 101$ then $\vec x^{\vec y} = x_11x_3$, so that $\prod \vec x^{\vec y} = x_1\wedge x_3$. Again, as there are $2^n$ independent terms in this decomposition, any semi-Boolean function can be written in this way.

The transformation of a semi-Boolean function written as sums of XOR terms to one written as sums of AND terms and back is what we call the Fourier transform of such a function.
As the X-spider allows us to compute the XOR of basis states, and the H-box allows us to compute the AND, it should come as no surprise that this Fourier transform relates these two generators.

This Fourier transform essentially boils down to one simple identity that holds for Boolean variables $x$ and $y$:
\begin{equation}\label{eq:Boolean-identity}
	x\oplus y = x + y - 2(x\cdot y)
\end{equation}
This equation allows us to write a 2-ary XOR term as a sum of AND terms. By iterating it, we can also reduce higher-arity XOR terms:
\begin{align*}
	(x\oplus y)\oplus z &= x\oplus z + y\oplus z - 2(x\cdot y)\oplus z
	\\
	&= x + z - 2(x\cdot z) + y + z - 2(y\cdot z) - 2(x\cdot y) - z + 4(x\cdot y\cdot z) \\
	&= x + y + z - 2(x\cdot y + x\cdot z + y\cdot z) + 4(x\cdot y\cdot z)
\end{align*}
We can also take the `inverse' of~\eqref{eq:Boolean-identity}:
\begin{equation}\label{eq:Boolean-identity-reverse}
	x\cdot y = \frac12 (x+y - x\oplus y)
\end{equation}
And again, this generalises:
\begin{equation}\label{eq:mult-xor-3}
	x\cdot y \cdot z = \frac14 (x + y + z -x\oplus y - x\oplus z - y\oplus z + x\oplus y \oplus z)
\end{equation}

For an example of how this relates to quantum gates, let us consider a simple example. The action of a CZ gate on a computational basis state $\ket{xy}$ can be written as $\text{CZ}\ket{xy} = (-1)^{x\cdot y}\ket{xy}$.
Using~\eqref{eq:Boolean-identity-reverse} and $-1 = e^{i\pi}$ we can then write $(-1)^{x\cdot y} = e^{i\frac\pi2 (x+y-x\oplus y)}$.
As diagrams, the equality of these two representations becomes the following identity:
\begin{equation}
	\input{CZ.tikz} \ \ = \ \ \input{CZ-decomposed.tikz}
\end{equation}
I.e.~on the left-hand side we see that the two input states get copied to the H-box in the middle, which gives a $-1$ phase if both inputs are $\ket{1}$. On the right-hand side we see two S gates, that supply conditional phases $e^{i\frac\pi2 x}$ and $e^{i\frac\pi2 y}$, while the central phase gadget (see~\eqref{eq:phase-gadget-unitary}), supplies $e^{-i\frac\pi2 x\oplus y}$.
Note that this equality can be easily derived diagrammatically from the decomposition of the Hadamard gate in~\eqref{eq:had-gadget}.
If instead of a CZ gate, we started with a CZ$(\alpha)$ gate~\eqref{eq:cz-alpha}, we would have end up with the decomposition~\eqref{eq:had-gadget-alpha}.

This process works very similarly for the CCZ gate. Indeed, $\text{CCZ}\ket{xyz} = (-1)^{x\cdot y\cdot z} \ket{xyz}$, and hence, using~\eqref{eq:mult-xor-3}, we expect the following equality:
\begin{equation}\label{eq:CCZ-decomposed}
	\input{CCZ.tikz} \ \ = \ \ \input{CCZ-decomposed.tikz}
\end{equation}
This identity can be proven diagrammatically, but requires quite some work to do in a general setting~\cite{GraphicalFourier2019}. By decomposing each of the phase gadgets in this diagram using CNOT and phase gates, we get the familiar decomposition of a CCZ gate into the Clifford+T gate set.

This process can be generalised to decompose H-boxes of arbitrary arity and carrying arbitrary complex phases. If the label is not a phase, but rather an arbitrary complex number, then a more complicated strategy is necessary analogous to the one described in Section~\ref{sec:scalars} for constructing arbitrary complex numbers as scalar ZX-diagrams, and analogous to the construction of the $\lambda$-boxes in Section~\ref{sec:completeness-oxford}.

\subsection{Optimising Toffoli gates}\label{sec:toffoli-optimise}

Using the identity~\eqref{eq:CCZ-decomposed} we can write the CCZ gate using 7 non-Clifford gates. By conjugating with Hadamards on one qubit we can then also write the Toffoli using 7 non-Clifford gates. In most models of error-corrected quantum computation, non-Clifford gates are significantly more expensive to implement than Clifford gates.%
\footnote{By the Eastin-Knill no-go theorem~\cite{eastin2009restrictions}, an error correcting code cannot have a universal transversal gateset, and hence some gates must be implemented in a different way, such as using magic state distillation~\cite{campbell2017roads}. In many popular fault-tolerant protocols, such as the surface code, the gate that must be distilled is the T gate, which is then the part of the computation that requires the majority of the resources~\cite{gorman2017quantum}.}
Hence there has been quite some work on reducing the number of non-Clifford gates, specifically T-gates, in a given quantum circuit~\cite{amy2014polynomial,amy2016t,heyfron2018efficient,zhang2019optimizing,nam2018automated} (some of which use the ZX-calculus~\cite{kissinger2019tcount,deBeaudrap2020Techniques}).

It has been shown that if one restricts to unitary gates in the Clifford+T gateset that at least 7 T gates are necessary to implement a Toffoli or CCZ~\cite{meet-in-the-middle2013,di2016parallelizing}. However, by allowing non-unitary constructions, such as the use of ancillae or measurement and classical control, this bound can be circumvented.
For instance, Jones showed~\cite{jones2013low} that the number of T gates needed for a Toffoli can be reduced to 4 if one is allowed to use ancillae and classical control, based on the result of Selinger~\cite{selinger2013quantum} that a combination of a Toffoli gate and a controlled $S^\dagger$ requires just 4 T gates.
Later, Gidney extended this result~\cite{Gidney2018halvingcostof} to show that a `compute-uncompute' pair of Toffoli gates can be implemented using classical control and just 4 T gates for the two Toffoli gates.

In this section we will demonstrate how these constructions can be presented and derived graphically. This material first appeared in~\cite{GraphicalFourier2019}.

First, Selinger's construction of the `Tof$^*$' gate consisting of a Toffoli and controlled $S^\dagger$ gate can be presented and simplified as follows:
\begin{equation}\label{eq:ccz-cs-simp}
	\input{ccz-cs-simp.tikz}
\end{equation}
To construct Jones' implementation of a Toffoli gate, we need an additional derived rewrite rule that says that an X phase gate when commuted through an H-box transforms into a controlled Z phase gate:
\begin{equation}\label{eq:had-phase-commute}
	\input{had-phase-commute.tikz}
\end{equation}
We can now rewrite the Toffoli gate to a more efficient construction:
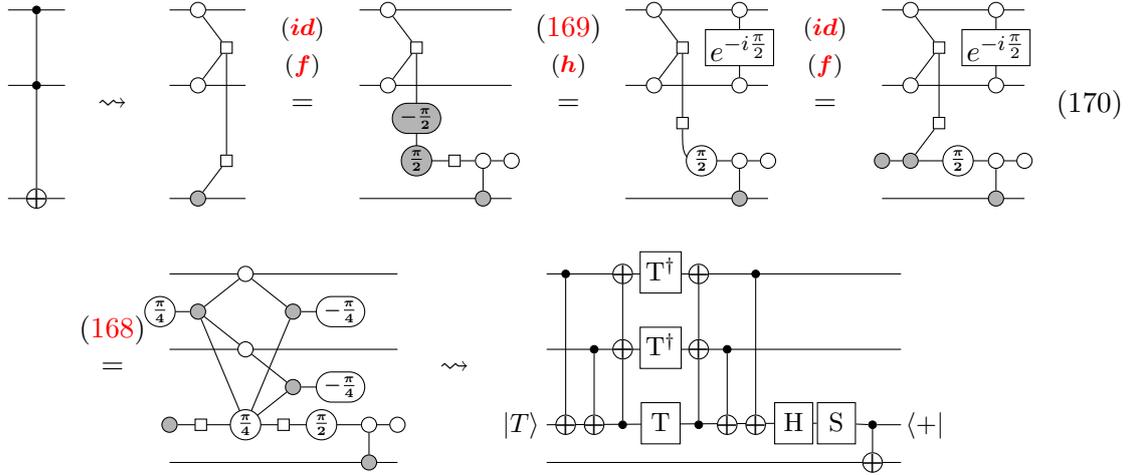
\begin{equation}\label{eq:tof-ancilla}
	\input{tof-ancilla.tikz}
\end{equation}
Here in the last step we transformed the ZX-diagram back into a circuit by decomposing all the phase gadgets into CNOT and T gates (cf.~Section~\ref{sec:phase-polynomial}). Note that it includes an ancilla prepared in the $\ket{T}:=\ket{0} + e^{i\frac\pi4}\ket{1}$ `magic state' and that this ancilla is post-selected to $\bra{+}$ at the end. In practice this `post-selection' will actually be a measurement in the X basis. When the outcome is $\bra{+}$ we get the desired Toffoli gate, but when the outcome is $\bra{-}$ we need to correct it by applying an additional gate. We can check what that gate is by first seeing what the effect is of the `wrong' measurement outcome:
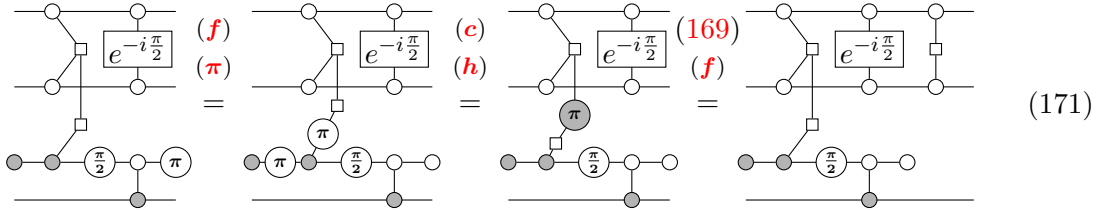
\begin{equation}\label{eq:tof-ancilla-correct}
	\input{tof-ancilla-correct.tikz}
\end{equation}
We see that the outcome $\bra{-}$ results in a circuit that has an additional CZ gate. Hence, to deterministically implement a Toffoli gate we need to apply the inverse of a CZ gate when the measurement outcome is $\bra{-}$. This inverse is itself a CZ gate.

Finally, let us similarly derive Gidney's efficient implementation of a pair of Toffoli gates that are acting on a qubit that only acts as a control:
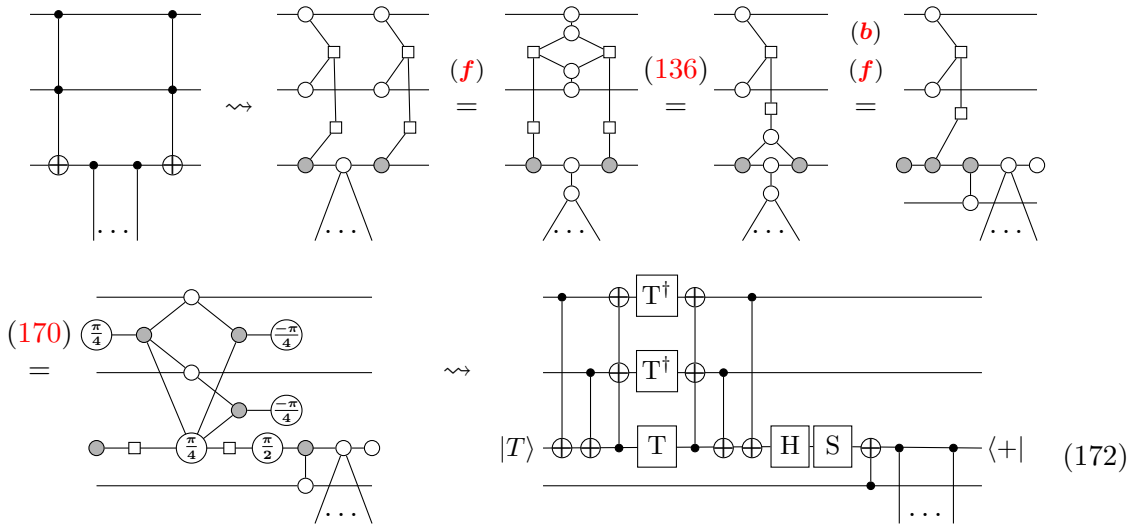
\begin{equation}
	\input{gidney-rewrite.tikz}
\end{equation}

Again, this diagram represents a post-selected measurement outcome for the ancillae. Using a similar procedure as in~\eqref{eq:tof-ancilla}, we can show that the wrong measurement outcome can be corrected by a CZ gate on the top 2 qubits, and a Z gate on the bottom qubit.

\section{Completeness}\label{sec:completeness}

In this section we will go into more detail about an issue that was the topic of many early ZX-calculus papers: completeness.
A graphical calculus is complete when its rewrite rules are powerful enough to prove any true equation. More formally, to each diagram we associate a linear map, and completeness says that if two diagrams represent the same linear map, then there should be a sequence of rewrites that transforms one diagram into the other.

Whether a calculus is complete depends on the specific rewrite rules we allow. So while we have been talking about `the' ZX-calculus in this paper, we should actually say `a' ZX-calculus, because there are several competing rule sets. The rules of Figure~\ref{fig:zx-rules} form the core rule set that are used in some variation in all the ZX-calculus papers. Most completeness proofs require at least one additional rule to prove completeness. 
For instance, it can be shown that the following equality is not provable using the rules of Figure~\ref{fig:zx-rules} for any $\alpha$ that is not a multiple of $\frac\pi2$~\cite{supplementarity}:
\begin{equation}\label{eq:supplementarity}
	\input{supplementarity.tikz}
\end{equation}
Note that this equation only holds up to a global scalar. For simplicity, we will ignore scalars in this section, just as we did in the other parts of the paper, but do note that scalars are often non-ignorable in completeness proofs.

As mentioned in Section~\ref{sec:clifford}, the rules of Figure~\ref{fig:zx-rules} are complete for the Clifford fragment~\cite{BackensCompleteness}, i.e.~where we restrict the phases on all the spiders to multiples of $\frac\pi2$.
Two other fragments where completeness has been studied intensively are the Clifford+T fragment, where the phases are multiples of $\frac\pi4$, and the universal fragment, where phases are arbitrary.
There have been two different approaches to completeness of the ZX-calculus that each have produced progressively simpler axiomatisations. We name these approaches after Oxford and Nancy as those were the places where the researchers working on these approaches were around 2017 when these results first appeared. The first completeness results contained some quite convoluted rules, that have since then been simplified. 

Those readers not interested in the history of completeness results or the specific rule sets involved can skip to Section~\ref{sec:euler} where we present arguably the simplest complete rule set, requiring just one additional rule.

As many of the completeness results appeared in quite a short time frame and in quite an entangled way, we give a brief history of completeness of the ZX-calculus in Section~\ref{sec:history-completeness}. Section~\ref{sec:ZW-calculus} introduces the ZW-calculus, which was instrumental in these early completeness results. Then in Sections~\ref{sec:completeness-oxford} and~\ref{sec:completeness-nancy} we summarise the Oxford and Nancy completeness results. Section~\ref{sec:euler} recalls a quite canonical completeness result that appeared after these results that requires just one additional rule beyond those of Figure~\ref{fig:zx-rules} to achieve completeness for the universal fragment. We end with Section~\ref{sec:other-fragments} where we give a brief rundown of other fragments for which completeness is known.

\subsection{History of completeness results}\label{sec:history-completeness}

The ZX-calculus was first introduced in a 2007 preprint by Coecke and Duncan~\cite{Coecke2007graphicalcalculus}, and more formally in 2008~\cite{CD1,CD2} (we will be using the date papers appeared on the ArXiv as their publication date). These results contained all the rules of Figure~\ref{fig:zx-rules} except that the Hadamard gate was seen as a generator instead of being decomposed into spiders as in~\eqref{eq:had-euler1} or~\eqref{eq:had-gadget}. That such a decomposition was necessary for completeness was realised by Duncan and Perdrix in 2009~\cite{DP1}.
The first milestone for completeness happened in 2013, when Backens proved that the rules of Figure~\ref{fig:zx-rules} suffice to prove completeness for the Clifford fragment~\cite{BackensCompleteness}.

In parallel to the development of the ZX-calculus was the development of the \emph{ZW-calculus} (which will be described in the next section). This was introduced in a 2010 paper by Coecke and Kissinger~\cite{CoeckeKissinger2010compositional}. Further work on the ZW-calculus was done in the next two years~\cite{Coecke2010rational,kissinger2012phd}, but then in the beginning of 2015 Hadzihasanovic found a complete ruleset for the ZW-calculus~\cite{hadzihasanovic2015diagrammatic}. This calculus was universal for integer matrices, and as such could describe Toffoli+Hadamard circuits. Since these gates are approximately universal for quantum computing~\cite{ShiToffoliHadamard,aharonov2003simple} this axiomatisation of the ZW-calculus became the first graphical calculus that could describe approximately universal quantum computation.

Then in May 2017, the Nancy researchers Jeandel, Perdrix and Vilmart published the first complete rule set of the Clifford+T fragment of the ZX-calculus~\cite{SimonCompleteness}. 
Their proof was based on an involved translation to and from Hadzihasanovic's ZW-calculus.
Inspired by these results, the Oxford researchers Ng and Wang published in June 2017 a completeness proof of the universal fragment of the ZX-calculus~\cite{HarnyCompleteness}. For this result they used a modified version of the ZW-calculus that was universal and complete for complex matrices, which had appeared in Hadzihasanovic's thesis which at that point had not yet appeared online~\cite{hadzihasanovic2017algebra}. By using this modified ZW-calculus, a more straightforward translation between the calculi was possible.
Now aware of this improved translation, the Nancy researchers managed to modify their results to also get a complete rule set for the universal fragment of the ZX-calculus which appeared January 2018~\cite{JPV-universal}. Completing the cycle, Ng and Wang published, also in January 2018, a complete rule set for the Clifford+T fragment that was based on their universal rule set~\cite{ng_completeness_2018}.

Based on the preprints~\cite{HarnyCompleteness,ng_completeness_2018} containing the Oxford completeness results, a slightly streamlined set of results was published in July 2018 by Ng, Wang and Hadzihasanovic~\cite{HarnyAmarCompleteness} which contains both the universal ZW-calculus as well as the Clifford+T and universal ZX-calculus completeness results.
The Nancy completeness results were also compiled in a longer 2019 paper~\cite{jeandel2019completeness}. Adding to the number of places where these completeness results have appeared, they were also included into each of the PhD theses of Vilmart, Ng and Wang~\cite{vilmart2019thesis,wang_completeness_2018,ng2018thesis}.

\subsection{W-states and the ZW-calculus}\label{sec:ZW-calculus}

The early completeness results of both Oxford and Nancy relied on a translation of the ZX-calculus into the related \emph{ZW-calculus}.
Before we go into the completeness results, let us therefore introduce this calculus.

Suppose we have $N$ parties, each of which has access to a single qubit of some (possibly) entangled state. Starting from the given state, which states can it be transformed into using just (stochastic) local qubit operations and classical communication? The resulting partitions of quantum states are known as \emph{SLOCC classes} (Stochastic Local Operations and Classical Control).
For $N=3$ it turns out that there are exactly two classes of states that have genuine three-party entanglement (i.e.~where the state does not entangle only two parties, or is a product state)~\cite{dur2000three}.
The canonical representatives of these two classes are the \emph{GHZ-state} and the \emph{W-state}:
\begin{equation}
	\ket{\text{GHZ}} = \ket{000} + \ket{111} \qquad \qquad \ket{\text{W}} = \ket{100} + \ket{010} + \ket{001}
\end{equation}
The GHZ-state can be easily represented by a 3-ary Z-spider, see Section~\ref{sec:ghz}. 
The representation of the W-state in terms of Z- and X-spiders is significantly more complex:
\begin{equation}
	\ket{\text{W}} \ = \ \input{W-state.tikz}
\end{equation}
Hence, just like was the case with the H-boxes of Section~\ref{sec:ZH}, even though we \emph{can} represent these states in the ZX-calculus, that does not mean that the ZX-calculus makes it necessarily easier to reason about them.

It was realised in~\cite{CoeckeKissinger2010compositional} that just like how the GHZ-state generalises to the Z-spider, we can generalise the W-state to a \emph{W-spider}:%
\footnote{The definition in~\cite{CoeckeKissinger2010compositional} is actually a bit different, as inputs and outputs are treated differently. We retrieve their definition by applying a NOT gate on all the outputs of our W-spider. While their definition does not enjoy the input/output symmetry our definition does, their notion does have an actual spider rule instead of the modified one we require; cf.~\eqref{eq:W-spider-rule}.}
\begin{equation}\label{eq:W-spider}
	\input{W-spider.tikz} \ \ = \ \ \ketbra{10\cdots 0}{0\cdots 0} + \cdots + \ketbra{0\cdots 0 1}{0\cdots 0} + \ketbra{0\cdots 0}{10\cdots 0} + \cdots + \ketbra{0\cdots 0}{0\cdots 0 1}
\end{equation}
I.e.~its linear map is the sum of all $\ketbra{x_1\cdots x_n}{y_1\cdots y_m}$ terms where precisely one of the $x_i$ and $y_j$ is $1$.
The 0-input 3-output W-spider is the W-state, while with a single input and output we get the NOT gate:
\begin{equation}
	\input{W-not.tikz} \ = \ \ketbra{1}{0} + \ketbra{0}{1}
\end{equation}

The W-spider~\eqref{eq:W-spider} has all the same symmetries that Z- and X-spiders have: we can permute inputs and outputs freely, and we can interchange inputs with outputs using cups and caps.

However, just as with the H-boxes of Section~\ref{sec:H-boxes}, it has a modified spider-fusion rule, which requires a 2-ary spider to be in the middle (cf.~\eqref{eq:Hbox-spider}):
\begin{equation}\label{eq:W-spider-rule}
	\input{W-spider-rule.tikz}
\end{equation}

The results of~\cite{CoeckeKissinger2010compositional} were extended in~\cite{hadzihasanovic2015diagrammatic} to a full graphical calculus for qubits. This calculus consists of W-spiders, Z-spiders which are allowed to have a phase of $\pi$ or $0$, and the `Fermionic swap':
\begin{equation}
	\input{fermion-swap.tikz} \ \ = \ \ 
	\begin{pmatrix}
	1&0&0&0\\
	0&0&1&0\\
	0&1&0&0\\
	0&0&0&-1
	\end{pmatrix}
\end{equation}
Note that in the ZX-calculus we can construct this Fermionic swap by composing a CZ gate with a SWAP.

\begin{figure}
\centering
\input{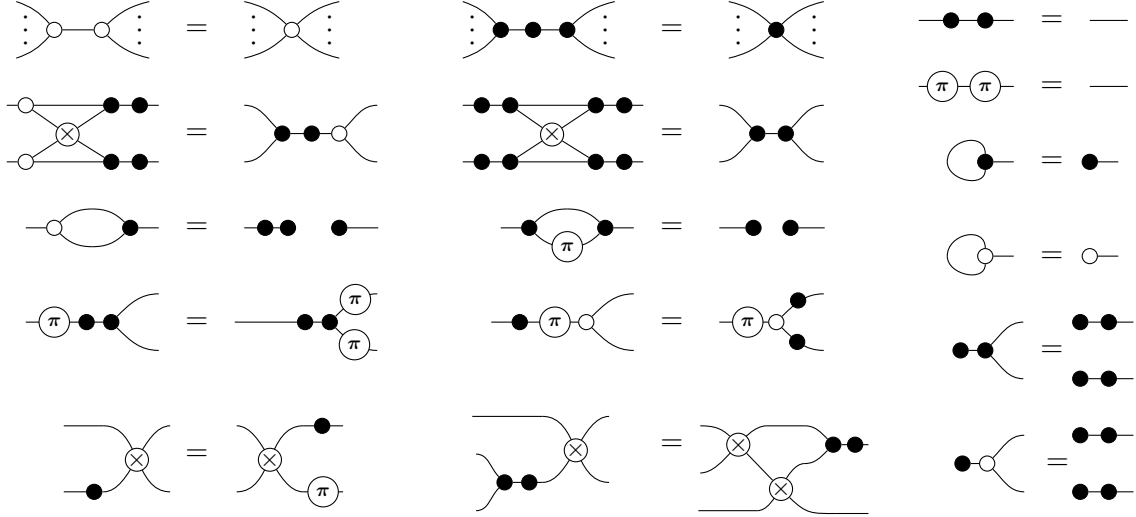}
\caption[Rules of the ZW-calculus]{
	The rules of the ZW-calculus as presented in~\cite{hadzihasanovic2015diagrammatic}, slightly modified to fit the conventions of this paper.
}
\label{fig:zw-rules}
\end{figure}

The rules of the ZW-calculus are presented in Figure~\ref{fig:zw-rules}.

Many of these rules should look familiar: we recognise rules analogous to \SpiderRule, \CopyRule, \PiRule, \BialgRule and \HopfRule. The only rules that have no direct similarity to those in the ZX-calculus are the rules involving the Fermionic swap on the bottom row.

As an aside, the reader might now wonder: ZX-calculus, ZH-calculus, ZW-calculus, how many Z*-calculi are there? The answer turns out to be that these are essentially the only ones. In~\cite{carette2020recipe}, Carette and Jeandel showed that if we have two spiders (i.e.~families of linear maps satisfying equations like~\eqref{eq:W-spider-rule}) acting on qubits, that interact according to a bialgebra rule like \BialgRule, then one of them must be the Z-spider (up to a global basis change), and the other must be the X-spider, H-box, or W-spider (up to some trivial modifications).%
\footnote{Complicating the picture somewhat, it is possible to get around this classification by focusing on non-spider-like maps, such as the calculus based in quaternions presented in~\cite{MillerBakewell2020ZQ}.}

The ZW-calculus is universal for matrices over the integers. This is an approximately universal fragment that includes the Toffoli and Hadamard gates (up to normalisation). The rules presented in Figure~\ref{fig:zw-rules} are also complete for this fragment.

In~\cite{HarnyAmarCompleteness} it was shown that for any commutative ring $R$, we can make an extended ZW$_R$-calculus that is universal for matrices over $R$. 
In the ZW$_R$-calculus, we allow the Z-spiders to be labelled by an arbitrary element $r\in R$:
\begin{equation}\label{eq:Z-spider-r}
\input{Zsp-r.tikz} \ \ :=\ \ \ketbra{0 \cdots 0}{0 \cdots 0} + r \ketbra{1 \cdots 1}{1 \cdots 1},
\end{equation}
The ruleset for ZW$_R$ presented in~\cite{HarnyAmarCompleteness} differs in quite a number of ways from those of Figure~\ref{fig:zw-rules}. However, the most significant changes are the additions of the following rules that allow arithmetic of the ring to be done in the calculus:%
\footnote{The other notable changes are that the Reidemeister moves for the Fermionic swap are explicitly included as rules, and that many of the rules of Figure~\ref{fig:zw-rules} now accept arbitrary labels on the Z-spiders.}
\begin{equation}
\input{ZW-rules-extended.tikz}
\end{equation}

Taking the ring to be the complex numbers, this gives a complete calculus for universal quantum computing, the first of its kind.

\subsection{The Oxford completeness results}\label{sec:completeness-oxford}

The work on the ZW-calculus, and especially its extension to arbitrary rings, raised the question whether those results couldn't be `ported' to the ZX-calculus.
Abstractly, suppose we have two languages $A$ and $B$ where $B$ is complete, and suppose we have translations $f:A\rightarrow B$ and $g:B\rightarrow A$ that are each others inverses. Then to prove two diagrams in $A$ are equal, we can transport them to $B$ using $f$, prove they are equal there, which is possible because $B$ is complete, and then transport the derivation back to $A$ using $g$.
To prove $A$ is complete it then suffices to show that each derivation in $B$ is still a valid derivation in $A$ when transported using $g$. In practice, this means you need to show that all the rules of $B$, when transported to $A$, are provable using the rules of $A$.

This means that in order to get a complete ZX-calculus we only need to give an encoding of ZX-diagrams into ZW-diagrams (which is easy), an encoding of ZW-diagrams into ZX-diagrams (this is harder), and then assume the rules of the ZW-calculus as rules in the ZX-calculus. This is essentially what the first completeness results for the ZX-calculus did. The crux is that the translated ZW-rules generally contain many generators as ZX-diagrams, and hence are quite unwieldy. To get a more elegant calculus, these rules must therefore be reduced to axioms that are smaller.

We present here the complete ZX-calculus for the universal fragment as proposed by Hadzihasanovic, Ng and Wang in~\cite{HarnyAmarCompleteness}. It should be noted that the earlier rulesets proposed by Ng and Wang for both the universal and Clifford+T fragment~\cite{HarnyCompleteness,ng_completeness_2018} are slightly more complex.

They start by introducing two new derived generators for the ZX-calculus, the \emph{triangle} and the \emph{$\lambda$-box}. Before we give their definition as ZX-diagrams, let us write the linear maps they represent:
\begin{equation}
	\input{triangle.tikz} \ \ = \ \ \ketbra{0}{0} + \ketbra{1}{0} + \ketbra{1}{1} \qquad \quad \input{lambda-box.tikz} \ \ = \ \ \ketbra{0}{0} + \lambda \ketbra{1}{1}
\end{equation}
Here $\lambda$ is a strictly positive real number $0<\lambda\in \R$. Note that the symbol of the triangle is intentionally not symmetric as the linear map it represents is not self-transpose: flipping the diagram horizontally results in a different linear map. The $\lambda$-box should not be confused with the H-boxes of Section~\ref{sec:ZH} that we will not use in this section.

The triangle is defined as a ZX-diagram as follows:%
\footnote{This is just one possible definition of the triangle as a ZX-diagram. Their completeness result works regardless of how the triangle is defined, although these technically give different rulesets.
This definition of the triangle first appeared in~\cite[Exercise 12.10]{CKbook}, although it was written using different notation.
In~\eqref{eq:triangle-def-2} we give an alternative definition of the triangle.}
\begin{equation}\label{eq:triangle-def-1}
	\input{triangle.tikz} \ \ = \ \ \input{triangle-as-zx.tikz}
\end{equation}

The definition of the $\lambda$-box is a bit more involved and requires case distinctions. Write $\lambda = \lambda_i + \lambda_f$ where $\lambda_i\in \mathbb N$ is the integer part of $\lambda$ and $\lambda_f \in [0,1)$ is the fractional part. Then we define
\begin{equation}
	\input{lambda-box.tikz} \ \ = \ \ \input{lambda-box-def.tikz}
\end{equation}
where the fractional part is defined directly, and the integer part by induction as follows:
\begin{equation}
	\input{lambda-box-fraction.tikz} \qquad \quad \input{lambda-box-integer.tikz} \qquad \quad \input{lambda-box-1.tikz}
\end{equation}
In the first equation $\alpha$ is related to $\lambda_f$ via $\lambda_f = e^{i\alpha} + e^{-i\alpha} = 2\cos(\alpha)$. Note that because of this definition that $\lambda$-boxes commute through Z-spiders.

\begin{figure}
\centering
\input{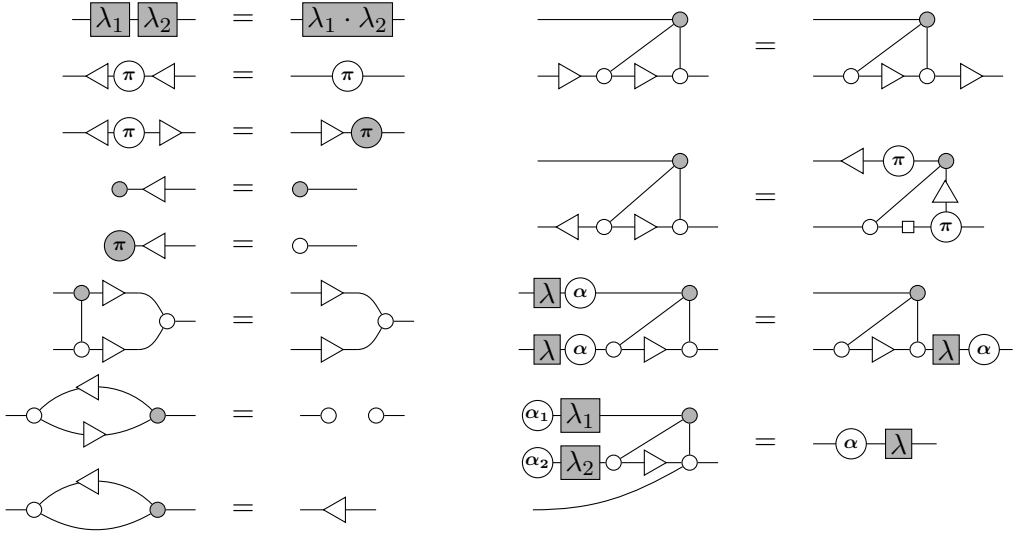}
\caption[Extended rules for ZX-calculus (Oxford)]{%
	The extended ruleset of the ZX-calculus as presented 
	in~\cite{hadzihasanovic2015diagrammatic}, slightly modified to fit the conventions of this paper.
	These rules hold for all $\lambda \in \R_{>0}$ and $\alpha\in \R$ and for all $\lambda_1,\lambda_2\in \R_{>0}$ and $\alpha_1,\alpha_2\in \R$ satisfying the side condition $\lambda e^{i\alpha} = \lambda_1e^{i\alpha_1} + \lambda_2 e^{i\alpha_2}$.
}
\label{fig:oxford-rules}
\end{figure}

Next to the standard ZX-calculus axioms of Figure~\ref{fig:zx-rules}, this completeness result requires a set of rules relating the triangle and $\lambda$-box to the standard spiders.%
\footnote{In this figure we do not include all their rules, as we view the triangle and $\lambda$-box as derived generators instead of actual generators, which makes some of their rules superfluous.}  
See Figure~\ref{fig:oxford-rules}.

The intuition behind most of these rules is that we identify a state $(1, z)^T$ with the complex number $z$. The triangle then acts as the successor function that maps it to $(1,z+1)^T$. A Z $\pi$-phase acts like negation, while an X $\pi$-phase corresponds (up to scalar) to multiplicative inverse. The repeating `gadget' in the right column of Figure~\ref{fig:zx-rules} acts as addition of two states. 

Note that the bottom-right rule has a side condition that requires $\lambda e^{i\alpha} = \lambda_1e^{i\alpha_1} + \lambda_2 e^{i\alpha_2}$, we call such rules \emph{non-linear} as the relation between the phases and labels in it aren't expressible as a linear function. It was argued in~\cite{JPV-universal} that any complete ruleset for the universal fragment of the ZX-calculus requires a non-linear rule.

By restricting the allowed phases in the spiders to multiples of $\frac\pi4$ this same ruleset is also complete for the Clifford+T fragment. In that case the non-linear rule reduces to a finite number of linear rules.

In~\cite{vilmartzxtriangle} Vilmart produced a variation on these completeness results. He promoted the triangle to a generator, and removed the $\lambda$-box. This resulted in a complete calculus for the Clifford+T fragment that required fewer axioms than those presented in Figure~\ref{fig:oxford-rules}. By adding an additional non-linear rule this calculus is complete for the universal fragment.

The ideas behind the Oxford completeness results led to the development of the \emph{algebraic} ZX-calculus, where the triangle is promoted to a generator and the spiders are allowed to be labelled by arbitrary complex numbers (giving essentially the same interpretation of Z-spiders as in the ZW-calculus, cf.~\eqref{eq:Z-spider-r})~\cite{algebraicZX}. The ruleset of this calculus consists of those in Figure~\ref{fig:zx-rules} and a simplified variation on the rules in Figure~\ref{fig:oxford-rules}. Its proof of completeness does not rely on the translation back and forth from ZW, instead proving that each diagram can be brought to a unique normal form~\cite{wang2020algebraic}.
Just like the ZW-calculus, the algebraic ZX-calculus can also be used as a complete calculus over an arbitrary ring, but in addition it can also describe matrices over semi-rings~\cite{ringZX}.

\subsection{The Nancy completeness results}\label{sec:completeness-nancy}

In parallel to the work on completeness done in Oxford, a group of researchers in Nancy was working on finding a complete ruleset for the ZX-calculus as well.
They also relied on a translation between the ZW-calculus and the ZX-calculus, but as they were not yet aware of the extension of the ZW-calculus to arbitrary rings, they required a more complicated translation that could encode the complex numbers appearing in the Clifford+T fragment into the real matrices of the ZW-calculus.

\begin{figure}
\centering
\input{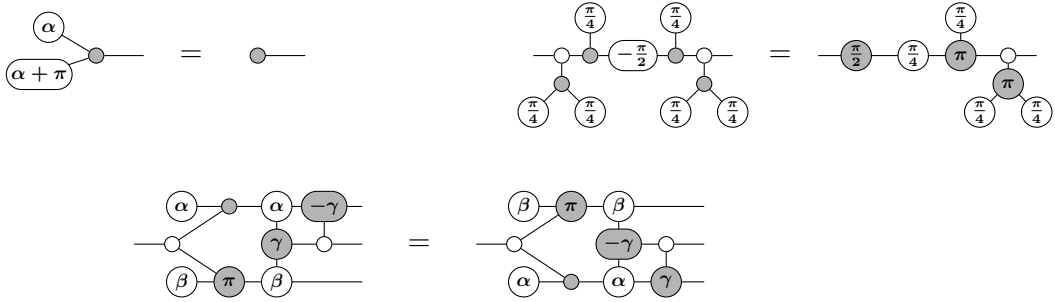}
\caption[Extended rules for ZX-calculus (Nancy)]{%
	The extended ruleset of the ZX-calculus as presented 
	in~\cite{SimonCompleteness}, slightly modified to fit the conventions of this paper.
	These rules hold for all $\alpha, \beta, \gamma\in \R$.
}
\label{fig:nancy-rules}
\end{figure}

The first completeness result of the Nancy group by Jeandel, Perdrix and Vilmart~\cite{SimonCompleteness} proved completeness of the Clifford+T fragment by adding three rules beyond the standard rules of Figure~\ref{fig:zx-rules}, see Figure~\ref{fig:nancy-rules}.

The first of these rules is the rule we saw in~\eqref{eq:supplementarity}, the other two look at first glance to be quite arbitrary. In~\eqref{eq:triangle-def-1} we saw how to write the triangle in terms of spiders. However, we can also write the triangle as a ZX-diagram as follows:
\begin{equation}\label{eq:triangle-def-2}
	\input{triangle.tikz} \ \ = \ \ \input{triangle-as-zx2.tikz}
\end{equation}
With this decomposition, the second rule of Figure~\ref{fig:nancy-rules} can be shown to be equivalent to the following rule that also appeared in Figure~\ref{fig:oxford-rules}:
\begin{equation}
	\input{BW-rule.tikz}
\end{equation}
As demonstrated in~\cite{SimonCompleteness}, the final rule is related to the property that two controlled unitaries commute when one is controlled on a qubit, while the other is `anti-controlled' on the same qubit (i.e.~fires when it is in the state $\ket{0}$). The rule seems to be necessary to prove this commutation when of these unitaries is a controlled Z-phase, while the other is a anti-controlled X-phase.

In a later paper~\cite{JPV-universal}, the same authors showed that using one additional non-linear rule, the calculus is complete for the universal fragment. This rule is the following:
\begin{equation}
	2e^{i\theta_3}\cos(\gamma) = e^{i\theta_1}\cos(\alpha) + e^{i\theta_2}\cos(\beta)\ \  \implies \ \ \input{JPV-nonlinear.tikz}
\end{equation}
While no straightforward interpretation of this rules is given in~\cite{JPV-universal}, it is necessary to relate the ring structure of the complex numbers to the group structure of the complex phases.
A more direct proof (not requiring a translation to ZW) that the rules of Figure~\ref{fig:zx-rules} and~\ref{fig:nancy-rules} are complete was given in~\cite{ZXNormalForm} by showing that each diagram can be brought to a unique normal form.

\subsection{Completeness from Euler decompositions}\label{sec:euler}

The Oxford completeness result requires many new small axioms (small in terms of the number of derived generators required to write them), while the Nancy completeness result requires only a few additional axioms but which are large and are harder to interpret. Hence, both axiomatisations are not fully satisfactory.
Luckily, since those first completeness results, a much simplified complete calculus has been found by Vilmart~\cite{vilmarteulercompleteness}. This completeness result requires just \emph{one} additional rule added to the standard ones of Figure~\ref{fig:zx-rules}. This non-linear rule simply expresses the equivalence of two possible Euler decompositions of a single-qubit rotation:
\begin{equation}\label{eq:euler-nonlinear}
	\input{nonlinear-euler.tikz}
\end{equation}
The crux of this rule is that the relationship between $\alpha_1,\beta_1,\gamma_1$ and $\alpha_2,\beta_2,\gamma_3$ is quite complicated, requiring iterated trigonometric functions to be fully expressed (see~\cite[Section~5.3]{wang_completeness_2018} for a derivation).

That such a rule would be required for completeness of the ZX-calculus was first realised in~\cite{Witt:2014aa}. Then in~\cite{coecke2018zx} it was shown that adding this rule to the standard rules allows one to prove completeness on two-qubit circuits (note that they did not require the specific relations between the phases, but only certain symmetries, such as that $\alpha_1=\gamma_1$ implies $\alpha_2=\gamma_2$). Then in~\cite{vilmarteulercompleteness}, Vilmart showed that the rule~\eqref{eq:euler-nonlinear}, in combination with the standard rules of Figure~\ref{fig:zx-rules} suffices to prove all four new rules of the Nancy completeness papers of Section~\ref{sec:completeness-nancy}, hence showing that \eqref{eq:euler-nonlinear} is sufficient to make the ZX-calculus complete.

This is quite pleasing, because the rules of Figure~\ref{fig:zx-rules} give completeness for the Clifford fragment and hence essentially encode all the behaviour of entanglement and the relations of the Pauli operators and then~\eqref{eq:euler-nonlinear} adds onto that the rotational structure of the Bloch sphere, retrieving a complete language for reasoning about qubit quantum computing.

That being said, the benefits of graphical reasoning are betrayed a bit by this rule because of the complicated side condition, and one might hope for a complete calculus not including any side conditions. As argued in~\cite{jeandel_rational_2018,JPV-universal} it however seems unlikely that all the relations between phases that are not rational multiples of $\pi$ can be captured by a finite set of rules.

An intuitive reason for why the ZX-calculus cannot seem to reason in a complete way about the universal fragment, is because it lacks a way to directly encode the complex numbers. Indeed, the extended ZW-calculus~\cite{HarnyAmarCompleteness}, the algebraic ZX-calculus~\cite{algebraicZX} and the ZH-calculus~\cite{backens2018zhcalculus} all do not require side conditions as they allow their generators to be labelled by arbitrary complex numbers and have rules directly encode the operations of addition and multiplication of complex numbers.


\subsection{Completeness for other fragments}\label{sec:other-fragments}

The previous sections focused on the Clifford+T fragment and the universal fragment of the ZX-calculus. These are obviously of interest as they correspond to commonly used gate sets in quantum computing. These are however not the only fragments for which completeness is known. In chronological order of proof of completeness:
\begin{itemize}
	\item Real Clifford: In 2013 Duncan and Perdrix~\cite{DP3} showed that restricting the phases in spiders to multiples of $\pi$ (which necessitates making the Hadamard a generator instead of letting it be a derived generator) with the rules of Figure~\ref{fig:zx-rules} gives a calculus that is complete for real Cliffords (i.e.~Clifford computation involving only real numbers).
	\item Clifford: Also in 2013, Backens~\cite{BackensCompleteness} showed that the $\frac\pi2$ fragment with the Figure~\ref{fig:zx-rules} rules is complete for the full Clifford fragment.
	\item Single-qubit Clifford+T: In 2014, Backens established that those same rules suffice for completeness of the single-qubit Clifford+T fragment~\cite{backens2014zx}.
	\item Fermionic quantum computing: In 2018, a variation on the ZW-calculus was proposed that is complete for Fermionic quantum computing~\cite{hadzihasanovic2018diagrammatic}.
	\item Toffoli+Hadamard: Quantum circuits consisting of Toffoli and Hadamard gates form an approximately universal model of quantum computation~\cite{ShiToffoliHadamard}. This model corresponds to matrices over $\mathbb Z[1/\sqrt{2}]$. Vilmart introduced in 2018 the ZX$\Delta$-calculus~\cite{vilmartzxtriangle} with a small number of new rules that adds the triangle as a generator. Restricting the phases in the spiders to multiples of $\pi$ gave a calculus that was complete for this fragment. In 2019 it was shown that a complete calculus for this fragment can also be constructed by restricting the phases in the ZH-calculus~\cite{zhphasefree}.
	\item Rational fragment: In 2018, Jeandel proved~\cite{jeandel_rational_2018} that the Nancy rules of Figure~\ref{fig:nancy-rules} together with one additional rule~\cite{cyclo} suffice to prove completeness for the rational fragment, where phases of spiders are allowed to be any rational multiple of $\pi$.
	\item Classical circuits: The Boolean semi-ring $\mathbb B = \{0,1\}$ has multiplication defined as usual, but with its addition satisfying $1+1 = 1$. Boolean circuits where we allow post-selection can be represented as matrices over $\mathbb B$. In 2020, Comfort showed~\cite{ZXand} that Z- and X-spiders augmented with an `AND-gate generator' and a natural set of rules reminiscent of Figure~\ref{fig:zh-rules} suffices to prove completeness for this fragment.
\end{itemize}

\section{Extensions of the language}\label{sec:extensions}

The ZX-calculus allows graphical reasoning about pure quantum processes on qubits. There are then seemingly three directions in which the language could be extended. The first is to allow the representation of \emph{impure}, i.e.~mixed or decohered, quantum processes. The second is to also allow the representation of classical processes together with the quantum processes. The third is to represent quantum systems of a different dimension, like qutrits or more general qudits.
In this section we will discuss these three extensions in turn.

\subsection{Mixed states, decoherence and discarding}\label{sec:mixed-state}

The easiest way to represent decoherence and classical measurement in ZX-diagrams is to add a generator to represent the process of \emph{discarding}. In pure quantum mechanics a system can never be destroyed, as information is always preserved. However, when we allow classical interaction with a system, such as the process of measurement, information can be lost.

In order to represent the discarding operation, we need to change the interpretation of ZX-diagrams. In the standard interpretation, a $n$-input, $m$-output ZX-diagram is seen as an element of $(\mathcal H)^{\otimes n}\otimes (\mathcal H^*)^{\otimes m}$, where $\mathcal H = \C^2$ and $\mathcal H^*$ denotes the conjugate space, which is what we would expect for a pure process.
For instance, in this interpretation the first of the 0-input, 1-output diagrams of~\eqref{eq:kets} is equal to the element $\ket{0}$ of $\mathcal H$.
In order to represent mixed processes, we need to lift these pure processes to operators on the Hilbert space instead, so that the state $\ket{0}$ is represented by the projector $\ketbra{0}{0}$ in $B(\mathcal H)$, the space of operators on $\mathcal H$. 
In general, a $n$-input, $m$-output ZX-diagram will correspond to a completely positive map from $B(\mathcal H)^{\otimes n}$ to $B(\mathcal H)^{\otimes m}$. The details of how this lifting works are not important for our purposes. We do show a way in which we can relate these `lifted' diagrams to `normal' ZX-diagrams in Section~\ref{sec:doubling}.

To make mixed ZX-diagrams we need an additional generator that we call \emph{discarding}, which we write with the `ground symbol':
\begin{equation}\label{eq:ground}
	\ground
\end{equation}
The operator associated to discarding is the (partial) trace $\tr: B(\mathcal H) \rightarrow \C$.
Using discarding we can easily represent the decoherence operator to both the Z-basis and the X-basis:
\begin{equation}\label{eq:decoherence}
	\input{decoherence-Z.tikz} \qquad \qquad \input{decoherence-X.tikz}
\end{equation}
That this is indeed the decoherence operator can be shown using the rewrite rules we will introduce now.

\begin{figure}
\centering
\input{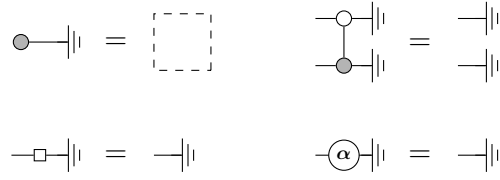}
\caption[Rules for the discard generator in the ZX-calculus]{%
	The rules for discarding in the ZX-calculus as presented 
	in~\cite{carette_completeness_2019}, slightly modified to fit the conventions of this paper.
	These rules hold for all $\alpha\in \R$. The dotted square represent an empty diagram. Because of the discard rule for the Hadamard these rules also hold with the colours interchanged.
}
\label{fig:discard-rules}
\end{figure}

Given a complete graphical calculus it turns out to be relatively straightforward to make the extended calculus where discarding is allowed complete as well~\cite{carette_completeness_2019}. Namely, you identity a set of diagrams that generate all the possible isometries and then you add rules that state that those diagrams are indeed discarded when the discard is applied to them.
We present a set of such rewrite rules for the universal fragment%
\footnote{The result of~\cite{carette_completeness_2019} that a complete ruleset can be extended to a complete set that includes the discard requires a technical property that is satisfied by the universal fragment, but is \emph{not} satisfied by the Clifford+T fragment. The rules we show here are still sound in the Clifford+T fragment, but are not sufficient to make it complete.} 
of the ZX-calculus in Figure~\ref{fig:discard-rules}.

Using these rules we can show that the decoherence operators of~\eqref{eq:decoherence} indeed act like decoherence. For instance, they are idempotent, in the sense that applying decoherence a second time does nothing:
\begin{equation}
	\input{decohere-twice.tikz}
\end{equation}

As can be easily shown, the Z-decoherence does not do anything when the input is a $\ket{0}$ or $\ket{1}$ (an X-spider), because these are in the preferred basis for the decoherence. However, inputting any `phase state' $\ket{0} + e^{i\alpha}\ket{1}$ gives the maximally mixed state:
\begin{equation}
	\input{plus-to-mixed.tikz}
\end{equation}
Here in the last step we write the rotated ground symbol for the transpose of discard, which corresponds to the maximally mixed state.

ZX-diagrams are universal for pure quantum operations. By Stinespring dilation, any mixed quantum process (that is, a completely-positive map) can be written as a pure process followed by a partial trace applied to some of its outputs.%
\footnote{Stinespring dilation first appeared in the context of C$^*$-algebras~\cite{stinespring1955positive}. In this infinite-dimensional setting it is useful to work with unital maps, instead of with trace-preserving maps. The original result therefore does not talk about partial traces. In a finite-dimensional setting our formulation can however be shown to be equivalent; see for instance~\cite[Theorem~6.61]{CKbook}.} 
Hence, ZX-diagrams augmented with the discarding generator can represent any mixed quantum process. This does of course not mean that all maps have a straightforward representation in this way.
For instance, there is currently no known representation of \emph{partial} decoherence in the ZX-calculus (the process that mixes decoherence and the identity process with some nontrivial probability).

\subsection{Classical control and doubling}\label{sec:doubling}

A maximally decohered quantum system only carries states that are in the basis preferred by the decoherence. For instance, the most general state on a qubit decohered in the Z-basis is of the form $p \ketbra{0}{0} + (1-p)\ketbra{1}{1}$, and hence is a classical probability distribution over the pure orthogonal states $\ketbra{0}{0}$ and $\ketbra{1}{1}$.

We can use this observation to encode quantum-classical interactions in the ZX-calculus: fix a basis, such as the Z-basis, and decohere every wire that is supposed to be classical to this basis. 
In this approach, a measurement in that same basis is also presented as decoherence. For instance, the teleportation protocol presented in Section~\ref{sec:teleportation} can also be presented using decoherence as measurement, and then we simplify it to show correctness:
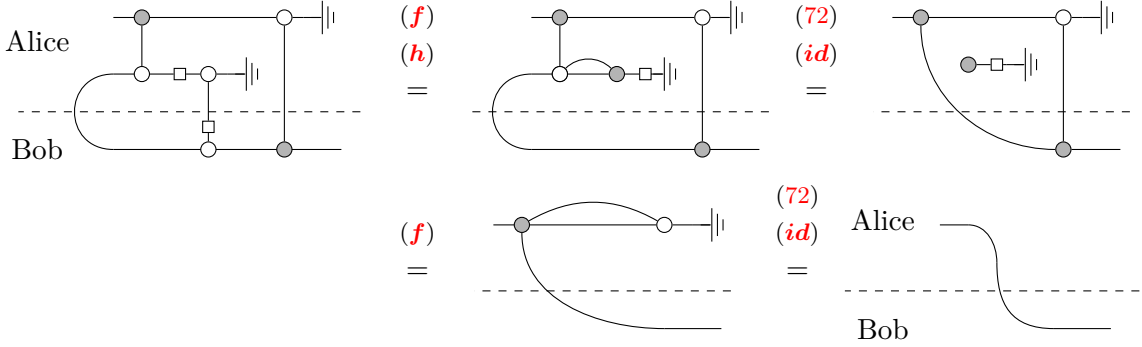
\ctikzfig{teleport-decoherence}
Here the vertical wires attached to the decoherence operator represent classical operations.

Admittedly, in this representation where classical wires are labelled by decoherence it is not immediately visually clear which wires carry classical information and which carry quantum information. In~\cite{coecke_paquette_pavlovic_2009,coecke2016categorical,CKbook} a different approach was suggested.
In this approach we write `quantum wires' as thick wires and classical wires as thin wires:
\begin{equation}
	\text{quantum:} \ \  \input{Zsp-a-dd.tikz} \qquad \qquad \text{classical:} \ \ \input{Zsp-a.tikz}
\end{equation}
The quantum maps can be seen as syntactic sugar for regular ZX-diagrams using a procedure known as \emph{doubling}. 
Analogous to how we represent a pure state $\ket{\psi}$ as the projector $\ketbra{\psi}{\psi}$, and a unitary $U$ by the conjugation map $\rho\mapsto U\rho U^\dagger$ when we are dealing with mixed states, we also need to `double' our ZX-diagrams when dealing with mixtures and decoherence.%
\footnote{Such a procedure can be defined for any dagger compact closed category, and is known as the \emph{CPM construction}~\cite{selinger2007dagger,coecke2010environment}. CPM stands for `completely positive maps' as the application of the CPM construction to the category of Hilbert spaces and linear maps gives the category of completely positive maps between the spaces of operators on Hilbert spaces.}

We define the quantum spider as a `doubled' spider by taking each quantum wire to correspond to two classical wires:
\begin{equation}
	\input{Zsp-a-dd.tikz} \ \ := \ \ \input{Zsp-a-undoubled.tikz}
\end{equation}
It is straightforward to verify that the standard rules of the ZX-calculus are still valid when working with quantum spiders defined in this way.

Using thick/thin wires, the discard operator acting on a quantum wire can be defined to be the cap from~\eqref{eq:cup-cap-def}:
\begin{equation}
	\input{ground-dd.tikz} \ \ := \ \ \input{cap-undoubled.tikz}
\end{equation}
We can then see what happens when we apply a discard to a quantum spider:
\begin{equation}
	\input{quantum-spider-discard.tikz}
\end{equation}
We end up with a classical spider, but with quantum wires coming out of it. This is indeed what we would expect from decoherence, as it destroys the quantum nature of the system, even though on the surface it still looks like a proper quantum system.

We can represent a Z-basis measurement as a Z-spider that takes a quantum wire as input and outputs a classical wire:
\begin{equation}
	\input{Z-measure.tikz}
\end{equation}
We can also define a spider that takes in a classical wire and outputs a quantum wire:
\begin{equation}
	\input{Z-encode.tikz}
\end{equation}
This spider corresponds to encoding a classical state into a quantum state (e.g.~the Boolean $0$ gets mapped to $\ket{0}$ and $1$ gets mapped to $\ket{1}$).

Using these measurement and encode operations (as well as the analogous ones in the X-basis) we can represent the teleportation protocol as follows:
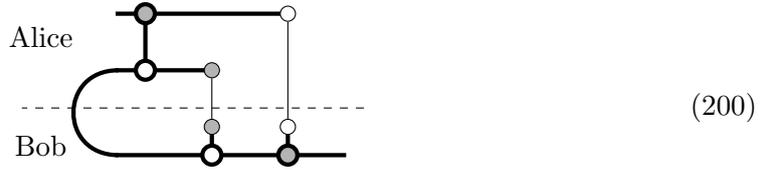
\begin{equation}
	\input{teleport-thick-thin.tikz}
\end{equation}
We leave it to the reader to verify that this indeed implements the state-teleportation protocol.

For more details on the thick-thin wires and how we can use them to represent quantum-classical processes see~\cite[Chapter~8]{CKbook}.

\subsection{ZX-calculus for qudits}\label{sec:qudits}

In this paper we have only worked with qubits, i.e.~two-dimensional Hilbert spaces. This is the setting in which the ZX-calculus was first developed and in which it works best. That being said, quite a lot of structure carries over into higher-dimensional spaces.

It is for instance possible to define spiders for Hilbert spaces of arbitrary (finite) dimension: pick any orthonormal basis $\ket{v_1},\ldots,\ket{v_n}$ and define
\begin{equation}
	\input{Zsp.tikz} \ \ = \ \ \sum_j \ketbra{v_j\cdots v_j}{v_j\cdots v_j}
\end{equation}
Then this family of linear maps satisfies the same spider fusion rule~\SpiderRule and identity rule~\IdRule as the spiders in the ZX-calculus do. While these spiders are still symmetric under swaps of inputs and swaps of outputs, they in general do not allow inputs and outputs to be interchanged using cups and caps as in~\eqref{eq:spider-cupcap}. This only holds when the basis of the spider is \emph{self-conjugate}, i.e.~when $\overline{\ket{v_i}} = \ket{v_i}$ for all basis vectors $\ket{v_i}$ (where $\overline{v}$ denotes the componentwise complex conjugate of the vector).

To define phases for spiders is a bit more complicated. A \emph{phase map} for a given orthonormal basis is a unitary that is diagonal in that basis. The set of phase maps for a given basis is a group called the \emph{phase group}.
For a two-dimensional space the phase group is isomorphic to $U(1)$ (when we ignore global phase in the unitaries). This is why we could parametrise the phases of spiders in the ZX-calculus with a single number.
In general, the phase group of a basis in an $n$-dimensional space will be isomorphic to $U(1)^{n-1}$, and hence we need to parametrise the phases of a spider by a vector $\vec \alpha$ of length $n-1$:
\begin{equation}\label{eq:Zsp-ndim}
	\input{Zsp-a-vec.tikz} \ \ = \ \ \ketbra{v_0\cdots v_0}{v_0 \cdots v_0} + \sum_j e^{i\alpha_j}\ketbra{v_j\cdots v_j}{v_j\cdots v_j}
\end{equation}
As in \SpiderRule, when higher-dimensional spiders fuse, their phases add, in this case pointwise:~$(\vec \alpha + \vec \beta)_j = \alpha_j + \beta_j$.

Much of the power of the ZX-calculus comes from the interaction between the Z-spiders and the X-spiders. Parts of that interaction can also be generalised to higher dimension.
We call a pair of spiders \emph{strongly complementary}~\cite{coecke2012strong} when they satisfy the generalised bialgebra equation~\eqref{eq:bialgebra-rule-many}:
\begin{equation}
	\input{bialgebra-rule-many.tikz}
\end{equation}
For brevity we will simply refer to such a pair as Z- and X-spiders.
It turns out that pairs of strongly complementary spiders in arbitrary-dimensional spaces have been fully classified~\cite{coecke2012strong}. 
Suppose the spiders are acting on an $n$-dimensional space and that the Z-spider is given by the basis $\ket{v_1},\ldots,\ket{v_n}$. Then there is a finite abelian group $G=\{g_1,\ldots,g_n\}$ such that
\begin{equation}\label{eq:strongcomp-spider}
	\input{Xsp.tikz} \ \ = \ \ \sum_{\sum_i g_i = \sum_j h_j} \ketbra{g_1\cdots g_n}{h_1\cdots h_m}
\end{equation}
where we define $\ket{g_i} := \ket{v_i}$. The sum in this equation is over all $g_1,\ldots, g_n,h_1,\ldots, h_m \in G$ such that $\sum_i g_i = \sum_j h_j$, where the sum denotes the group operation. In this definition, the 2-input 1-output X-spider corresponds to the linear map defined by $\ket{g}\otimes \ket{h} \mapsto \ket{g+h}$.

This relation generalises the construction for Z- and X-spiders as for $n=2$ the only possible group is $\mathbb Z_2$, so that the X-spider acts as XOR on the basis states, exactly as we saw was the case in the ZX-calculus.

Note that the converse for strong complementarity is also true: if we fix a spider defined according to a basis $\ket{v_1},\ldots,\ket{v_n}$ and we fix an $n$-element abelian group $G$, then defining a family of maps according to~\eqref{eq:strongcomp-spider} gives a spider that is strongly complementarity to the first spider~\cite{coecke2012strong}. As there exist for each $n\in \N$ an abelian group with $n$ elements, pairs of strongly complementary spiders are possible in every dimension.

The $\pi$ commutation rule \PiRule and the state copy rule \CopyRule also have natural generalisations to higher dimensions where now in dimension $n$, there are $n$ phases that can be commuted/copied through the spider of the opposite colour~\cite{coecke2012strong}.

The only rules of Figure~\ref{fig:zx-rules} that do not have a straightforward translation to higher dimensions are the ones involving Hadamard gates for which we need to know more about the specific group.
A finite abelian group $G$ can always be split up into a product of cyclic groups of prime order. If the group $G$ is a non-trivial product then spiders defined according to~\eqref{eq:Zsp-ndim} and~\eqref{eq:strongcomp-spider} can be `split' along this product into spiders acting on lower-dimensional spaces, hence we can restrict ourselves to thinking about $\mathbb Z_p$ for $p$ prime.
In that case we write the basis for the Z-spider as $\ket{0},\ldots,\ket{p-1}$ and then we can define the complementary spider by the `Fourier-transformed' basis $\ket{f_j} = \sum_k \omega^{jk} \ket{k}$ where~$\omega = e^{i\pi/p}$ is the $p$th root of unity. 
The corresponding Hadamard gate is then the quantum Fourier transform $F$ and satisfies $F^4 = \id$. This unitary is not self-transpose and not self-inverse. 
So while it does give a type of colour-change rule between the two spiders, it does not work as elegantly as in the two-dimensional setting. In particular,  the meta-rule that all rules hold with the colours interchanged no longer holds. For instance, if the basis of the first spider is self-conjugate, then the complementary basis is not, so while the first spider enjoys the symmetries of~\eqref{eq:spider-cupcap}, the second spider does not.

ZX-diagrams over any dimension can be shown to be universal~\cite{wang2014qutrit}. That is, every linear map between qudits of dimension $p$ can be written as a diagram consisting of the two complementary spiders related by the Fourier transform described above, where all the spiders are allowed to be labelled by a vector of phases $\vec\alpha = (\alpha_1,\ldots,\alpha_{p-1})$.

For qutrits a ruleset analogous to that of Figure~\ref{fig:zx-rules} has been shown to be complete for the qutrit Clifford fragment~\cite{EPTCS266.3}. The corresponding statement for general qudits is still unknown. A general complete ruleset for qutrits and qudits is also unknown.

\section{Concluding remarks}

In this paper we gave a general introduction to the ZX-calculus.
We showed how the ZX-calculus can be used to graphically reason about many parts of quantum computation.
We covered several aspects of the language, including the representation and simplification of Clifford circuits and Toffoli gates, the categorical origins of the ZX-calculus and the matter of completeness, and we ended with several known extensions that allow ZX-diagrams to reason about mixed processes and qudits.

Since the publication of the completeness results of the ZX-calculus in 2017~\cite{HarnyAmarCompleteness,SimonCompleteness}, the numbers of works using the ZX-calculus has increased dramatically (in the first decade after the first 2007 preprint on the ZX-calculus~\cite{Coecke2007graphicalcalculus} around 60 preprints, papers and theses that used the ZX-calculus appeared, about the same number of works that have appeared in the period 2018--2020\footnote{See the chronological list of publications at \url{http://zxcalculus.com/publications.html}.}).
Many of the recent works focus on specific applications such as MBQC~\cite{kissinger2017MBQC,wetering-gflow}, quantum circuit optimisation~\cite{cliffsimp,deBeaudrap2020Techniques,kissinger2019tcount,phaseGadgetSynth} and as a tool for reasoning about surface code quantum computing~\cite{horsman2011quantum,deBeaudrap2020Paulifusion,Gidney2019efficientmagicstate,horsman2017surgery,hanks2019effective}. 
It is too early yet to tell in which domain the ZX-calculus will be most useful. Hazarding a guess, it would make sense that areas where the circuit model is not suitable would stand the most to gain from using the ZX-calculus. Indeed, many recent papers using the ZX-calculus deal with non-unitary models of quantum computation such as MBQC and surface code quantum computing.

The ZX-calculus could also prove a valuable tool in learning about quantum computing, as the graphical rewrite rules make it easy to remember the various identities and properties of computations.

\medskip
\noindent\textbf{Acknowledgments}: The author is supported by a Rubicon fellowship financed by the Dutch Research Council (NWO). The author wishes to thank Aleks Kissinger in helping find some missing references, Quanlong Wang, Cole Comfort, and Renaud Vilmart for proofreading parts of the paper, and Alex Cowtan and Lia Yeh for various suggestions and proofreading.

\phantomsection
\addcontentsline{toc}{section}{References}
\bibliographystyle{plainnat}
\bibliography{../bibliography}

\appendix


\clearpage
\section{ZX-calculus cheatsheets}
In these appendices we present several pages of cheatsheets meant for reference and for printing separately.

\subsection{Generators and their matrices}

\begin{align*}
\text{Z-spider:}& \ \ \ \ \quad \input{Zsp-a.tikz} \ \ &:=&\ \ \ketbra{0 \cdots 0}{0 \cdots 0} + e^{i \alpha} \ketbra{1 \cdots 1}{1 \cdots 1} \\
\text{X-spider:}& \ \ \ \ \quad \input{Xsp-a.tikz} \ \ &:=&\ \ \ketbra{+ \cdots +}{+ \cdots +} + e^{i \alpha} \ketbra{- \cdots -}{- \cdots -}
\end{align*}
\begin{align*}
\input{ket0.tikz}\ \ &=\ \ \ket{+} + \ket{-} \ = \sqrt{2}\ket{0}
\qquad\qquad
\input{ketplus.tikz}\ \ =\ \ \ket{0} + \ket{1} \ = \sqrt{2}\ket{+} \\
\input{ket1.tikz}\ \ &=\ \ \ket{+} - \ket{-} \ = \sqrt{2}\ket{1}
\qquad\qquad\!
\input{ketminus.tikz}\ \ =\ \ \ket{0} - \ket{1} \ = \sqrt{2}\ket{-}
\end{align*}

\begin{equation*}
\input{Z-a.tikz} \ \ =\ \ \begin{pmatrix} 1 & 0 \\ 0& e^{i\alpha}\end{pmatrix}
\qquad \quad \input{X-a.tikz} \ \ =\ \ \frac12\begin{pmatrix} 1+e^{i\alpha}&   1-e^{i\alpha} \\ 1-e^{i\alpha} & 1+e^{i\alpha}\end{pmatrix}
\end{equation*}
\begin{equation*}
\input{Z-id.tikz}\ \ = \ \ \input{X-id.tikz} \ \ = \ \ \begin{pmatrix}1&0\\0&1\end{pmatrix}
\end{equation*}
\begin{equation*}
\input{Zsp-3.tikz}\ \ = \ \ \begin{pmatrix}1&0\\0&0\\0&0\\0&1\end{pmatrix}
\qquad \quad 
\input{Xsp-3.tikz}\ \ = \ \ \frac{1}{\sqrt{2}}\begin{pmatrix}1&0&0&1\\0&1&1&0\end{pmatrix}
\end{equation*}

\begin{equation*}
	\input{had-euler2.tikz}
\end{equation*}

\begin{equation*}
\input{scalars.tikz}
\end{equation*}

\begin{equation*}
	\input{swap.tikz} \ \ = \ \ \begin{pmatrix}1&0&0&0\\0&0&1&0\\0&1&0&0\\0&0&0&1\end{pmatrix}
\end{equation*}

\begin{equation*}
	\input{cup.tikz} \ \ =\  \ \begin{pmatrix}1\\0\\0\\1\end{pmatrix} \qquad \qquad \qquad\input{cap.tikz} \ \ =\ \ \begin{pmatrix}1&0&0&1\end{pmatrix}
\end{equation*}

\clearpage
\subsection{Unitaries}

Here follows a list of unitaries and their representation in the ZX-calculus. The first column specifies one or several common names for this unitary. The adjoint of the unitary is given by horizontally flipping the diagram, negating all the phases and conjugating the global scalar factor (if applicable).

\begin{centering}
\begin{tabular}{l|c|r}
Name & Diagram & Matrix \\
\hline
\hline
&&\\[-0.25cm]
identity & \input{id.tikz} & $\begin{pmatrix}1&0\\0&1\end{pmatrix}$ 
\\&& \\[-0.25cm] \hline && \\[-0.25cm]
Pauli Z & \input{Z-pi.tikz} & $\begin{pmatrix}1&0\\0&-1\end{pmatrix}$
\\&& \\[-0.25cm] \hline && \\[-0.25cm]
\parbox{2cm}{Pauli X\\ NOT gate} & \input{X-pi.tikz} & $\begin{pmatrix}0&1\\1&0\end{pmatrix}$
\\&& \\[-0.25cm] \hline && \\[-0.25cm]
Pauli Y & $i$ \ \input{Y.tikz} & $\begin{pmatrix}0&-i\\i&0\end{pmatrix}$
\\&& \\[-0.25cm] \hline && \\[-0.25cm]
Hadamard gate & \input{had.tikz} & $\frac{1}{\sqrt{2}}\begin{pmatrix}1&1\\1&-1\end{pmatrix}$
\\&& \\[-0.25cm] \hline && \\[-0.25cm]
S gate &\input{S.tikz} & $\begin{pmatrix}1&0\\0&i\end{pmatrix}$
\\&& \\[-0.25cm] \hline && \\[-0.25cm]
V gate &\input{V.tikz} & $\frac12\begin{pmatrix} 1+i&   1-i \\ 1-i & 1+i\end{pmatrix}$
\\&& \\[-0.25cm] \hline && \\[-0.25cm]
T gate &\input{T.tikz} & $\begin{pmatrix}1&0\\0&e^{i\frac\pi4}\end{pmatrix}$
\\&& \\[-0.25cm] \hline && \\[-0.25cm]
\parbox{2cm}{CNOT gate\\ CX gate} & $\sqrt{2}$ \  \input{CNOT-ZX.tikz} & $\begin{pmatrix}1&0&0&0\\0&1&0&0\\0&0&0&1\\0&0&1&0\end{pmatrix}$ 
\\&& \\[-0.25cm] \hline && \\[-0.25cm]
CZ gate & $\sqrt{2}$ \ \input{CZ.tikz} & $\begin{pmatrix}1&0&0&0\\0&1&0&0\\0&0&1&0\\0&0&0&-1\end{pmatrix}$
\\&& \\[-0.25cm] \hline && \\[-0.25cm]
\parbox{3cm}{Ising interaction\\ phase gadget} & $\sqrt{2}e^{-i\frac\alpha2}$ \  \input{phase-gadget-2.tikz} & \parbox{3cm}{$\exp(-i\frac{\alpha}{2} Z\otimes Z) = $ 
\\ $e^{-i\frac{\alpha}{2}}\begin{pmatrix}1&0&0&0\\0&e^{i\alpha}&0&0\\0&0&e^{i\alpha}&0\\0&0&0&1\end{pmatrix}$}
\end{tabular}
\end{centering}

\clearpage
\subsection{Basic Rewrite rules}

Below is a table of rewrite rules for ZX-diagrams. All the rules also hold with the colours interchanged and the inputs and outputs permuted arbitrarily. See also Figure~\ref{fig:zx-rules}.

\begin{changemargin}{-1cm}{-1cm}
\begin{tabular}{l|c|c}
Name & Rewrite rule & Description \\
\hline
\hline
&&\\
Spider fusion & \input{spider-fusion-Z.tikz} & \parbox{5cm}{Adjacent spiders of the same colour fuse and their phases add}
\\&& \\[0cm] \hline && \\
Identity removal & \input{id-removal-Z.tikz} & \parbox{5cm}{A phasefree spider of arity 2 can be removed.}
\\&& \\[0cm] \hline && \\
\parbox{3cm}{Hadamard-cancellation} & \input{had-had-cancel.tikz} & \parbox{5cm}{Two Hadamard gates in a row cancel each other.}
\\&& \\[0cm] \hline && \\
$\pi$ commutation & \input{pi-copy-rule.tikz} & \parbox{5cm}{A $\pi$ phase copies through a spider of the opposite colour and flips its phase.}
\\&& \\[0cm] \hline && \\
State copy & \input{keta-spider-copy.tikz} & \parbox{5cm}{Copies a computational basis state, $\ket{0}$ or $\ket{1}$, through a spider.}
\\&& \\[0cm] \hline && \\
Colour change & \input{colour-change.tikz} & \parbox{5cm}{The two spiders are related to each other by Hadamard gates. Can also be seen as a rule for commuting a Hadamard gate through a spider.}
\\&& \\[0cm] \hline && \\
\parbox{3cm}{Bialgebra\\Strong complementarity} & \input{bialgebra-rule-many.tikz} & \parbox{5cm}{An adjacent pair of phase-free Z- and X-spiders can be commuted past one another at the cost of potentially introducing many more spiders.}
\\&& \\[0cm] \hline && \\
Hopf & \input{hopf-rule.tikz} & \parbox{5cm}{When spiders of opposite colour are connected by more than one wire, we can remove those excess wires pairwise.}
\end{tabular}

\end{changemargin}

\clearpage
\subsection{Derived rewrite rules}

Below is a table of rewrite rules for ZX-diagrams. All the rules also hold with the colours interchanged and the inputs and outputs permuted arbitrarily.

\begin{changemargin}{-3cm}{-3cm}
\begin{tabular}{l|c|c}
Name & Rewrite rule & Description \\
\hline
\hline
&&\\
\parbox{3cm}{Hadamard self-loop removal} & \input{had-self-loop-short.tikz} & \parbox{5cm}{A Hadamard gate connected twice to the same spider is absorbed by introducing a $\pi$ phase. Cf.~\eqref{eq:had-self-loop-removal}.}
\\&& \\[0cm] \hline && \\
\parbox{3cm}{Hopf for Hadamard edges} & \input{hopf-had.tikz} & \parbox{5cm}{Spiders of the same type connected multiple times via a Hadamard-edge disconnect. Cf.~\eqref{eq:remove-double-edge}.}
\\&& \\[0cm] \hline && \\
Y-state identity & \input{Y-state-identity.tikz} & \parbox{5cm}{Relates two ways of writing the Pauli Y eigenstates. Cf.~\eqref{eq:s-state-equality}.}
\\&& \\[0cm] \hline && \\
\parbox{3cm}{phase gadget fusion} & \input{phase-gadget-fusion.tikz} & \parbox{5cm}{Two phase gadgets connected to the same set of spiders fuse together. Cf.~\eqref{eq:phase-gadget-fusion}.} 
\\&& \\[0cm] \hline && \\
\parbox{3cm}{Local\\complementation} & \input{lc-simp.tikz} & \parbox{5cm}{A $\frac\pi2$ spider can be removed by complementing the connectivity amongst its neighbours. Cf.~\eqref{eq:lc-simp}.}
\\&& \\[0cm] \hline && \\
Pivot & \input{pivot-simp-smaller.tikz} & \parbox{5cm}{A connected pair of $a\pi$ spiders can be removed by complementing the connectivity amongst their neighbours. Cf.~\eqref{eq:pivot-simp}.}
\end{tabular}

\end{changemargin}

\clearpage
\subsection{ZX-calculus full cheatsheet}
The following rewrite rules hold for all $\alpha,\beta, \alpha_i,\beta_j,\gamma_k\in \R$ and $a\in \{0,1\}$ (up to global non-zero scalar).
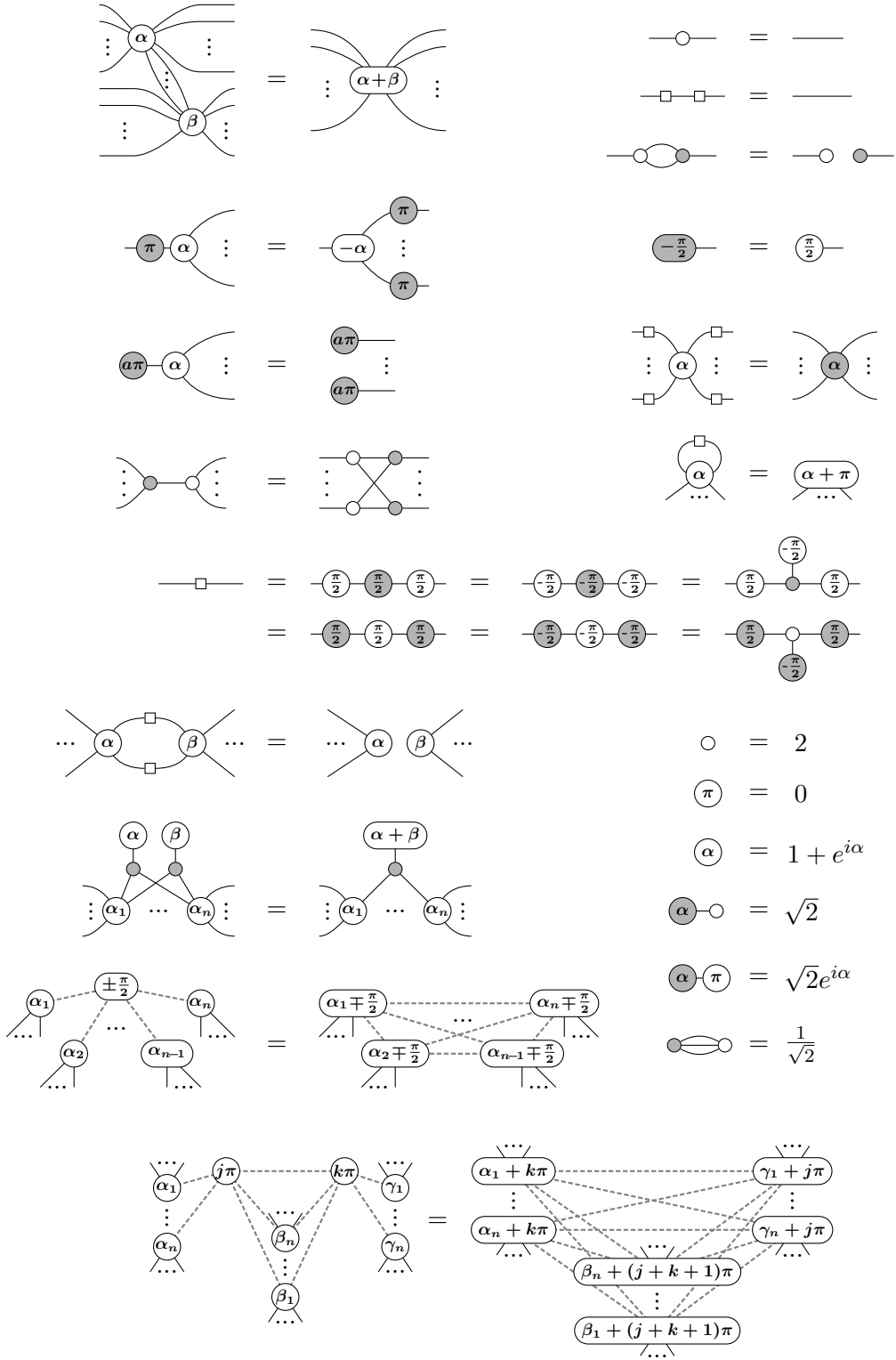
\ctikzfig{ZX-cheatsheet}

\clearpage
\subsection{Circuit identities}\label{app:circuit-identities}
The following circuit identities among X, Z, S, V, H, CNOT, CZ gates and phase gadgets hold (up to global non-zero scalar).
These rules hold for any $\alpha,\beta\in\R$ and also hold with the colours (white and grey) interchanged, with the inputs and outputs switched, or with the phases negated.
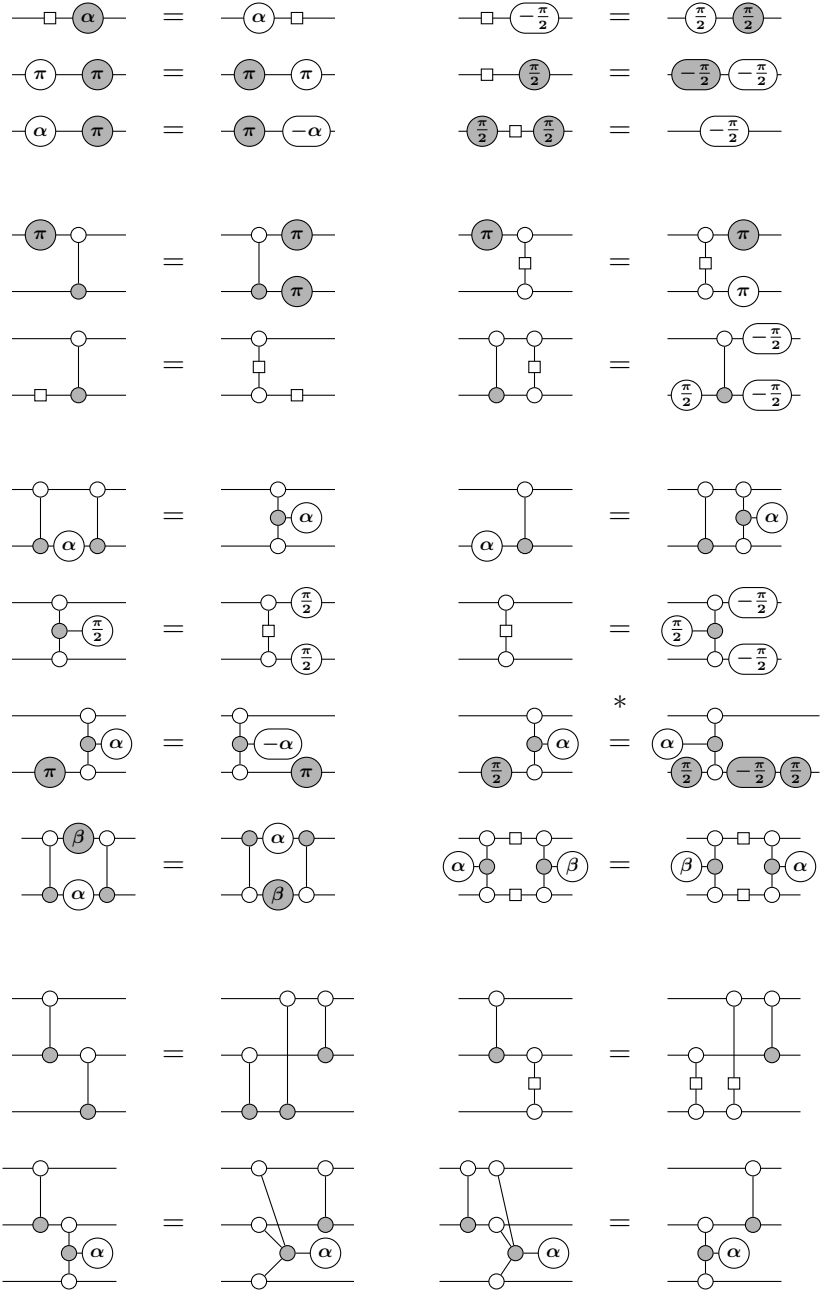
\ctikzfig{circuit-commutation-rules}
The rule * might not look particularly useful, but note that the left part of the diagram on the right expresses the `Pauli gadget' $\exp(-i\frac\alpha2 Z\otimes Y)$.
For more on such rules and Pauli gadgets see~\cite{phaseGadgetSynth}.

\section{Extended generators and rules}

The following additional generators are sometimes considered for ZX-diagrams:
\begin{align*}
	\text{Triangle:}& \ \ \quad \input{triangle.tikz} \ \ &:=& \ \ \ketbra{0}{0} + \ketbra{1}{0} + \ketbra{1}{1} \ \ &= \ \ \begin{pmatrix}1&0\\1&1\end{pmatrix}\\
	\text{$\lambda$-box:}& \ \ \quad \input{lambda-box.tikz} \ \ &:=& \ \ \ketbra{0}{0} + \lambda \ketbra{1}{1} \ \ &= \ \ \begin{pmatrix}1&0\\0&\lambda \end{pmatrix} \\
	\text{H-box:}& \ \ \ \ \input{H-spider.tikz} \ \ &:=& \ \ \sum_{\vec x, \vec y} (-1)^{x_1\ldots x_n y_1\ldots y_m} \ket{x_1\ldots x_n}\bra{y_1\ldots y_m} & \\
	\text{W-spider:}& \ \ \ \ \input{W-spider.tikz} \ \ &:=& \ \ \sum_{\lvert\vec x\rvert + \lvert \vec y\rvert = 1} \ketbra{x_1\cdots x_n}{y_1\cdots y_m}
\end{align*}
For the triangle and $\lambda$-box see Section~\ref{sec:completeness-oxford}, for the H-box see Section~\ref{sec:H-boxes} and for the W-spider see Section~\ref{sec:ZW-calculus}. In the definition of the H-box and the W-spider $\vec x$ and $\vec y$ are bitstrings. In the H-box definition we sum over all bitstrings while in the W-spider definition we sum over all bitstrings where precisely one of the components of the concatenated bitstring $\vec x\vec y$ is $1$ (and thus the rest is $0$).

Rules for the triangle and $\lambda$-box are presented in Figure~\ref{fig:oxford-rules}, rules for the H-box in Figure~\ref{fig:zh-rules} and rules for the W-spider in Figure~\ref{fig:zw-rules}.
The following relations hold between these derived generators and the Z- and X-spiders.

\begin{equation*}
	\input{triangle.tikz} \ = \ \sqrt{2} \ \input{triangle-as-zx.tikz} \ = \ \input{triangle-as-zx2.tikz} \!\! = \ \frac12\ \input{triangle-as-ZH.tikz} \ = \ \input{triangle-as-ZH0.tikz} \ = \ \input{triangle-as-ZW.tikz}
\end{equation*}
\begin{equation*}
	\input{lambda-box.tikz} \ = \ \input{lambda-box-as-zh.tikz} \ = \ \input{lambda-box-as-zw.tikz}
\end{equation*}
\begin{equation*}
	\input{W-state-3.tikz} \ = \ -i\sqrt{\frac{8}{3}}\ \ \input{W-state.tikz} \ = \ \frac{1}{\sqrt{2}}\ \input{W-state-as-ZH.tikz} \ = \ \frac{1}{2\sqrt{2}} \ \input{W-state-as-triangle.tikz}
\end{equation*}
\begin{equation*}
	\input{H-box-3.tikz} \ \ = \ \ \sqrt{2}^5\ \ \input{H-box-3-fourier.tikz} \ \ = \ \ \input{H-box-3-triangle.tikz}
\end{equation*}

\clearpage
\subsection{ZH-calculus cheatsheet}
The following useful identities involving H-boxes hold (up to global non-zero scalar) for any $\alpha\in\R$ and $a,b\in \C$.
\ctikzfig{ZH-cheatsheet}

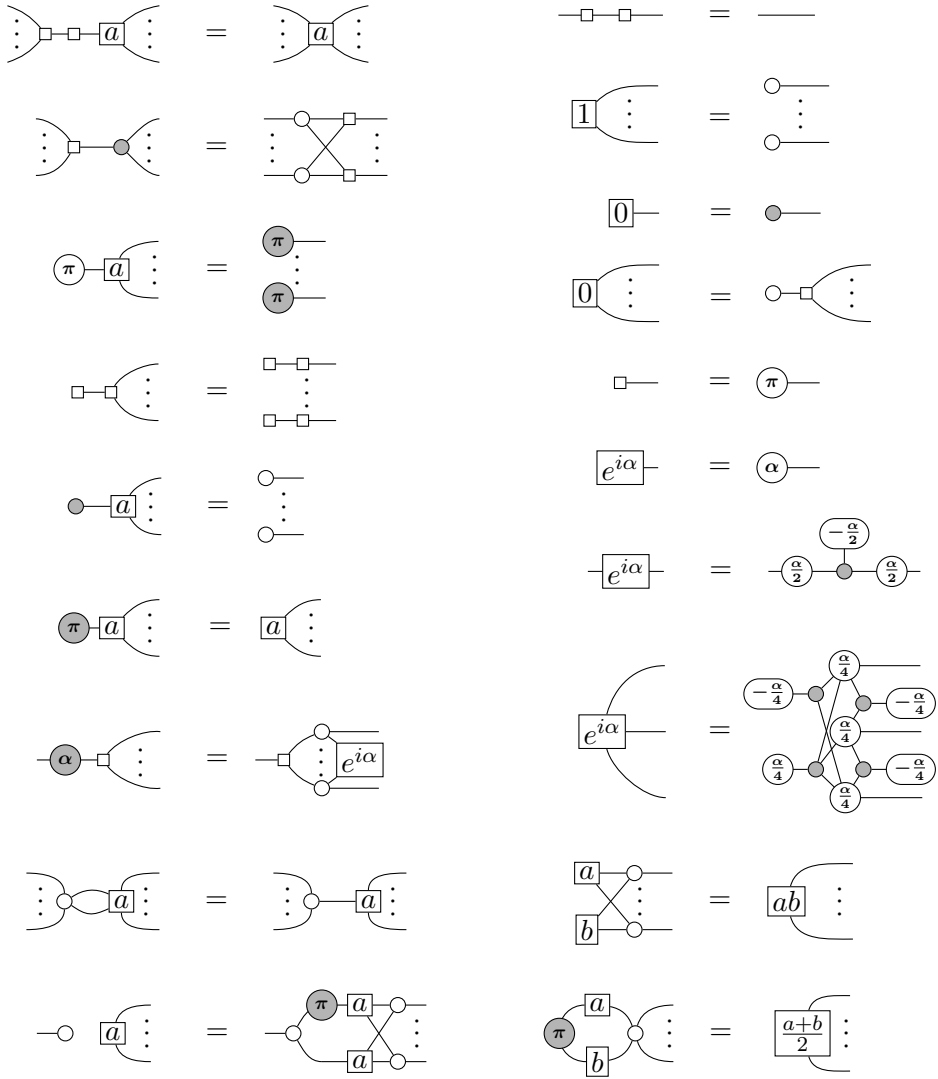
\end{document}

%% file: preamble.tex
\usepackage[utf8]{inputenc}
\usepackage[english]{babel}
\usepackage{amsthm,amsmath}
\usepackage{hyperref}
\usepackage[numbers,sort&compress]{natbib}
\usepackage[all]{hypcap}
\usepackage{stmaryrd}
\usepackage{graphicx}
\usepackage{keycommand}
\usepackage{multicol}
\usepackage{array}

\theoremstyle{definition}

\newtheorem{example*}[theorem]{Example*}
\newtheorem{examples*}[theorem]{Examples*}

\newtheorem{remark*}[theorem]{Remark*}

\newtheorem*{theorem*}{Theorem}
\newtheorem*{corollary*}{Corollary}
\newtheorem*{lemma*}{Lemma}
\newtheorem*{proposition*}{Proposition}

\newenvironment{changemargin}[2]{%
\begin{list}{}{%
\setlength{\topsep}{0pt}%
\setlength{\leftmargin}{#1}%
\setlength{\rightmargin}{#2}%
\setlength{\listparindent}{\parindent}%
\setlength{\itemindent}{\parindent}%
\setlength{\parsep}{\parskip}%
}%
\item[]}{\end{list}}

\usepackage{tikzit}
\input{zh.tikzdefs}

\input{zh.tikzstyles}

\usepackage[all]{hypcap} 

\makeatletter
\newcommand\etc{etc\@ifnextchar.{}{.\@}\xspace}

\makeatother

\newcommand{\intf}[1]{\left\llbracket #1 \right\rrbracket} 

\usepackage{bm}

\newcommand{\abs}[1]{\ensuremath{|#1|}}
\newcommand{\C}{\mathbb{C}}
\newcommand{\R}{\mathbb{R}}
\newcommand{\N}{\mathbb{N}}
\newcommand{\id}{\text{id}}
\newcommand{\tr}{\text{Tr}}

%% file: zh.tikzstyles
\tikzstyle{dot}=[inner sep=0.3mm, minimum width=2mm, minimum height=2mm, draw, shape=circle, font={\footnotesize}, tikzit fill=magenta]
\tikzstyle{white dot}=[dot, fill=white, text depth=-0.2mm, tikzit category=ZH-pf, draw=black]
\tikzstyle{white phase dot}=[minimum size=5mm, font={\footnotesize\boldmath}, shape=rectangle, rounded corners=2mm, inner sep=0.2mm, outer sep=-2mm, scale=0.8, tikzit shape=circle, draw=black, fill=white, tikzit category=ZH-pf, tikzit fill=white, tikzit draw=blue]
\tikzstyle{gray dot}=[dot, fill={rgb,255: red,180; green,180; blue,180}, text depth=-0.2mm, tikzit category=ZH-pf]
\tikzstyle{gray phase dot}=[white phase dot, tikzit shape=circle, tikzit draw=blue, fill={rgb,255: red,180; green,180; blue,180}, font={\footnotesize\boldmath}]
\tikzstyle{hadamard}=[fill=white, draw, inner sep=0.6mm, minimum height=1.5mm, minimum width=1.5mm, shape=rectangle, tikzit shape=rectangle, tikzit category=ZH-pf]
\tikzstyle{small hadamard}=[hadamard]
\tikzstyle{lambda}=[hadamard, fill={rgb,255: red,180; green,180; blue,180}, tikzit shape=rectangle]
\tikzstyle{halfscalar}=[star, fill=black, draw=black, minimum size=8pt, inner sep=0pt]
\tikzstyle{box}=[shape=rectangle, text height=1.5ex, text depth=0.25ex, yshift=0.2mm, fill=white, draw=black, minimum height=3mm, minimum width=5mm, font={\small}]
\tikzstyle{Z dot}=[inner sep=0mm, minimum size=2mm, shape=circle, draw=black, fill={zx_green}, tikzit fill=green]
\tikzstyle{Z phase dot}=[minimum size=5mm, font={\footnotesize\boldmath}, shape=rectangle, rounded corners=2mm, inner sep=0.2mm, outer sep=-2mm, scale=0.8, tikzit shape=circle, draw=black, fill={zx_green}, tikzit draw=blue, tikzit fill=green]
\tikzstyle{X dot}=[Z dot, shape=circle, draw=black, fill={zx_red}, tikzit fill=red]
\tikzstyle{X phase dot}=[Z phase dot, tikzit shape=circle, tikzit draw=blue, fill={zx_red}, font={\footnotesize\color{black}\boldmath}, tikzit fill=red]
\tikzstyle{H box}=[hadamard]
\tikzstyle{st}=[star, star points=5, fill=white, draw=black, inner sep=1.2pt, line width=1.2pt, tikzit fill=blue, tikzit draw=red, tikzit category=ZH-pf]
\tikzstyle{triangle}=[regular polygon, regular polygon sides=3, fill=white, draw=black, inner sep=0pt, minimum width=1em, tikzit draw=blue, tikzit category=ZH-pf, tikzit fill=cyan]
\tikzstyle{not}=[fill={rgb,255: red,180; green,180; blue,180}, draw=black, shape=circle, font={$\neg$}, dot]
\tikzstyle{vertex}=[inner sep=0mm, minimum size=1mm, shape=circle, draw=black, fill=black]
\tikzstyle{vertex set}=[inner sep=0mm, minimum size=1mm, shape=circle, draw=black, fill=white, font={\footnotesize\boldmath}]
\tikzstyle{wide point}=[fill=white, draw, shape=isosceles triangle, shape border rotate=-90, isosceles triangle stretches=true, inner sep=0pt, minimum width=1.5cm, minimum height=6.12mm, yshift=-0.0mm]
\tikzstyle{medium gray box}=[semilarge box, fill={rgb,255: red,180; green,180; blue,180}]
\tikzstyle{small box}=[rectangle, inline text, fill=white, draw, minimum height=5mm, yshift=-0.5mm, minimum width=5mm, font={\small}]
\tikzstyle{small gray box}=[small box, fill={rgb,255: red,180; green,180; blue,180}]
\tikzstyle{medium box}=[rectangle, inline text, fill=white, draw, minimum height=5mm, yshift=-0.5mm, minimum width=8mm, font={\small}]
\tikzstyle{ddot}=[line width=1.6pt, inner sep=0mm, minimum width=2.5mm, minimum height=2.5mm, draw, shape=circle]
\tikzstyle{dd white}=[ddot, fill=white, tikzit draw=green]
\tikzstyle{dd white phase}=[white phase dot, line width=1.6pt, tikzit draw=yellow]
\tikzstyle{dd gray}=[ddot, fill={rgb,255: red,180; green,180; blue,180}, tikzit draw=green]
\tikzstyle{dd gray phase}=[gray phase dot, line width=1.6pt, tikzit draw=yellow]

\tikzstyle{simple}=[-]
\tikzstyle{hadamard edge}=[-, dashed, dash pattern=on 2pt off 1pt, thick, draw=gray]
\tikzstyle{gray}=[-, draw={blue!60!white}, tikzit draw=blue]
\tikzstyle{blue}=[-, draw={blue!60!white}, tikzit draw=blue]
\tikzstyle{brace edge}=[-, tikzit draw=blue, decorate, decoration={brace,amplitude=1mm,raise=-1mm}]
\tikzstyle{diredge}=[->]
\tikzstyle{not edge}=[-, dashed, dash pattern=on 2pt off 1.5pt, thick, draw={rgb,255: red,255; green,68; blue,68}]
\tikzstyle{double edge}=[-, double, shorten <=-1mm, shorten >=-1mm, double distance=2pt]
\tikzstyle{boldedge}=[-, line width=1.6pt, shorten <=-0.17mm, shorten >=-0.17mm, tikzit draw=blue]

%% file: figures/Z-a.tikz
\begin{tikzpicture}
	\begin{pgfonlayer}{nodelayer}
		\node [style=white phase dot] (0) at (0, 0) {$\alpha$};
		\node [style=none] (1) at (-1.25, 0) {};
		\node [style=none] (2) at (1.25, 0) {};
	\end{pgfonlayer}
	\begin{pgfonlayer}{edgelayer}
		\draw (1.center) to (0);
		\draw (0) to (2.center);
	\end{pgfonlayer}
\end{tikzpicture}

%% file: figures/X-a.tikz
\begin{tikzpicture}
	\begin{pgfonlayer}{nodelayer}
		\node [style=gray phase dot] (0) at (0, 0) {$\alpha$};
		\node [style=none] (1) at (-1.25, 0) {};
		\node [style=none] (2) at (1.25, 0) {};
	\end{pgfonlayer}
	\begin{pgfonlayer}{edgelayer}
		\draw (1.center) to (0);
		\draw (0) to (2.center);
	\end{pgfonlayer}
\end{tikzpicture}

%% file: figures/Z-id.tikz
\begin{tikzpicture}
	\begin{pgfonlayer}{nodelayer}
		\node [style=none] (73) at (-1.25, 0.025) {};
		\node [style=none] (74) at (1.25, 0.025) {};
		\node [style=white dot] (75) at (0, 0.025) {};
	\end{pgfonlayer}
	\begin{pgfonlayer}{edgelayer}
		\draw (73.center) to (74.center);
	\end{pgfonlayer}
\end{tikzpicture}

%% file: figures/X-id.tikz
\begin{tikzpicture}
	\begin{pgfonlayer}{nodelayer}
		\node [style=none] (73) at (-1.25, 0.025) {};
		\node [style=none] (74) at (1.25, 0.025) {};
		\node [style=gray dot] (75) at (0, 0.025) {};
	\end{pgfonlayer}
	\begin{pgfonlayer}{edgelayer}
		\draw (73.center) to (74.center);
	\end{pgfonlayer}
\end{tikzpicture}

%% file: figures/id.tikz
\begin{tikzpicture}
	\begin{pgfonlayer}{nodelayer}
		\node [style=none] (73) at (-1.25, 0.025) {};
		\node [style=none] (74) at (1.25, 0.025) {};
	\end{pgfonlayer}
	\begin{pgfonlayer}{edgelayer}
		\draw (73.center) to (74.center);
	\end{pgfonlayer}
\end{tikzpicture}

%% file: figures/ket0.tikz
\begin{tikzpicture}
	\begin{pgfonlayer}{nodelayer}
		\node [style=gray dot] (15) at (0, 0) {};
		\node [style=none] (16) at (1.25, 0) {};
	\end{pgfonlayer}
	\begin{pgfonlayer}{edgelayer}
		\draw (15) to (16.center);
	\end{pgfonlayer}
\end{tikzpicture}

%% file: figures/ketplus.tikz
\begin{tikzpicture}
	\begin{pgfonlayer}{nodelayer}
		\node [style=white dot] (15) at (0, 0) {};
		\node [style=none] (16) at (1.25, 0) {};
	\end{pgfonlayer}
	\begin{pgfonlayer}{edgelayer}
		\draw (15) to (16.center);
	\end{pgfonlayer}
\end{tikzpicture}

%% file: figures/ket1.tikz
\begin{tikzpicture}
	\begin{pgfonlayer}{nodelayer}
		\node [style=gray phase dot] (15) at (0, 0) {$\pi$};
		\node [style=none] (16) at (1.25, 0) {};
	\end{pgfonlayer}
	\begin{pgfonlayer}{edgelayer}
		\draw (15) to (16.center);
	\end{pgfonlayer}
\end{tikzpicture}

%% file: figures/ketminus.tikz
\begin{tikzpicture}
	\begin{pgfonlayer}{nodelayer}
		\node [style=white phase dot] (15) at (0, 0) {$\pi$};
		\node [style=none] (16) at (1.25, 0) {};
	\end{pgfonlayer}
	\begin{pgfonlayer}{edgelayer}
		\draw (15) to (16.center);
	\end{pgfonlayer}
\end{tikzpicture}

%% file: figures/Zsp-nolegs.tikz
\begin{tikzpicture}
	\begin{pgfonlayer}{nodelayer}
		\node [style=white phase dot] (0) at (0, 0) {$\alpha$};
	\end{pgfonlayer}
\end{tikzpicture}

%% file: figures/Xsp-nolegs.tikz
\begin{tikzpicture}
	\begin{pgfonlayer}{nodelayer}
		\node [style=gray phase dot] (0) at (0, 0) {$\alpha$};
	\end{pgfonlayer}
\end{tikzpicture}

%% file: figures/cup.tikz
\begin{tikzpicture}
	\begin{pgfonlayer}{nodelayer}
		\node [style=none] (12) at (0, -0.75) {};
		\node [style=none] (14) at (0, 0.75) {};
		\node [style=none] (15) at (0, 0.75) {};
		\node [style=none] (16) at (0, -0.75) {};
	\end{pgfonlayer}
	\begin{pgfonlayer}{edgelayer}
		\draw [bend left=90, looseness=1.75] (16.center) to (15.center);
	\end{pgfonlayer}
\end{tikzpicture}

%% file: figures/cap.tikz
\begin{tikzpicture}
	\begin{pgfonlayer}{nodelayer}
		\node [style=none] (12) at (0, -0.75) {};
		\node [style=none] (14) at (0, 0.75) {};
		\node [style=none] (15) at (0, 0.75) {};
		\node [style=none] (16) at (0, -0.75) {};
	\end{pgfonlayer}
	\begin{pgfonlayer}{edgelayer}
		\draw [bend right=90, looseness=1.75] (16.center) to (15.center);
	\end{pgfonlayer}
\end{tikzpicture}

%% file: figures/had.tikz
\begin{tikzpicture}
	\begin{pgfonlayer}{nodelayer}
		\node [style=none] (73) at (-1.25, 0) {};
		\node [style=none] (74) at (1.25, 0) {};
		\node [style=hadamard] (75) at (0, 0) {};
	\end{pgfonlayer}
	\begin{pgfonlayer}{edgelayer}
		\draw (73.center) to (74.center);
	\end{pgfonlayer}
\end{tikzpicture}

%% file: spider-fusion-Z.tikz
\begin{tikzpicture}
	\begin{pgfonlayer}{nodelayer}
		\node [style=white phase dot] (0) at (-2.5, -1.25) {$\beta$};
		\node [style=none] (1) at (-1.25, -0.25) {};
		\node [style=none] (2) at (-4.25, -0.25) {};
		\node [style=none] (3) at (-4.25, -2.25) {};
		\node [style=none] (4) at (-1.25, -2.25) {};
		\node [style=none] (5) at (-1.25, -0.75) {};
		\node [style=none] (6) at (-4.25, -0.75) {};
		\node [style=none, rotate=90] (7) at (-4.5, -1.5) {...};
		\node [style=none, rotate=90] (8) at (-1.5, -1.5) {...};
		\node [style=none] (9) at (-5.25, 1.75) {};
		\node [style=white phase dot] (10) at (-4, 1.25) {$\alpha$};
		\node [style=none] (11) at (-2.25, 2.25) {};
		\node [style=none] (12) at (-5.25, 2.25) {};
		\node [style=none] (13) at (-2.25, 0.25) {};
		\node [style=none, rotate=90] (14) at (-2, 1) {...};
		\node [style=none] (15) at (-2.25, 1.75) {};
		\node [style=none] (16) at (-5.25, 0.25) {};
		\node [style=none, rotate=90] (17) at (-5, 1) {...};
		\node [style=none] (18) at (-1.25, 0.25) {};
		\node [style=none] (19) at (-1.25, 2.25) {};
		\node [style=none] (20) at (-1.25, 1.75) {};
		\node [style=none] (21) at (-5.25, -0.75) {};
		\node [style=none] (22) at (-5.25, -2.25) {};
		\node [style=none] (23) at (-5.25, -0.25) {};
		\node [style=none] (24) at (0, 0) {$=$};
		\node [style=none, rotate=90] (25) at (-3.25, 0) {...};
		\node [style=none] (26) at (1, 1.5) {};
		\node [style=none] (27) at (5, 1) {};
		\node [style=none, rotate=90] (28) at (1.5, -0.25) {...};
		\node [style=none] (29) at (5, -1.5) {};
		\node [style=none, rotate=90] (30) at (4.75, -0.25) {...};
		\node [style=none] (31) at (1, 1) {};
		\node [style=white phase dot] (32) at (3, 0) {$\ \alpha\!+\!\beta\ $};
		\node [style=none] (33) at (5, 1.5) {};
		\node [style=none] (34) at (1, -1.5) {};
	\end{pgfonlayer}
	\begin{pgfonlayer}{edgelayer}
		\draw [style=none, in=-141, out=0, looseness=0.75] (3.center) to (0);
		\draw [style=none, in=180, out=-39, looseness=0.75] (0) to (4.center);
		\draw [style=none, in=180, out=22, looseness=0.75] (0) to (5.center);
		\draw [style=none, in=180, out=39, looseness=0.75] (0) to (1.center);
		\draw [style=none, in=0, out=180, looseness=1.00] (0) to (6.center);
		\draw [style=none, in=150, out=0, looseness=1.00] (2.center) to (0);
		\draw [style=none, in=-141, out=0, looseness=0.75] (16.center) to (10);
		\draw [style=none, in=180, out=-15, looseness=1.00] (10) to (13.center);
		\draw [style=none, in=180, out=22, looseness=1.00] (10) to (15.center);
		\draw [style=none, in=180, out=39, looseness=0.75] (10) to (11.center);
		\draw [style=none, in=0, out=158, looseness=0.75] (10) to (9.center);
		\draw [style=none, in=141, out=0, looseness=0.75] (12.center) to (10);
		\draw [style=none, bend left=15, looseness=1.00] (10) to (0);
		\draw [style=none] (11.center) to (19.center);
		\draw [style=none] (15.center) to (20.center);
		\draw [style=none] (13.center) to (18.center);
		\draw [style=none] (23.center) to (2.center);
		\draw [style=none] (21.center) to (6.center);
		\draw [style=none] (22.center) to (3.center);
		\draw [style=none, bend left=15, looseness=1.00] (0) to (10);
		\draw [style=none, in=-120, out=0, looseness=1.00] (34.center) to (32);
		\draw [style=none, in=180, out=-60, looseness=1.00] (32) to (29.center);
		\draw [style=none, in=180, out=45, looseness=0.75] (32) to (27.center);
		\draw [style=none, in=180, out=60, looseness=1.00] (32) to (33.center);
		\draw [style=none, in=0, out=135, looseness=0.75] (32) to (31.center);
		\draw [style=none, in=120, out=0, looseness=1.00] (26.center) to (32);
	\end{pgfonlayer}
\end{tikzpicture}

%% file: spider-fusion-X.tikz
\begin{tikzpicture}
	\begin{pgfonlayer}{nodelayer}
		\node [style=gray phase dot] (0) at (-2.5, -1.25) {$\beta$};
		\node [style=none] (1) at (-1.25, -0.25) {};
		\node [style=none] (2) at (-4.25, -0.25) {};
		\node [style=none] (3) at (-4.25, -2.25) {};
		\node [style=none] (4) at (-1.25, -2.25) {};
		\node [style=none] (5) at (-1.25, -0.75) {};
		\node [style=none] (6) at (-4.25, -0.75) {};
		\node [style=none, rotate=90] (7) at (-4.5, -1.5) {...};
		\node [style=none, rotate=90] (8) at (-1.5, -1.5) {...};
		\node [style=none] (9) at (-5.25, 1.75) {};
		\node [style=gray phase dot] (10) at (-4, 1.25) {$\alpha$};
		\node [style=none] (11) at (-2.25, 2.25) {};
		\node [style=none] (12) at (-5.25, 2.25) {};
		\node [style=none] (13) at (-2.25, 0.25) {};
		\node [style=none, rotate=90] (14) at (-2, 1) {...};
		\node [style=none] (15) at (-2.25, 1.75) {};
		\node [style=none] (16) at (-5.25, 0.25) {};
		\node [style=none, rotate=90] (17) at (-5, 1) {...};
		\node [style=none] (18) at (-1.25, 0.25) {};
		\node [style=none] (19) at (-1.25, 2.25) {};
		\node [style=none] (20) at (-1.25, 1.75) {};
		\node [style=none] (21) at (-5.25, -0.75) {};
		\node [style=none] (22) at (-5.25, -2.25) {};
		\node [style=none] (23) at (-5.25, -0.25) {};
		\node [style=none] (24) at (0, 0) {$=$};
		\node [style=none, rotate=90] (25) at (-3.25, 0) {...};
		\node [style=none] (26) at (1, 1.5) {};
		\node [style=none] (27) at (5, 1) {};
		\node [style=none, rotate=90] (28) at (1.5, -0.25) {...};
		\node [style=none] (29) at (5, -1.5) {};
		\node [style=none, rotate=90] (30) at (4.75, -0.25) {...};
		\node [style=none] (31) at (1, 1) {};
		\node [style=gray phase dot] (32) at (3, 0) {$\ \alpha\!+\!\beta\ $};
		\node [style=none] (33) at (5, 1.5) {};
		\node [style=none] (34) at (1, -1.5) {};
	\end{pgfonlayer}
	\begin{pgfonlayer}{edgelayer}
		\draw [style=none, in=-141, out=0, looseness=0.75] (3.center) to (0);
		\draw [style=none, in=180, out=-39, looseness=0.75] (0) to (4.center);
		\draw [style=none, in=180, out=22, looseness=0.75] (0) to (5.center);
		\draw [style=none, in=180, out=39, looseness=0.75] (0) to (1.center);
		\draw [style=none, in=0, out=180, looseness=1.00] (0) to (6.center);
		\draw [style=none, in=150, out=0, looseness=1.00] (2.center) to (0);
		\draw [style=none, in=-141, out=0, looseness=0.75] (16.center) to (10);
		\draw [style=none, in=180, out=-15, looseness=1.00] (10) to (13.center);
		\draw [style=none, in=180, out=22, looseness=1.00] (10) to (15.center);
		\draw [style=none, in=180, out=39, looseness=0.75] (10) to (11.center);
		\draw [style=none, in=0, out=158, looseness=0.75] (10) to (9.center);
		\draw [style=none, in=141, out=0, looseness=0.75] (12.center) to (10);
		\draw [style=none, bend left=15, looseness=1.00] (10) to (0);
		\draw [style=none] (11.center) to (19.center);
		\draw [style=none] (15.center) to (20.center);
		\draw [style=none] (13.center) to (18.center);
		\draw [style=none] (23.center) to (2.center);
		\draw [style=none] (21.center) to (6.center);
		\draw [style=none] (22.center) to (3.center);
		\draw [style=none, bend left=15, looseness=1.00] (0) to (10);
		\draw [style=none, in=-120, out=0, looseness=1.00] (34.center) to (32);
		\draw [style=none, in=180, out=-60, looseness=1.00] (32) to (29.center);
		\draw [style=none, in=180, out=45, looseness=0.75] (32) to (27.center);
		\draw [style=none, in=180, out=60, looseness=1.00] (32) to (33.center);
		\draw [style=none, in=0, out=135, looseness=0.75] (32) to (31.center);
		\draw [style=none, in=120, out=0, looseness=1.00] (26.center) to (32);
	\end{pgfonlayer}
\end{tikzpicture}

%% file: figures/X-pi.tikz
\begin{tikzpicture}
	\begin{pgfonlayer}{nodelayer}
		\node [style=gray phase dot] (0) at (0, 0) {$\pi$};
		\node [style=none] (1) at (-1.25, 0) {};
		\node [style=none] (2) at (1.25, 0) {};
	\end{pgfonlayer}
	\begin{pgfonlayer}{edgelayer}
		\draw (1.center) to (0);
		\draw (0) to (2.center);
	\end{pgfonlayer}
\end{tikzpicture}

%% file: figures/Z-pi.tikz
\begin{tikzpicture}
	\begin{pgfonlayer}{nodelayer}
		\node [style=white phase dot] (0) at (0, 0) {$\pi$};
		\node [style=none] (1) at (-1.25, 0) {};
		\node [style=none] (2) at (1.25, 0) {};
	\end{pgfonlayer}
	\begin{pgfonlayer}{edgelayer}
		\draw (1.center) to (0);
		\draw (0) to (2.center);
	\end{pgfonlayer}
\end{tikzpicture}

%% file: figures/bra-Z-alpha.tikz
\begin{tikzpicture}
	\begin{pgfonlayer}{nodelayer}
		\node [style=none] (73) at (0, 0) {};
		\node [style=white phase dot] (76) at (1, 0) {$\alpha$};
	\end{pgfonlayer}
	\begin{pgfonlayer}{edgelayer}
		\draw (73.center) to (76);
	\end{pgfonlayer}
\end{tikzpicture}

%% file: figures/bra-Z-alpha-pi.tikz
\begin{tikzpicture}
	\begin{pgfonlayer}{nodelayer}
		\node [style=none] (73) at (0, 0) {};
		\node [style=white phase dot] (76) at (1.5, 0) {$\ \alpha+\pi~$};
	\end{pgfonlayer}
	\begin{pgfonlayer}{edgelayer}
		\draw (73.center) to (76);
	\end{pgfonlayer}
\end{tikzpicture}

%% file: figures/bra-X-alpha.tikz
\begin{tikzpicture}
	\begin{pgfonlayer}{nodelayer}
		\node [style=none] (73) at (0, 0) {};
		\node [style=gray phase dot] (76) at (1, 0) {$\alpha$};
	\end{pgfonlayer}
	\begin{pgfonlayer}{edgelayer}
		\draw (73.center) to (76);
	\end{pgfonlayer}
\end{tikzpicture}

%% file: figures/bra-X-alpha-pi.tikz
\begin{tikzpicture}
	\begin{pgfonlayer}{nodelayer}
		\node [style=none] (73) at (0, 0) {};
		\node [style=gray phase dot] (76) at (1.5, 0) {$\ \alpha+\pi~$};
	\end{pgfonlayer}
	\begin{pgfonlayer}{edgelayer}
		\draw (73.center) to (76);
	\end{pgfonlayer}
\end{tikzpicture}

%% file: figures/bang-box-example0.tikz
\begin{tikzpicture}
	\begin{pgfonlayer}{nodelayer}
		\node [style=white dot] (0) at (-0.75, -0.5) {};
		\node [style=gray dot] (7) at (0.75, -0.5) {};
	\end{pgfonlayer}
\end{tikzpicture}

%% file: figures/bang-box-example1.tikz
\begin{tikzpicture}
	\begin{pgfonlayer}{nodelayer}
		\node [style=white dot] (0) at (-0.75, -0.5) {};
		\node [style=white dot] (1) at (0, 0.5) {};
		\node [style=none] (2) at (0, 1) {};
		\node [style=gray dot] (7) at (0.75, -0.5) {};
	\end{pgfonlayer}
	\begin{pgfonlayer}{edgelayer}
		\draw (2.center) to (1);
		\draw (1) to (0);
		\draw (7) to (1);
	\end{pgfonlayer}
\end{tikzpicture}

%% file: figures/Z-state.tikz
\begin{tikzpicture}
	\begin{pgfonlayer}{nodelayer}
		\node [style=white dot] (332) at (0, 0) {};
		\node [style=none] (333) at (1, 0) {};
	\end{pgfonlayer}
	\begin{pgfonlayer}{edgelayer}
		\draw (332) to (333.center);
	\end{pgfonlayer}
\end{tikzpicture}

%% file: figures/Z-effect.tikz
\begin{tikzpicture}
	\begin{pgfonlayer}{nodelayer}
		\node [style=white dot] (332) at (0, 0) {};
		\node [style=none] (333) at (-1, 0) {};
	\end{pgfonlayer}
	\begin{pgfonlayer}{edgelayer}
		\draw (332) to (333.center);
	\end{pgfonlayer}
\end{tikzpicture}

%% file: figures/Hbox-1-1.tikz
\begin{tikzpicture}[decoration=brace]
	\begin{pgfonlayer}{nodelayer}
		\node [style=hadamard] (3) at (0, 0) {$a$};
		\node [style=none] (4) at (-1, 0) {};
		\node [style=none] (7) at (1, 0) {};
	\end{pgfonlayer}
	\begin{pgfonlayer}{edgelayer}
		\draw (3) to (4.center);
		\draw (3) to (7.center);
	\end{pgfonlayer}
\end{tikzpicture}

%% file: figures/Hbox-1-1-1.tikz
\begin{tikzpicture}[decoration=brace]
	\begin{pgfonlayer}{nodelayer}
		\node [style=hadamard] (3) at (0, 0) {$-1$};
		\node [style=none] (4) at (-1.25, 0) {};
		\node [style=none] (7) at (1.25, 0) {};
	\end{pgfonlayer}
	\begin{pgfonlayer}{edgelayer}
		\draw (3) to (4.center);
		\draw (3) to (7.center);
	\end{pgfonlayer}
\end{tikzpicture}

%% file: figures/H-a-state.tikz
\begin{tikzpicture}
	\begin{pgfonlayer}{nodelayer}
		\node [style=hadamard] (0) at (0, 0) {$e^{i\alpha}$};
		\node [style=none] (2) at (1.5, 0) {};
	\end{pgfonlayer}
	\begin{pgfonlayer}{edgelayer}
		\draw (0) to (2.center);
	\end{pgfonlayer}
\end{tikzpicture}

%% file: figures/Z-a-state.tikz
\begin{tikzpicture}
	\begin{pgfonlayer}{nodelayer}
		\node [style=white phase dot] (0) at (0, 0) {$\alpha$};
		\node [style=none] (2) at (1.25, 0) {};
	\end{pgfonlayer}
	\begin{pgfonlayer}{edgelayer}
		\draw (0) to (2.center);
	\end{pgfonlayer}
\end{tikzpicture}

%% file: figures/H-state-1.tikz
\begin{tikzpicture}
	\begin{pgfonlayer}{nodelayer}
		\node [style=hadamard] (0) at (0, 0) {$1$};
		\node [style=none] (2) at (1, 0) {};
	\end{pgfonlayer}
	\begin{pgfonlayer}{edgelayer}
		\draw (0) to (2.center);
	\end{pgfonlayer}
\end{tikzpicture}

%% file: figures/H-state.tikz
\begin{tikzpicture}
	\begin{pgfonlayer}{nodelayer}
		\node [style=hadamard] (0) at (0, 0) {};
		\node [style=none] (2) at (1, 0) {};
	\end{pgfonlayer}
	\begin{pgfonlayer}{edgelayer}
		\draw (0) to (2.center);
	\end{pgfonlayer}
\end{tikzpicture}

%% file: figures/Z-pi-state.tikz
\begin{tikzpicture}
	\begin{pgfonlayer}{nodelayer}
		\node [style=white phase dot] (0) at (0, 0) {$\pi$};
		\node [style=none] (2) at (1.25, 0) {};
	\end{pgfonlayer}
	\begin{pgfonlayer}{edgelayer}
		\draw (0) to (2.center);
	\end{pgfonlayer}
\end{tikzpicture}

%% file: figures/triangle.tikz
\begin{tikzpicture}
	\begin{pgfonlayer}{nodelayer}
		\node [style=triangle, rotate=270] (0) at (0, 0) {};
		\node [style=none] (1) at (-1, 0) {};
		\node [style=none] (2) at (1, 0) {};
	\end{pgfonlayer}
	\begin{pgfonlayer}{edgelayer}
		\draw (1.center) to (2.center);
	\end{pgfonlayer}
\end{tikzpicture}

%% file: figures/lambda-box.tikz
\begin{tikzpicture}
	\begin{pgfonlayer}{nodelayer}
		\node [style=none] (73) at (-1.25, 0) {};
		\node [style=none] (74) at (1.25, 0) {};
		\node [style=lambda] (75) at (0, 0) {$\lambda$};
	\end{pgfonlayer}
	\begin{pgfonlayer}{edgelayer}
		\draw (73.center) to (74.center);
	\end{pgfonlayer}
\end{tikzpicture}

%% file: figures/ground-dd.tikz
\begin{tikzpicture}
	\begin{pgfonlayer}{nodelayer}
		\node [style=none] (0) at (0.5, 0) {\ground};
		\node [style=none] (4) at (-0.5, 0) {};
		\node [style=none] (5) at (0.45, 0) {};
	\end{pgfonlayer}
	\begin{pgfonlayer}{edgelayer}
		\draw [style=boldedge] (4.center) to (5.center);
	\end{pgfonlayer}
\end{tikzpicture}

%% file: figures/Y.tikz
\begin{tikzpicture}
	\begin{pgfonlayer}{nodelayer}
		\node [style=white phase dot] (0) at (-0.5, 0) {$\pi$};
		\node [style=none] (1) at (-1.25, 0) {};
		\node [style=none] (2) at (1.25, 0) {};
		\node [style=gray phase dot] (3) at (0.5, 0) {$\pi$};
	\end{pgfonlayer}
	\begin{pgfonlayer}{edgelayer}
		\draw (1.center) to (0);
		\draw (0) to (2.center);
	\end{pgfonlayer}
\end{tikzpicture}

%% file: figures/S.tikz
\begin{tikzpicture}
	\begin{pgfonlayer}{nodelayer}
		\node [style=white phase dot] (0) at (0, 0) {$\frac\pi2$};
		\node [style=none] (1) at (-1.25, 0) {};
		\node [style=none] (2) at (1.25, 0) {};
	\end{pgfonlayer}
	\begin{pgfonlayer}{edgelayer}
		\draw (1.center) to (0);
		\draw (0) to (2.center);
	\end{pgfonlayer}
\end{tikzpicture}

%% file: figures/V.tikz
\begin{tikzpicture}
	\begin{pgfonlayer}{nodelayer}
		\node [style=gray phase dot] (0) at (0, 0) {$\frac\pi2$};
		\node [style=none] (1) at (-1.25, 0) {};
		\node [style=none] (2) at (1.25, 0) {};
	\end{pgfonlayer}
	\begin{pgfonlayer}{edgelayer}
		\draw (1.center) to (0);
		\draw (0) to (2.center);
	\end{pgfonlayer}
\end{tikzpicture}

%% file: figures/T.tikz
\begin{tikzpicture}
	\begin{pgfonlayer}{nodelayer}
		\node [style=white phase dot] (0) at (0, 0) {$\frac\pi4$};
		\node [style=none] (1) at (-1.25, 0) {};
		\node [style=none] (2) at (1.25, 0) {};
	\end{pgfonlayer}
	\begin{pgfonlayer}{edgelayer}
		\draw (1.center) to (0);
		\draw (0) to (2.center);
	\end{pgfonlayer}
\end{tikzpicture}

%% file: figures/triangle-as-ZH0.tikz
\begin{tikzpicture}
	\begin{pgfonlayer}{nodelayer}
		\node [style=none] (1) at (-1, 0) {};
		\node [style=none] (2) at (1.75, 0) {};
		\node [style=hadamard] (3) at (0, 0) {$0$};
		\node [style=gray phase dot] (5) at (1, 0) {$\pi$};
	\end{pgfonlayer}
	\begin{pgfonlayer}{edgelayer}
		\draw (1.center) to (2.center);
	\end{pgfonlayer}
\end{tikzpicture}

%% file: figures/lambda-box-as-zh.tikz
\begin{tikzpicture}
	\begin{pgfonlayer}{nodelayer}
		\node [style=none] (73) at (-1.25, 0) {};
		\node [style=none] (74) at (1.25, 0) {};
		\node [style=white dot] (75) at (0, 0) {};
		\node [style=hadamard] (76) at (0, -1) {$\lambda$};
	\end{pgfonlayer}
	\begin{pgfonlayer}{edgelayer}
		\draw (73.center) to (74.center);
		\draw (76) to (75);
	\end{pgfonlayer}
\end{tikzpicture}

%% file: figures/lambda-box-as-zw.tikz
\begin{tikzpicture}
	\begin{pgfonlayer}{nodelayer}
		\node [style=none] (73) at (-1.25, 0) {};
		\node [style=none] (74) at (1.25, 0) {};
		\node [style=white phase dot] (75) at (0, 0) {$\lambda$};
	\end{pgfonlayer}
	\begin{pgfonlayer}{edgelayer}
		\draw (73.center) to (74.center);
	\end{pgfonlayer}
\end{tikzpicture}